# Tuning from unipolar (p-type or n-type) to ambipolar charge transport efficiency in bowl-shaped perylene-derivatives: a DFT study

Suryakanti Debata,[†] Nataliya N Karaush,[‡] and Sridhar Sahu[*,†]

[†]*High Performance Computing lab, Department of Physics, Indian Institute of Technology (Indian School of Mines) Dhanbad, India*

[‡]*Department of Chemistry and Nanomaterials Science, Bohdan Khmelnytsky National University, Cherkasy, Ukraine*

E-mail: sridharsahu@iitism.ac.in

## Abstract

A series of bowl-shaped dicyclopenta perylene (DCPP) derivatives have been theoretically constructed by indeno-substitution at the peri-positions of DCPP, and suitably functionalizing with aza-, fluoride and imide-groups to enhance the electron transport behavior in the materials. To further ensure the solubility and stability of these organic compounds, we incorporated triethylsilylethynyl (TES) groups in the designed structures. The factors such as degree of aromaticity, electronic structure, molecular packing motif, intermolecular charge coupling, and charge transfer rate are essential in determining the charge transporting ability. The low-lying LUMO-levels (< -4.0 eV) and high electron affinities (> 3.0 eV) of a few DCPPs ensure efficient electron injection from the metal electrodes. These molecules are arranged in bowl-in-bowl columnar packing, which is suitable for facilitating the intermolecular charge transport in the crystal. As a result, we observed enhanced hole-transport behavior in DCPP-9 ($\mu_h$ = 6.296 $cm^2V^{-1}s^{-1}$), electron transport in DCPP-TES-6 ($\mu_e$ = 0.142 $cm^2V^{-1}s^{-1}$) and ambipolar nature of DCPP-12 and DCPP-TES-12. The DCPP-derivatives are also optically active in the UV-visible region, which is confirmed from the TD-DFT analysis. Inspired from their non-centrosymmetric molecular geometry and optical activity, we also investigated their non-linear optical (NLO) responses, which may pave their way towards applications in photonics and optoelectronics.

## 1. Introduction

Bowl-shaped polycyclic aromatic hydrocarbons (PAHs) (commonly known as buckybowls or $\pi$-bowls) are the open shell structures containing both the convex and concave surfaces in its curved geometry. In the last two decades, use of corannulene and sumanene have been extensively studied in organic electronics (OE) research. These two important structures being the wreckages of $C_{60}$ fullerenes, behave in the similar way as the parent fullerenes in terms of their physicochemical properties.[1–4] For instance, corannulene exhibits a dipole moment of 2.1 debye, that helps in boosting the interfacial charge accumulation and hence the mobility of the system.[5–7] Furthermore, the electronic and charge transport properties of the bowl-shaped organic semiconductors (OSCs) can be modulated by altering their molecular packing motifs.[1] Therefore, the bowl conformation is advantageous as compared to the planar PAHs for its ability of controllable molecular packing in building solids, bowl-to-bowl inversion (BWBI) feature, supramolecular assemblies, chiroptical activity, small HOMO-LUMO gap, and electronic conductivity. These interesting properties are originated from the distorted $\pi$-conjugation in the non-planar buckybowls, which gives rise to a minor overlapping of p-orbitals and emergence of s-orbital characteristic in the material.[8,9]

Previously, the stereochemical properties of OSCs in OE were overlooked. However, the past few years witness the increasing interest towards contorted PAH-based OE.[10–13] In this pursuit, a comparative study has been carried out by Fu *et al.*[14] , where the bowl-shaped (CorDI) compound was found to possess high electronic conductivity (mobility ~ 0.05 $cm^2V^{-1}s^{-1}$). This is accredited to the low lying LUMO level of CorDI (-3.50 eV; experimentally), which is comparatively lower than that of the planar PyDI, indicating a stable charge injection into the CorDI crystal. The electronic conductivity is also attributed to the strong $\pi$-$\pi$ stacking along the bowl-in-bowl and the convex-convex faces in its crystal structure. Similarly, Yin *et al.*[15] designed the bowl-shaped PDI dimers by N-annulation at the bridge sites, which showed enhanced intramolecular charge transport as compared to the dimeric molecule of planar PDIs. Such bowl conformation resulted from the stress strain and the heteroannulation strain along both the N-annulated bay regions.

Besides, bowl-shaped compounds have intriguing optical characteristics that could be beneficial in a variety of optoelectronic applications. As studied by Zhu *et al.*,[16] the heteroatomdoped buckybowls show excellent optoelectronic property, exhibiting optical absorption in the range of 375-490 nm and green fluorescence in the range of 484-496 nm with high quantum yield. In this case, both UV-vis absorption and fluorescence spectra of the heteroatom-doped compounds were blue-shifted due to weaker $\pi$-conjugation in the molecules as compared to the undoped counterpart. Similarly, Armakovic *et al.*[17] revealed the optical activity of Band N-doped sumanenes in the visible region of solar spectrum. Notwithstanding, only a handful of work on bowl-shaped OSCs are reported in literature. Besides, different functional groups such as halogens and imide were reported to have substantial influences on the optoelectronic and charge transport properties of the organic semiconducting materials.[18,19] In a recent review article, Haupt and Lentz[20] summarized the structure-property relationship in a series of coronene derivatives and inferred that the electron-withdrawing substituents had direct influence on the electron affinity (EA) of the material. Moreover, OSCs functionalized with selective functional groups represent remarkable n-type characteristics. For example, Sanyal *et al.*[21] investigated the pristine corannulene and its imide functionalized compound; pentafluorophenylcorannulene-5-monoimide, and observed that both possessed n-type semiconducting behavior with the charge mobilities of $\mu_{electron}$ = 0.1498 $cm^2V^{-1}s^{-1}$ and $\mu_{hole}$ = 0.0031 $cm^2V^{-1}s^{-1}$. Shi *et al.*[22] reported an imide-fused corannulene showing n-type property.

As compared to corannulene and sumanene, the dicyclopenta[ghi, pqr] perylene (DCPP), being a substructure of $C_{70}$ and higher fullerenes, has been less studied due to its synthetic inaccessibility. A single DCPP structure contains one perylene unit with twenty-two $sp^2$ hybridized carbon atoms, including two five-membered carbon rings fused along the bay positions. Incorporation of these five-membered carbon rings develop global anti-aromaticity that forces the molecule to become diradicaloid by retrieving the aromaticity of the perylene core. This problem has been solved by crowding the side carbon atoms of the cyclopenta ring with the bulky chains. Zou *et al.*[23] have reported the synthesis of aryl-substituted dicyclopenta-[4,3,2,1-ghi:4',3',2',1'-pqr]perylene, which carries a paratropic current through the outer ring, consequently showing a global antiaromatic behavior. However, the dicationoic and dianionic counterparts show aromatic behavior, which is verified from the NICS(1)$_{zz}$ calculations. Alongside, the DFT calculation also revealed a low BWBI barrier (6.07 kcal $mol^{-1}$) for the DCPP core, thereby recognizing the advantages of DCPP over the other bowl-shaped structures. Recently, Gao *et al.*[24] have demonstrated the BWBI structures of meta substituted diindeno [4,3,2,1-fghi:4',3',2',1'-opqr] perylene (DIP) compounds, out of which the triethylsilyl-ethynyl (TES) functionalized DIP showed p-type characteristics with the mobility value reaching to 0.31 $cm^2V^{-1}s^{-1}$. Whereas, this compound exhibited ambipolar charge transfer characteristics after co-crystallization with $C_{70}$ fullerene (mobility of hole: 0.40 $cm^2V^{-1}s^{-1}$ and electron: 0.07 $cm^2V^{-1}s^{-1}$), which was further validated from the electronic coupling between the two fragments.

In this work, we present the tuning of hole-to-electron-transporting and ambipolar activity in DCPP-based OSCs through carefully designing their molecular structures. These modelled compounds are roughly classified into three groups as illustrated in **Fig. 1**. Additionally, TES-groups are added to the molecules, concerning their solubility and stability in the operational environment. The structure-property relationship of these DCPP-derivatives is estimated through density functional theory (DFT). Their hypothetical crystal structures are also predicted to estimate the molecular packing arrangements in solid-state. Several factors, such as molecular orbitals, aromaticity, reorganization energies, packing motifs, electronic coupling etc. are taken into consideration when analyzing the charge transport properties of the DCPP-derivatives. The effect of structural changes on non-linear optical behavior of the designed compounds are also investigated.

## 2. Theory and computational details

The optimal functioning of optoelectronic devices based on OSCs is concomitantly aided by efficient charge injection and high mobility of charge carriers. At room temperature, the incoherent hopping model proposed by the Marcus–Hush equation is used to estimate the charge transfer process of π-conjugated organic semiconductors.[25–27] The charge hopping rate (K) using the non-adiabatic Marcus theory is described as follows.

$$K = \frac{V_{ij}}{\hbar}\left(\frac{\pi}{\lambda K_B T}\right)^{\frac{1}{2}} \exp\left(-\frac{(\Delta G_{ij}-\lambda)^2}{4\lambda K_B T}\right) \quad [1]$$

Where $\hbar$, $\lambda$, $K_B$ and T represent the reduced Plank's constant, reorganization energy, Boltzmann constant, and absolute temperature respectively. $V_{ij}$ and $\Delta G_{ij}$ are the charge transfer integral and Gibbs free energy difference between the initial and final states, respectively. However, $\Delta G_{ij}$ goes off to zero in case of self-exchange reactions, making the Marcus-Hush expression free from the $\Delta G$ term.

The following formula can be used to calculate the reorganization energy (λ) using Nelsen's four-point technique.

$$\lambda_{h/e} = \left(E^*_{+/-} - E_{+/-}\right) + \left(E^*_{h/e} - E\right) \quad [2]$$

Where, $\lambda_{h/e}$ refer to the hole/electron reorganization energies respectively. $E^*_{+/-}$, $E_{+/-}$, $E^*_{h/e}$ and E correspond to the energies of charged states in the neutral geometry, charged states in their own relaxed geometries, neutral states in the geometries of cation/anion species, and ground-state energy of the neutral species respectively. Here, we have considered only the internal reorganization energy as the external contribution, which is due to the repolarization of surrounding medium is negligible as compared to the internal geometrical relaxation.

Based on the molecular orbitals of the conjugated organic materials, the charge transfer integral/ electron coupling ($V_{ij}$) with respect to different molecular orbitals of the OSC can be calculated by the formula:

$$V_{ij} = \left| \frac{J_{ij} - \frac{1}{2} S_{ij}(e_i + e_j)}{1 - S_{ij}^2} \right| \quad [3]$$

Where, $e_i$ and $e_j$ are the site energies of the $i^{th}$ and $j^{th}$ molecules, mathematically expressed as: $e_{i/j} = \langle \psi_{i/j} | \hat{H} | \psi_{i/j} \rangle$. For a dimer consisting of two nearest neighboring molecules, the charge transfer integrals ($J_{ij}$) and spatial overlap ($S_{ij}$) matrix elements can be written as; $J_{ij} = \langle \psi_i | \hat{H} | \psi_j \rangle$ and $S_{ij} = \langle \psi_i | \psi_j \rangle$ respectively, $\hat{H}$ being the Kohn-Sham Hamiltonian of that particular dimer, and the states $\psi_{i/j}$ correspond to the HOMO/LUMO (for hole/electron transport respectively) of the isolated dimers i and j.[28]

After determining the charge transfer rate, the Einstein-Smoluchowski relation can be used to calculate the charge-hopping mobilities (μ) in the zero-field limit as depicted in the equation below.

$$\mu = \frac{eD}{K_B T} \quad [4]$$

Here, e and D are the electronic charge and diffusion coefficient of the charge carriers (hole/electron) respectively.

The charge transport in an organic crystal with columnar packing is almost one-dimensional. As a result, the diffusion coefficient (D) of charge carriers has a straightforward relationship with K as follows.

$$D = \frac{1}{2} r^2 K \quad [5]$$

Where, r is the intermolecular distance of the dimer under consideration[29].

Two molecular descriptors are often employed to evaluate the effectiveness of the process of charge injection from an electrode to an OSC. First is the HOMO and LUMO energy alignment w.r.t the work function (ϕ) of the metal electrode, and secondly, the ionization potential (IP) and electron affinity (EA) of the organic molecules. The metal–semiconductor interface being commonly described as a Mott–Schottky barrier, decides the charge injection process, where the barrier height can be determined by the difference between ϕ and the HOMO/LUMO energy of the OSC. Contrarily, IP and EA of OSC are also the important characteristics for predicting the effectiveness of charge injection from the electrodes, as well as the oxidation and reduction process upon exposed to air. The adiabatic/vertical electron affinities (AEAs/VEAs) and ionization potentials (AIP/VIP) of the OSCs can be evaluated as;

$$\mathrm{AIP(VIP)} = \mathrm{E_{+}(E_{+}^{*})} - \mathrm{E} \quad [6]$$

$$\mathrm{AEA(VEA)} = \mathrm{E} - \mathrm{E_{-}(E_{-}^{*})} \quad [7]$$

We use the exchange-correlation hybrid functionals including B3LYP, and PBE0; and dispersion corrected functionals such as B3LYP-D3, and B3LYP-D3(BJ) within the framework of density functional theory (DFT), for optimizing the initial geometries of the molecules. The 6-311G(d,p) triple zeta basis set was used for all the calculations, which is considered as a suitable basis set in balancing the accuracy and computational cost.[30,31] To ensure whether the molecules have reached their real local minima on the potential energy (PE) surface, the vibrational frequencies are calculated. The absence of imaginary frequencies supports the fact that the molecules have reached their stable equilibria.[32,33] The charge reorganization energies are computed using four-point method on the adiabatic PE surface with the above mentioned DFT formalism. Furthermore, the electronic structure parameters such as frontier molecular orbitals (FMOs), adiabatic/vertical electron affinities (AEAs/VEAs) and ionization potentials (AIP/VIP) are calculated in the similar fashion. All the calculations have been performed on Gaussian 09 (E.01) package,[34] except that, the charge transfer integrals ($V_h$/ $V_e$) along different hopping channels are computed on the Amsterdam Density Functional (ADF) package with PW91/TZP level of theory.[35–37] The UV-visible absorption patterns of the DCPP compounds have been simulated in the TD-DFT formalism using polarizable continuum model (PCM), employing TD-CAM-B3LYP/6-311G(d,p) level of theory.[38,39] We considered twenty states of charge transition in the dichloromethane solvent for this calculation.

# 3. Results and discussion

## 3.1. Geometrical analysis

A series of 24 bowl-shaped structures, as shown in **Fig. 1** are designed by taking DCPP as the initial geometry. The designing of these structures was inspired from the report of Huang *et. al.*,[40] where they have demonstrated the synthesis and optical properties of p-trifluoromethylphenyl (TFMP) substituted DCPP-derivatives. The geometrical information of these compounds (DCPP-TEMP-1, 2, and 3) are obtained from the Cambridge crystallographic data centre (CCDC # 1886505, 1886506, 1886507). These structures are further relaxed using the aforesaid computational methodologies (**Table S1** of the supporting information). By comparing the geometrical and electronic properties of the optimized structures with the available experimental data, it is found that the B3LYP/6-311G(d,p) method is more suitable in terms of accuracy and computational cost. Therefore, we employed B3LYP/6-311G(d,p) for the study of the other compounds in this series (**Table S2, S3**). Keeping in mind that the bulky TFMP groups in the bay-positions of DCPP-TFMPs hinder the intermolecular charge transport, we replaced these groups with hydrogen atoms in our designed structures. As a result, the DCPP-1 takes a bowl-shape in contrast to the planar geometry of the experimentally available DCPP-TEMP-1 counterpart. Further, the structures of DCPP-2 and DCPP-3 are obtained by fusing indeno-groups at the peri positions of DCPP-1. Also, we performed the end functionalization in DCPP-3 by fluorination or imidation to arrive at DCPP-4 and 5 respectively (Group-I in **Fig. 1**), whereas, DCPP-6, 7 and 8 are designed by aza-substitution at the bay-positions of DCPP-3, 4, and 5 respectively (Group-II). DCPP-9 to 12 (Group-III) are similarly constructed by integrating hexagonal rings at the ortho-positions of the respective DCPP-structures, as shown in **Fig. 1** and **Table S2**. Also, to ensure solubility and stability of the DCPP compounds, the triethylsilylethynyl (TES) groups are substituted at the meta positions. This substitution may influence the geometry, aromaticity, electronic structures and crystal packing motifs of DCPP-derivatives,[1,41] which will be explored in the succeeding sections.

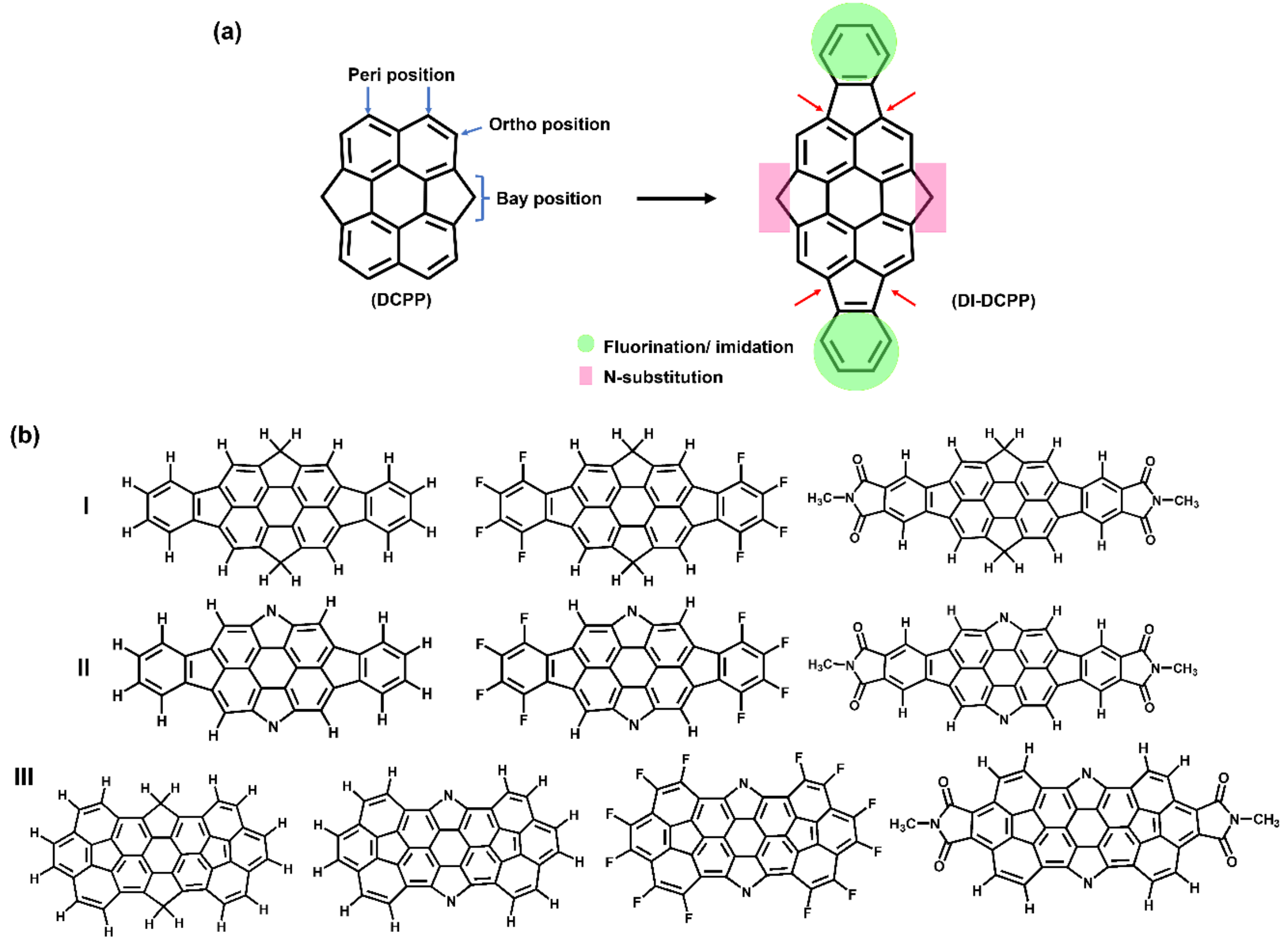

**Figure 1**. Schematic representation of (a) molecular engineering of DCPP-derivatives, (b) classification of the molecules into three groups; pure, fluoride- and imide-substituted diindeno-DCPPs represented as group-I, their aza-substituted counterparts (group-II), and DCPP-derivatives with ring-annulation at the ortho positions (group-III).

The structural deviation of these bowl-shaped DCPP derivatives can be quantified from the bowl-depth; $D_{LS}$ and cone angle; $\alpha = 2\tan^{-1}\left\{{(R - r)}/{D_{LS}}\right\}$ as reported by Roch *et al.*,[2] where, r and R are the radii of the circles associated with two extreme centroids, i.e., base and top of the molecular bowls. The bowl-depths and cone angles of the corresponding DCPPs and DCPP-TESs are listed in Table S4. The $D_{LS}$ values undoubtedly increased with the structural changes from DCPP-1 to 5, moreover the aza-substituted DCPPs (group-II) possess higher $D_{LS}$ values as compared to the undoped molecules. The shapes of group-III molecular bowls are quite different from the group-I and II compounds with higher $D_{LS}$. The α-values were found to decrease with increasing curvature of the molecules. It is also evident that, except DCPP-7, the other compounds do not undergo any significant structural change upon TES-substitution.

## 3.2. Bowl-to-bowl inversion barrier

## 3.3. Aromaticity of the DCPP derivatives (NICS1zz and ACID plot)

### 3.4. Frontier molecular orbitals (FMOs)

The ease of charge injection and air-stability are essential factors for the practical applications of all high-performance OSCs. Therefore, determining the energy levels of FMOs (HOMO and LUMO) is important in estimating the type of materials and their charge injection barriers. The LUMO energy onset for stable electron injection in n-type OSCs is estimated to be -4.0 eV to -4.1 eV.[42,43] However, OSCs with excessively low LUMO energy may create chemical instability in the air. Hence, with reference to the calcium (φ = 2.87 eV to 2.90 eV), magnesium (φ = 3.68 eV) and gold (φ = 5.1 eV) electrodes, which are the best choices for n-type OSCs, the LUMO levels of the organic materials are reported to be in the range of -3.0 eV to -4.4 eV.[44,45] Notwithstanding, there exist few reports on the airstable n-type organic materials possessing LUMO level as low as -4.3 eV (in few naphthalene diimide derivatives),[46] -4.6 eV (for cyano-substituted perylene diimides),[47,48] -4.9 eV (for fluorinated copper pthalocyanines; $F_{16}CuPC$)[42,49,50] etc. Based on the above evidences, we considered the LUMO for air-stable electron injection to be in the range of -4.0 eV to -4.9 eV. As observed in DCPP-1, 2 and 3 **(Fig. 2(a)** and **Table S4**), after indeno-substitutions in the bare DCPP, the LUMO levels are gradually shifted from -1.919 eV to -2.661 eV. With fluorination (DCPP-4) and imidation (DCPP-5) at the ends of DCPP-3, the LUMO levels are further shifted to -3.267 eV and -3.408 eV respectively. This is due to to the electron withdrawing capability of F- and imide-groups in the OSCs. We also observed a similar trend in the aza-substituted compounds (DCPP-6, 7 and 8), where the LUMO levels are down-shifted significantly as compared to their undoped counterparts (DCPP-3, 4 and 5 respectively). The HOMO and LUMO alignments of DCPP-9 differ slightly from those of DCPP-3, resulting in a wider HOMO-LUMO gap. This demonstrates that, although ring fusion at the ortho-positions of DCPP-3 has no substantial effect on its LUMO level, it does improve its stability. Therefore, we moved further with aza-substitution (DCPP-10) and subsequent fluorination and imidation (DCPP-11 and 12 respectively) in the structure of DCPP-9. After the N-substitution, we observe a significant lowering of LUMO level in DCPP-10, as represented in **Fig. 2(a)**. However, after further functionalization, the LUMO levels were shifted much lower, and it is remarkably lower in DCPP-11 (-5.157 eV). Such a low-lying LUMO level in DCPP-11 is regarded to be unsuitable for OSCs due to chemical instability. DCPP-11 was also found to possess the lowest HOMO-LUMO gap (1.413 eV) as compared to the other molecules in the series, where the $E_g$ varied from 1.4-3.1 eV. While comparing the $E_g$ values of group-I, II and III compounds, it is found that the aza-substituted DCPPs (group-II) possess lower $E_g$ as compared to group-I molecules. There is a further decrease of $E_g$ with extending π-conjugation in group-III molecules. A similar trend is also followed by the TES-substituted DCPPs (DCPP-TES-1 to 12) as shown in **Fig. 2(b)** and **Table S4**.

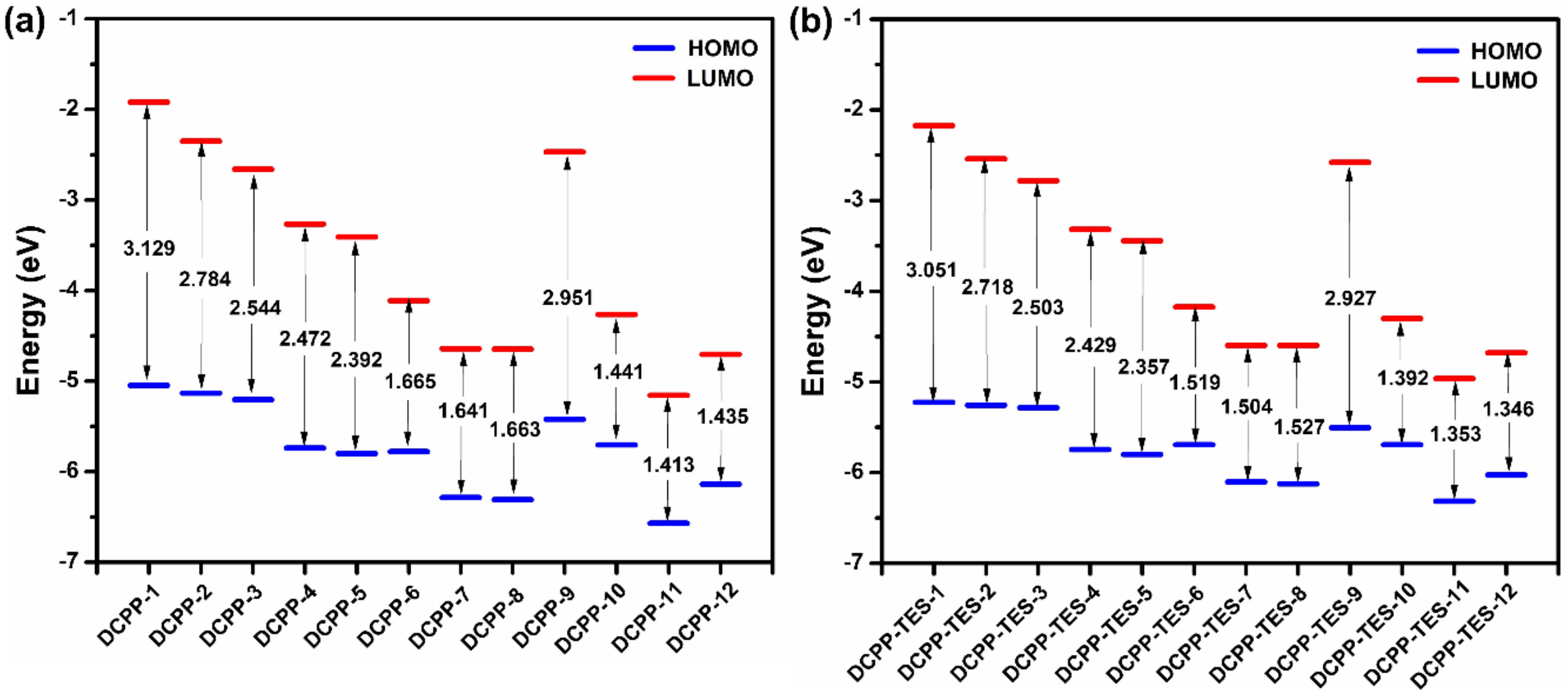

**Figure 2**. HOMO, LUMO energy levels and HOMO-LUMO gaps of (a) DCPP and (b) DCPP-TES derivatives.

The spatial distribution of HOMO and LUMO orbitals over the DCPPs and the DCPP-TES-molecules are presented in **Table S5** and **S6** respectively. In DCPP-1 o 5 it is seen that, both HOMO and LUMO orbitals are delocalized over the entire molecular backbone, indicating the better charge coupling with the neighboring molecules in the crystal. However, after the aza-substitution, the LUMOs are more localized on the DCPP core, which may be the consequence of the electron withdrawing property of N-dopants. After the ring fusion at the ortho-positions of DCPP compounds, the delocalization of the molecular orbitals is apparently visible, which are emerged from π-conjugation in the systems. Similarly, in some DCPP-TES derivatives, we could find the delocalized molecular orbitals along the side chains also. Thus, the TES-substitution may favor the inter-molecular charge transfer in the organic crystal.

### 3.5. IP, EA and reorganization energy

The hole and electron injection properties of an OSC (p-type/n-type) can be related to the IP (the amount of energy it takes to remove an electron in gas-phase) and EA (the amount of energy liberated when one electron is added to a molecule in gas-phase) values of their monomers. The adiabatic/vertical IP and EA (AIP/VIP and AEA/VEA) of the DCPPs and DCPP-TESs are listed in **Table 1**. It is widely believed that the EA should be at least 3.0 eV to allow for facile electron injection to OSC, but no higher than 4.0 eV since negative charges might react with ambient oxidants like water or oxygen. However, the stability can be restrained by the other parameters such as crystal packing and film morphology. Likewise, low IP is necessary for the operation of p-type OSCs[51]. For DCPP compounds in group-I (DCPP-1 to 5), the AIP/VIP and AEA/VEA values are found to be in the range of 6.330-6.839 eV/6.396-6.922 eV and 0.618-2.471 eV/0.511-2.340 eV respectively. Here, the IP values are lowered with the indeno-substitution and then increased with the peripheral F-doping or imide-annulation. However, the EA values gradually increased with indeno- and subsequent peripheral substitutions, favoring electron injection in the modified compounds. After N-doping in the bay-positions of the DCPP bowls (group-II: DCPP-6 to 8), the EA values increased significantly, which are exceeding 3 eV (see **Table 1**). Such high EA values would improve the air-stable electron injection ability of the compounds. As we incorporate more number of rings at the ortho-positions of DCPPs (group-III: DCPP-9 to 12), the EA values of the respective compounds are increased. This indicates the fact that the electron injection is more favored with enhancing π-conjugation. For TES-substituted DCPPs, a similar trend was also followed in the IP and EA values. However, we must be aware that DCPP-11 and DCPP-TES-11 have AEA values > 4 eV, which indicates that their stability may be compromised.

Conversely, the charge transporting capacity of an OSC depends on the reorganization energies ($\lambda_h/\lambda_e$: hole/electron reorganization energies), which are dominated by its structural and electronic characteristics. Among the various factors, the longer π-conjugated chain, non-bonding FMOs, and molecular rigidity are significant in lowering $\lambda_h/\lambda_e$ which is insisted by the Marcus theory (Eq.1) for getting favorable charge transport in the OSC.[52,53] Also, lower $\lambda_e$ improves the electron-transporting ability of the material, which is desirable in our study. The as calculated $\lambda_h$ and $\lambda_e$ values of the DCPP-derivatives are presented in **Table 1**. Out of these, DCPP-TES-10, 11, 12 possess low $\lambda_e$ as compared to $\lambda_h$, suggesting electron-transporting nature of the compounds upon TES substitution in the parent molecules. However, we got equal values of $\lambda_h$ and $\lambda_e$ in DCPP-TES-9 (183.92 meV), which designates the fact that DCPP-TES-9 molecule is favorable in both hole- and electron-transportation. When compared the reorganization energies of DCPP-10, it is found that, there is a close association between $\lambda_e$ (331.81 meV) and $\lambda_h$ (326.93 meV). The charge transport characteristics of the aforementioned DCPP-derivatives can be further verified from their transfer integrals (V), which are to be discussed in the following section.

**Table 1**. Internal reorganization energies ($\lambda_h$ and $\lambda_e$), adiabatic and vertical ionization potentials (AIP and VIP) and electron affinities (AEA and VEA) of the DCPP compounds, computed at B3LYP/6-311G(d,p) level of theory.

| Compound | AIP (eV) | VIP (eV) | AEA (eV) | VEA (eV) | $\lambda_h$ (meV) | $\lambda_e$ (mev) |
|---|---|---|---|---|---|---|
| DCPP-1 | 6.402 | 6.473 | 0.618 | 0.511 | 143.70 | 211.69 |
| DCPP-2 | 6.358 | 6.428 | 1.187 | 1.068 | 140.14 | 237.93 |
| DCPP-3 | 6.330 | 6.396 | 1.599 | 1.474 | 133.37 | 251.17 |
| DCPP-4 | 6.839 | 6.922 | 2.223 | 2.078 | 165.74 | 289.02 |
| DCPP-5 | 6.837 | 6.913 | 2.471 | 2.340 | 152.59 | 262.73 |
| DCPP-6 | 6.905 | 7.028 | 3.054 | 2.855 | 273.61 | 611.41 |
| DCPP-7 | 7.390 | 7.524 | 3.624 | 3.395 | 286.22 | 614.25 |
| DCPP-8 | 7.364 | 7.489 | 3.720 | 3.481 | 258.25 | 611.22 |
| DCPP-9 | 6.498 | 6.581 | 1.427 | 1.324 | 167.10 | 206.11 |
| DCPP-10 | 6.773 | 6.895 | 3.199 | 3.070 | 326.93 | 331.81 |
| DCPP-11 | 7.565 | 7.740 | 4.131 | 3.975 | 410.21 | 379.84 |
| DCPP-12 | 7.122 | 7.265 | 3.743 | 3.582 | 384.24 | 402.05 |
| DCPP-TES-1 | 6.351 | 6.443 | 1.101 | 0.983 | 184.92 | 233.56 |
| DCPP-TES-2 | 6.305 | 6.391 | 1.545 | 1.421 | 172.85 | 245.19 |
| DCPP-TES-3 | 6.269 | 6.348 | 1.853 | 1.730 | 159.98 | 246.99 |
| DCPP-TES-4 | 6.712 | 6.802 | 2.404 | 2.262 | 181.07 | 283.16 |
| DCPP-TES-5 | 6.717 | 6.800 | 2.600 | 2.472 | 166.93 | 256.03 |
| DCPP-TES-6 | 6.582 | 6.781 | 3.296 | 3.086 | 266.09 | 402.05 |
| DCPP-TES-7 | 6.952 | 7.175 | 3.746 | 3.525 | 336.00 | 436.03 |
| DCPP-TES-8 | 6.943 | 7.153 | 3.802 | 3.580 | 343.50 | 450.22 |
| DCPP-TES-9 | 6.466 | 6.558 | 1.640 | 1.536 | 183.92 | 183.92 |
| DCPP-TES-10 | 6.581 | 6.746 | 3.362 | 3.240 | 427.67 | 278.86 |
| DCPP-TES-11 | 7.144 | 7.325 | 4.046 | 3.914 | 432.19 | 327.22 |
| DCPP-TES-12 | 6.813 | 7.006 | 3.841 | 3.682 | 393.08 | 318.23 |

### 3.6. Molecular electrostatic potential (ESP) surface analysis

The charge transport properties and molecular packing in an organic crystal depends on the chemical composition and intermolecular interactions. The ESP surface analysis is a qualitative approach of estimating the non-covalent interactions among the molecules in the organic crystal. The structure dependent ESP profiles of DCPPs (Fig. 3) and DCPP-TESs (Fig. S5) are displayed in color spectra. Here, the red and blue regions represent the negative and positive ESP respectively. In DCPP-1, the negative charges are found to be localized on the ring centers, whereas, these are more concentrated on the indeno-positions in DCPP-2 and DCPP-3. However, the electron withdrawing fluorine atoms create positive ESP in DCPP-4, which is apparent from the electron deficient blue regions of DCPP skeleton. It is believed that, electron delocalization is highly favored by the positive ESP values at the ring centers.[54] With imide substitution (DCPP-5) also, the core DCPP gets positive ESP, while the negative charges are accumulated on the oxygen atoms of both imide-groups. In aza-substituted compounds (group-II), the negative charges are indeed concentrated at the nitrogen centers, and of course on oxygen atoms of the imide groups of DCPP-8. This trend is also followed by the group-III compounds. The TES-substitution in the DCPPs (**Fig. S5**) also maintains similar ESP profiles in their respective compounds.

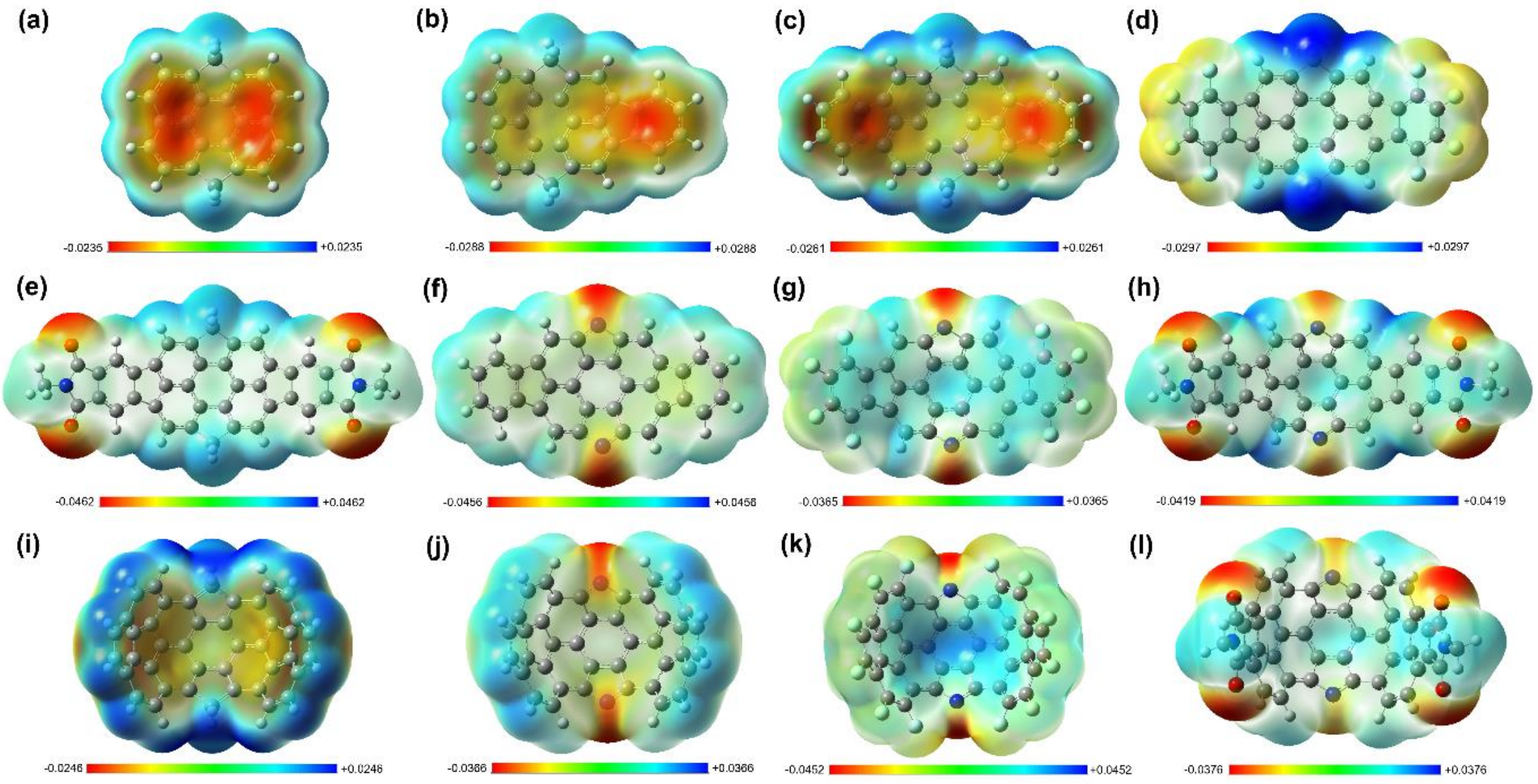

**Figure 3.** Electrostatic potential (ESP) mapping of the DCPP compounds.

### 3.7. Crystal structures and charge transfer

Molecular packing motif, being one of the key factors determining the charge transfer parameters are predicted using Materials Studio program as detailed in **Section S1** of the supporting information. During the calculation, the geometrical configurations of the molecules are maintained rigid, whereas, only the unit cell parameters are optimized to avoid any molecular deformation. The crystal structures of DCPPs and DCPP-TESs showing different hopping pathways, and their corresponding lattice parameters are presented in **Table S8** and **S9** respectively. It is evident that all the compounds are packed in either monoclinic or orthorhombic space groups. In terms of their molecular packing, except for DCPP-2, the other DCPP-compounds in group-I are packed into bowl-in-bowl columnar packing with anti-parallel orientation, which is also followed by DCPP-TES-3 and 4. All the group-II compounds and their TES-structures, except the DCPP-TES-8 follow a similar kind of bowl-in-bowl columnar packing. Also, the deep-bowls of group-III DCPPs follow the same trend, which is an exception in DCPP-9. For both DCPP-9 and DCPP-TES-9, we could find the staggered orientation of molecular bowls, and the consecutive dimers are jointly packed in anti-parallel columnar stacking. Both DCPP-TES-5 and 8 crystals are also exhibiting stagger-like packing with high molecular disorder. It is considered that the columnar crystal packing of bowl-shaped organic molecules is beneficial in improving the transportation of charge carriers.[55]

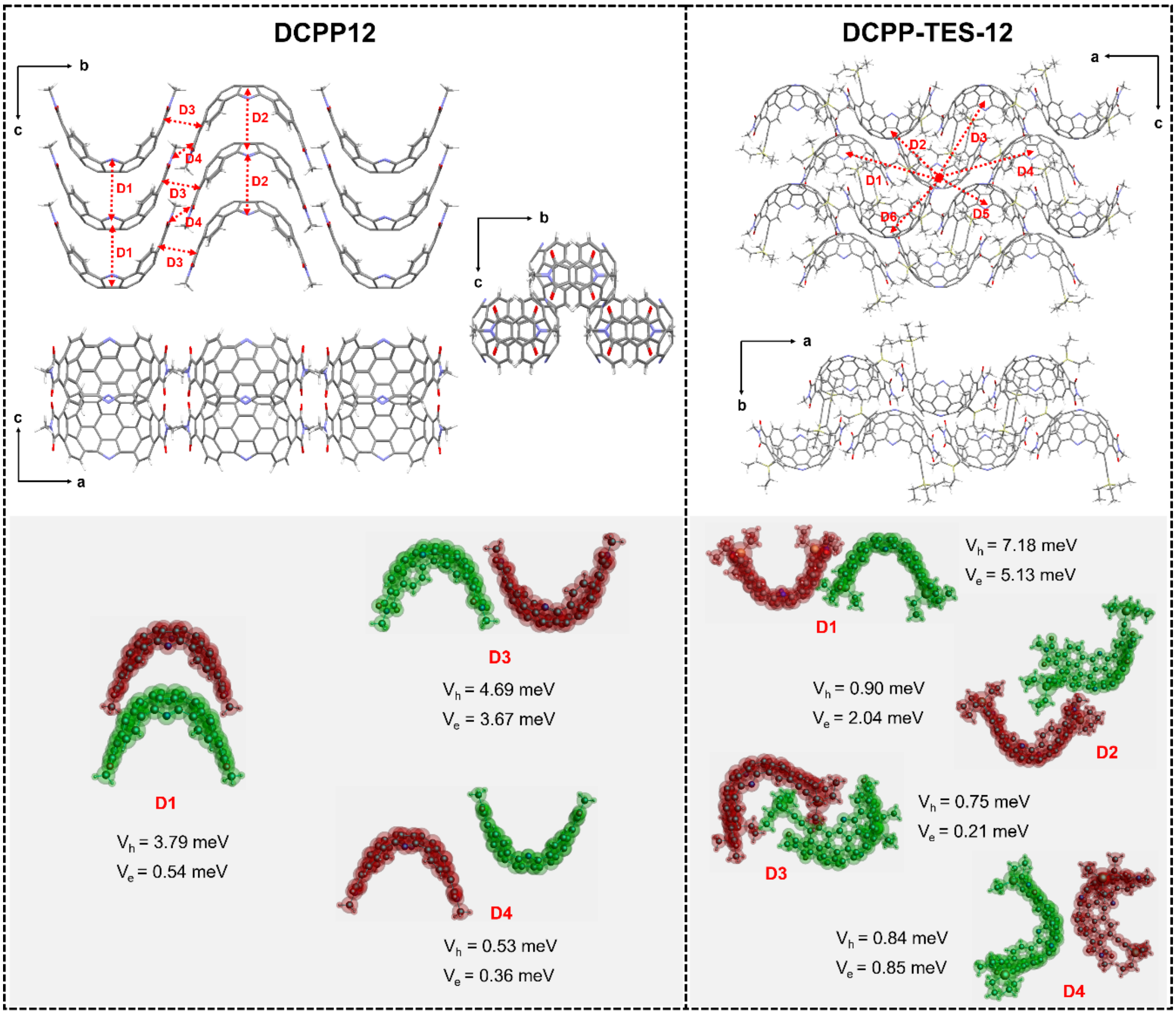

**Figure 4.** Molecular packing motifs viewed along different planes of the crystals (upper panels), and the intermolecular charge couplings ($V_h$ and $V_e$) corresponding to the selective dimers (lower panels) of DCPP-12 (left panel) and DCPP-TES-12 (right panel).

The intermolecular electronic coupling is an important factor in determining the charge transport property of an OSC, that depends on the relative positioning and orientation of the neighboring molecules in the crystal. **Table 2** gathers the charge transport parameters of DCPP- and DCPP-TES-derivatives: hopping pathways, intermolecular distances (r), the corresponding hole/electron couplings ($V_h/V_e$), hole/electron transfer rates ($K_h/K_e$), and diffusion mobilities for holes/electrons ($\mu_h/\mu_e$). The hopping pathways D1, D2, D3, … are presented by the dotted red-lines in the crystal structures of the studied compounds (see **Fig. 4** and **Fig S3-S26**). The electronic couplings of these DCPP-based structures as calculated using ADF: PW91/TZP level were found to be as high as ~85 meV (for $V_h$) and ~78 meV (for $V_e$), which are close to various previously reported figures. For example, the hole transporting characteristics of biphenylene-containing acenes ($V_h$=0-50.04 meV computed at PW91/TZP level)[56], fluoroarene-oligothiophene ($V_h$=26-53 meV at PW91/TZP level)[57], aryl-substituted anthracene ($V_h$=57 meV at PW91/TZP level)[37], benzothiophenes ($V_h$=6.88-69.41 meV at PW91PW91/ def2-TZVP level)[58], octaseleno[8]circulene ($V_h$= 73.61 meV at PW91PW91/6-31G(d,p) level)[59] have been reported. Similarly, the electron couplings of several n-type OSCs are investigated: imide-fused corannulene derivative ($V_e$= 2.28-131 meV at PBE0/DZP level)[12], naphthodithiophene diimide ($V_e$= 12.6-120.1 meV at PW91/TZ2P level)[18], perylene diimide ($V_e$= 0-110.29 meV at PW91PW91/ 6-31G(d))[60], aza[6]helicene ($V_e$= 40-70 meV at B3LYP/6-31G(d))[61], perfluoropentacene ($V_e$= 0-79.36 meV), TIPS-PENTCN$_4$ ($V_e$= 0-44.61 meV) calculated at PW91/TZ2P level[62], and different other OSCs. It is evident from the obtained results that the intermolecular charge coupling is stronger in the DCPPs and DCPP-TESs with bowl-in-bowl concave-convex packing and stacking along convex-convex faces.

Once the information on $V_h/V_e$ and $\lambda_h/\lambda_e$ are obtained, the charge transfer integrals for holes and electrons can be computed using Eq. 1. The calculated values of $K_h$ and $K_e$ as presented in **Table 2** suggest that DCPP-TES-1 has the highest $K_h$ (= $4.70\times10^{13}$ $s^{-1}$) for hole hopping among the other compounds. Nevertheless, the $K_h$ for DCPP-9 and DCPP-TES-4 are found to be in the same order (i.e., $10^{13}$ $s^{-1}$). Besides, DCPP-1, DCPP-2, DCPP-3, DCPP-TES-2 and DCPP-TES-3, which are particularly well suited for hole injection, also have $K_h$ values in the range of $10^{11}$-$10^{12}$ $s^{-1}$, signifying the hole transporting ability of these compounds. When the electron transfer rates of DCPP-derivatives are compared, the maximum $K_e$ for the compounds showing favorable electron injection character (DCPP-6, 7, 8, 10, 12, and the TES-analogues) are found in the order of $10^{8}$-$10^{12}$ $s^{-1}$. However, the DCPP-TES-derivatives (DCPP-TES-6, 7, 8 and 10) reveal higher $K_e$ value as compared to $K_h$, which supports the n-type characteristics in the materials. Furthermore, we observed a balanced hole and electron transport found in DCPP-12 and DCPP-TES-12, which may result in the ambipolar characteristics in the material. The charge transport properties of the DCPP-derivatives are finally validated by their charge carrier mobilities, which are determined using Eq. 4. The p-type characteristics are found in DCPP-1, 2, 3, their TES-derivatives, and DCPP-9 with appropriate hole injection as well as hole transport behavior. The hole mobility of DCPP-9 ($\mu_h$= 6.296 $cm^2V^{-1}s^{-1}$) is found to be the highest in the series. Similarly, the highest $\mu_e$= 0.142 $cm^2V^{-1}s^{-1}$ was discovered in DCPP-TES-6, which is better than many reported values in bowl-shaped OSCs.[12,14,22] The electron-mobilities of DCPP-TES-7, 8 and 10 are found to be $1.40\times10^{-3}$, $3.36\times10^{-5}$ and $4.27\times10^{-2}$ $cm^2V^{-1}s^{-1}$ respectively, revealing the ability of N-doping to induce n-type characteristic in these OSCs. On the other hand, both DCPP-12 ($\mu_h$ / $\mu_e$= $4.90\times10^{-3}$ / $2.50\times10^{-3}$ $cm^2V^{-1}s^{-1}$) and DCPP-TES-12 ($\mu_h$ / $\mu_e$= $1.04\times10^{-2}$ / $1.22\times10^{-2}$ $cm^2V^{-1}s^{-1}$) exhibit balanced hole and electron mobilities, where, the mobility is enhanced by one order of magnitude after TES-substitution. Hence, the charge mobility of DCPP-based compounds can be tuned from unipolar (p-type or n-type) to ambipolar by using appropriate molecular designing techniques and formulations.

**Table 2**. Computed values of charge transport parameters of the DCPP- and DCPP-TES-derivatives: intermolecular distances (r) along different hopping pathways, intermolecular hole/electron couplings ($V_h$/$V_e$), hole/electron transfer rates ($K_h$/$K_e$), and diffusion mobilities for hole/electron ($\mu_h$/$\mu_e$).

| **Compound** | **Pathways** | **r (Å)** | **$\lvert V_h \rvert$ (meV)** | **$\lvert V_e \rvert$ (meV)** | **$K_h$ ($s^{-1}$)** | **$K_e$ ($s^{-1}$)** | **$\mu_h$ ($cm^2V^{-1}s^{-1}$)** | **$\mu_e$ ($cm^2V^{-1}s^{-1}$)** | **$\mu_h^{max}$ ($cm^2V^{-1}s^{-1}$)** | **$\mu_e^{max}$ ($cm^2V^{-1}s^{-1}$)** |
|---|---|---|---|---|---|---|---|---|---|---|
| DCPP-1 | D1, D2 | 3.669 | 3.80 | 77.96 | $1.57\times10^{11}$ | $2.82\times10^{13}$ | $4.10\times10^{-3}$ | 0.738 | $2.04\times10^{-2}$ | 0.738 |
| | D3 | 10.788 | 5.51 | 9.35 | $3.31\times10^{11}$ | $4.05\times10^{11}$ | $7.50\times10^{-2}$ | $9.18\times10^{-2}$ | | |
| | D4 | 10.992 | 2.82 | 0.89 | $8.67\times10^{10}$ | $3.67\times10^{9}$ | $2.04\times10^{-2}$ | $8.63\times10^{-4}$ | | |
| DCPP-2 | D1, D2 | 3.904 | 26.25 | 32.27 | $7.92\times10^{12}$ | $3.54\times10^{12}$ | 0.235 | 0.105 | 0.409 | 0.301 |
| | D3, D5 | 10.883 | 0.58 | 2.65 | $3.87\times10^{9}$ | $2.38\times10^{10}$ | $8.92\times10^{-4}$ | $5.50\times10^{-3}$ | | |
| | D4 | 11.328 | 11.94 | 18.84 | $1.64\times10^{12}$ | $1.21\times10^{12}$ | 0.409 | 0.301 | | |
| | D6 | 10.085 | 2.78 | 1.98 | $8.88\times10^{10}$ | $1.33\times10^{10}$ | $1.76\times10^{-2}$ | $2.60\times10^{-3}$ | | |
| | D7, D8 | 14.513 | 3.60 | 5.48 | $1.49\times10^{11}$ | $1.02\times10^{11}$ | $6.11\times10^{-2}$ | $4.18\times10^{-2}$ | | |
| DCPP-3 | D1, D2 | 4.015 | 9.27 | 0.52 | $1.08\times10^{12}$ | $7.88\times10^{8}$ | $3.40\times10^{-2}$ | $2.47\times10^{-5}$ | 0.167 | $2.41\times10^{-2}$ |
| | D3 | 11.309 | 6.67 | 3.07 | $5.62\times10^{11}$ | $2.75\times10^{10}$ | 0.140 | $6.80\times10^{-3}$ | | |
| | D4 | 11.602 | 0.02 | 0.38 | $5.05\times10^{6}$ | $4.21\times10^{8}$ | $1.32\times10^{-6}$ | $1.10\times10^{-4}$ | | |
| | D5 | 17.184 | 2.50 | 3.79 | $7.89\times10^{10}$ | $4.18\times10^{10}$ | $4.54\times10^{-2}$ | $2.41\times10^{-2}$ | | |
| DCPP-4 | D1, D2 | 4.047 | 25.31 | 21.04 | $5.25\times10^{12}$ | $8.30\times10^{11}$ | 0.167 | $2.65\times10^{-2}$ | 0.167 | $2.65\times10^{-2}$ |
| | D3 | 12.184 | 0.75 | 1.19 | $4.61\times10^{9}$ | $2.66\times10^{9}$ | $1.30\times10^{-3}$ | $7.68\times10^{-4}$ | | |
| | D4 | 11.617 | 1.48 | 7.02 | $1.79\times10^{10}$ | $9.24\times10^{10}$ | $4.70\times10^{-3}$ | $2.43\times10^{-2}$ | | |
| | D5 | 17.992 | 0.97 | 0.85 | $7.71\times10^{9}$ | $1.35\times10^{9}$ | $4.90\times10^{-3}$ | $8.54\times10^{-4}$ | | |
| DCPP-5 | D1, D2 | 4.090 | 15.57 | 7.28 | $2.35\times10^{12}$ | $1.34\times10^{11}$ | $7.65\times10^{-2}$ | $4.40\times10^{-3}$ | 0.016 | 0.225 |
| | D3 | 14.066 | 2.04 | 15.20 | $4.03\times10^{10}$ | $5.85\times10^{11}$ | 0.016 | 0.225 | | |
| | D4 | 14.353 | 0.88 | 0.58 | $7.50\times10^{9}$ | $8.52\times10^{8}$ | 0.003 | $3.42\times10^{-4}$ | | |
| | D5 | 24.516 | 0 | 0 | 0 | 0 | 0 | 0 | | |
| DCPP-6 | D1, D2 | 4.268 | 63.80 | 45.58 | $9.07\times10^{12}$ | $1.17\times10^{11}$ | 0.322 | 0.004 | 0.322 | 0.004 |
| | D3 | 11.734 | 0.26 | 1.31 | $1.51\times10^{8}$ | $9.63\times10^{7}$ | $4.04\times10^{-5}$ | $2.58\times10^{-5}$ | | |
| | D4 | 12.351 | 1.25 | 0.01 | $3.48\times10^{9}$ | $5.61\times10^{3}$ | 0.001 | $1.67\times10^{-9}$ | | |
| | D5 | 15.812 | 1.99 | 1.08 | $8.83\times10^{9}$ | $6.55\times10^{7}$ | $4.30\times10^{-3}$ | $3.19\times10^{-5}$ | | |
| DCPP-7 | D1, D2 | 4.240 | 4.49 | 6.82 | $3.91\times10^{10}$ | $2.53\times10^{9}$ | $1.40\times10^{-3}$ | $8.85\times10^{-5}$ | 0.145 | $2.05\times10^{-4}$ |
| | D3 | 12.204 | 0.90 | 0.33 | $1.57\times10^{6}$ | $5.92\times10^{6}$ | $4.56\times10^{-4}$ | $1.72\times10^{-6}$ | | |

| | | | | | | | | | |
|---|---|---|---|---|---|---|---|---|---|
| | D4 | 11.883 | 16.48 | 3.70 | $5.27\times10^{8}$ | $7.44\times10^{8}$ | 0.145 | $2.05\times10^{-4}$ | | |
| DCPP-8 | D1, D2 | 4.399 | 32.37 | 10.46 | $2.81\times10^{12}$ | $6.14\times10^{9}$ | 0.106 | $2.31\times10^{-4}$ | 0.106 | $2.31\times10^{-4}$ |
| | D3 | 13.181 | 0.88 | 2.11 | $2.08\times10^{9}$ | $2.50\times10^{8}$ | $7.03\times10^{-4}$ | $8.45\times10^{-5}$ | | |
| | D4 | 13.622 | 0.73 | 0.52 | $1.43\times10^{9}$ | $1.52\times10^{7}$ | $5.17\times10^{-4}$ | $5.48\times10^{-6}$ | | |
| DCPP-9 | D1, D2 | 6.264 | 3.72 | 19.44 | $1.12\times10^{11}$ | $1.88\times10^{12}$ | $8.60\times10^{-3}$ | 0.144 | 6.296 | 3.050 |
| | D3 | 13.084 | 0 | 0.09 | 0 | $4.03\times10^{7}$ | 0 | $1.35\times10^{-5}$ | | |
| | D4 | 10.279 | 61.50 | 54.54 | $3.06\times10^{13}$ | $1.48\times10^{13}$ | 6.296 | 3.050 | | |
| | D5, D6 | 8.873 | 3.72 | 14.76 | $1.12\times10^{11}$ | $1.09\times10^{12}$ | $1.72\times10^{-2}$ | 0.166 | | |
| DCPP-10 | D1, D2 | 10.388 | 9.84 | 5.97 | $1.18\times10^{11}$ | $4.10\times10^{10}$ | $2.48\times10^{-2}$ | $8.60\times10^{-3}$ | $3.36\times10^{-2}$ | $1.75\times10^{-2}$ |
| | D3 | 10.781 | 11.04 | 4.06 | $1.48\times10^{11}$ | $1.90\times10^{10}$ | $3.36\times10^{-2}$ | $4.30\times10^{-3}$ | | |
| | D4 | 7.658 | 8.84 | 11.53 | $9.52\times10^{10}$ | $1.53\times10^{11}$ | $1.09\times10^{-2}$ | $1.75\times10^{-2}$ | | |
| DCPP-11 | D1, D2 | 10.798 | 3.77 | 7.29 | $6.89\times10^{9}$ | $3.58\times10^{10}$ | $1.60\times10^{-3}$ | $8.10\times10^{-3}$ | $0.59\times10^{-2}$ | $8.10\times10^{-3}$ |
| | D3 | 11.205 | 7.07 | 1.33 | $2.42\times10^{10}$ | $1.19\times10^{9}$ | $0.59\times10^{-2}$ | $2.92\times10^{-4}$ | | |
| | D4 | 8.245 | 2.15 | 3.83 | $2.24\times10^{9}$ | $9.89\times10^{9}$ | $2.97\times10^{-4}$ | $1.30\times10^{-3}$ | | |
| DCPP-12 | D1, D2 | 7.831 | 3.79 | 0.54 | $9.27\times10^{9}$ | $1.54\times10^{10}$ | $1.10\times10^{-3}$ | $1.80\times10^{-3}$ | $4.90\times10^{-3}$ | $2.50\times10^{-3}$ |
| | D3 | 13.298 | 4.69 | 3.67 | $1.42\times10^{10}$ | $7.13\times10^{9}$ | $4.90\times10^{-3}$ | $2.50\times10^{-3}$ | | |
| | D4 | 15.915 | 0.53 | 0.36 | $1.81\times10^{8}$ | $6.86\times10^{7}$ | $8.94\times10^{-5}$ | $3.38\times10^{-5}$ | | |
| DCPP-TES-1 | D1, D5 | 21.026 | 0.03 | 0.03 | $5.81\times10^{6}$ | $3.20\times10^{6}$ | $5.00\times10^{-6}$ | $2.76\times10^{-6}$ | 1.810 | 0.476 |
| | D2, D8 | 20.629 | 0.03 | 0.07 | $5.81\times10^{6}$ | $1.75\times10^{7}$ | $4.81\times10^{-6}$ | $1.45\times10^{-5}$ | | |
| | D3, D7 | 4.447 | 85.36 | 58.90 | $4.70\times10^{13}$ | $1.24\times10^{13}$ | 1.810 | 0.476 | | |
| | D4, D6 | 22.321 | 0 | 0 | 0 | 0 | 0 | 0 | | |
| DCPP-TES-2 | D1, D5 | 18.541 | 0.01 | 0 | $7.50\times10^{5}$ | 0 | $5.02\times10^{-7}$ | 0 | $8.70\times10^{-3}$ | $9.22\times10^{-2}$ |
| | D2, D8 | 18.867 | 0.08 | 0.05 | $4.80\times10^{7}$ | $7.82\times10^{6}$ | $3.33\times10^{-5}$ | $5.42\times10^{-6}$ | | |
| | D3, D7 | 4.015 | 6.08 | 30.66 | $2.77\times10^{11}$ | $2.94\times10^{12}$ | $8.70\times10^{-3}$ | $9.22\times10^{-2}$ | | |
| | D4, D6 | 19.074 | 0.07 | 0.27 | $3.67\times10^{7}$ | $2.28\times10^{8}$ | $2.60\times10^{-5}$ | $1.61\times10^{-4}$ | | |
| DCPP-TES-3 | D1, D2 | 4.172 | 7.59 | 21.26 | $5.10\times10^{11}$ | $1.38\times10^{12}$ | 0.017 | 0.047 | 0.217 | 0.125 |
| | D3 | 16.486 | 6.81 | 8.81 | $4.10\times10^{11}$ | $2.37\times10^{11}$ | 0.217 | 0.125 | | |
| | D4 | 17.011 | 0.59 | 1.02 | $3.08\times10^{9}$ | $3.18\times10^{9}$ | $1.70\times10^{-3}$ | $1.80\times10^{-3}$ | | |
| DCPP-TES-4 | D1, D2 | 7.600 | 1.10 | 0.03 | $8.21\times10^{9}$ | $1.81\times10^{6}$ | $9.23\times10^{-4}$ | $2.03\times10^{-7}$ | 6.178 | 0.776 |
| | D3 | 9.406 | 25.65 | 40.44 | $4.46\times10^{12}$ | $3.29\times10^{12}$ | 0.769 | 0.566 | | |
| | D4 | 10.322 | 66.27 | 43.15 | $2.98\times10^{13}$ | $3.74\times10^{12}$ | 6.178 | 0.776 | | |
| | D5 | 18.073 | 1.24 | 1.05 | $1.04\times10^{10}$ | $2.22\times10^{9}$ | $6.60\times10^{-3}$ | $4.60\times10^{-4}$ | | |
| DCPP-TES-5 | D1, D2 | 13.686 | 0.01 | 0.04 | $8.09\times10^{5}$ | $4.40\times10^{6}$ | $2.95\times10^{-7}$ | $1.60\times10^{-6}$ | $2.02\times10^{-6}$ | $1.96\times10^{-5}$ |
| | D3 | 14.711 | 0 | 0.02 | 0 | $1.10\times10^{6}$ | 0 | $4.63\times10^{-7}$ | | |
| | D4 | 11.948 | 0.03 | 0.16 | $7.28\times10^{6}$ | $7.03\times10^{7}$ | $2.02\times10^{-6}$ | $1.96\times10^{-5}$ | | |
| DCPP-TES-6 | D1, D2 | 9.742 | 12.05 | 38.11 | $3.55\times10^{11}$ | $7.69\times10^{11}$ | $6.56\times10^{-2}$ | 0.142 | $6.56\times10^{-2}$ | 0.142 |
| | D3 | 12.948 | 0.01 | 0.02 | $2.45\times10^{5}$ | $2.12\times10^{5}$ | $7.98\times10^{-8}$ | $6.91\times10^{-8}$ | | |
| | D4, D5 | 12.372 | 0.11 | 1.03 | $2.96\times10^{7}$ | $5.61\times10^{8}$ | $8.82\times10^{-6}$ | $1.67\times10^{-4}$ | | |
| DCPP-TES-7 | D1, D2 | 8.190 | 0.36 | 0.89 | $1.43\times10^{8}$ | $2.89\times10^{8}$ | $1.86\times10^{-5}$ | $3.78\times10^{-5}$ | $2.52\times10^{-4}$ | $1.40\times10^{-3}$ |
| | D3, D4 | 9.000 | 0.84 | 1.75 | $7.77\times10^{8}$ | $1.12\times10^{9}$ | $1.22\times10^{-4}$ | $1.76\times10^{-4}$ | | |
| | D5, D6 | 10.537 | 1.03 | 4.26 | $1.17\times10^{9}$ | $6.62\times10^{9}$ | $2.52\times10^{-4}$ | $1.40\times10^{-3}$ | | |
| | D7, D8 | 17.384 | 0.06 | 0.01 | $3.96\times10^{6}$ | $3.65\times10^{4}$ | $2.33\times10^{-6}$ | $2.15\times10^{-8}$ | | |
| DCPP-TES-8 | D1, D2 | 21.546 | 0 | 0.01 | 0 | $3.13\times10^{4}$ | 0 | $2.83\times10^{-8}$ | $1.25\times10^{-6}$ | $3.36\times10^{-5}$ |
| | D3, D4 | 18.330 | 0 | 0 | 0 | 0 | 0 | 0 | | |
| | D5, D6 | 11.421 | 0.07 | 0.65 | $4.93\times10^{6}$ | $1.32\times10^{8}$ | $1.25\times10^{-6}$ | $3.36\times10^{-5}$ | | |
| DCPP-TES-9 | D1, D3 | 10.778 | 0.90 | 0.05 | $5.29\times10^{9}$ | $1.63\times10^{6}$ | $1.20\times10^{-3}$ | $3.69\times10^{-6}$ | $3.25\times10^{-2}$ | 4.52 |
| | D2 | 15.271 | 0.23 | 0.56 | $3.46\times10^{8}$ | $2.05\times10^{9}$ | $1.57\times10^{-4}$ | $9.30\times10^{-4}$ | | |
| | D4 | 12.028 | 4.20 | 49.57 | $1.15\times10^{11}$ | $1.61\times10^{13}$ | $3.25\times10^{-2}$ | 4.52 | | |
| | D5, D6 | 17.607 | 0.04 | 0.46 | $1.05\times10^{7}$ | $1.38\times10^{9}$ | $6.31\times10^{-6}$ | $8.34\times10^{-4}$ | | |
| DCPP-TES-10 | D1, D2 | 13.351 | 1.40 | 7.65 | $7.81\times10^{8}$ | $1.23\times10^{11}$ | $2.71\times10^{-4}$ | $4.27\times10^{-2}$ | $2.70\times10^{-3}$ | $4.27\times10^{-2}$ |
| | D3 | 10.395 | 5.64 | 0.87 | $1.27\times10^{10}$ | $1.59\times10^{9}$ | $2.70\times10^{-3}$ | $3.35\times10^{-4}$ | | |
| | D4 | 12.790 | 0.17 | 5.33 | $1.15\times10^{7}$ | $5.98\times10^{10}$ | $3.67\times10^{-6}$ | $1.90\times10^{-2}$ | | |
| DCPP-TES-11 | D1, D2 | 15.462 | 0.07 | 0.09 | $1.87\times10^{6}$ | $9.86\times10^{6}$ | $8.70\times10^{-7}$ | $4.59\times10^{-6}$ | $5.80\times10^{-3}$ | $6.30\times10^{-2}$ |
| | D3, D4 | 7.910 | 0.01 | 4.99 | $3.81\times10^{4}$ | $3.03\times10^{10}$ | $4.64\times10^{-9}$ | $3.70\times10^{-3}$ | | |
| | D5, D6 | 12.975 | 0.52 | 3.96 | $1.03\times10^{8}$ | $1.91\times10^{10}$ | $3.38\times10^{-5}$ | $6.30\times10^{-2}$ | | |
| | D7, D8 | 12.881 | 6.85 | 2.17 | $1.79\times10^{10}$ | $5.73\times10^{9}$ | $5.80\times10^{-3}$ | $1.90\times10^{-3}$ | | |
| | D9, D10 | 13.872 | 0.12 | 0.10 | $5.49\times10^{6}$ | $1.22\times10^{7}$ | $2.06\times10^{-6}$ | $4.56\times10^{-6}$ | | |
| DCPP-TES-12 | D1, D4 | 13.308 | 7.18 | 5.13 | $3.01\times10^{10}$ | $3.55\times10^{10}$ | $1.04\times10^{-2}$ | $1.22\times10^{-2}$ | $1.04\times10^{-2}$ | $1.22\times10^{-2}$ |
| | D2, D6 | 12.063 | 0.90 | 2.04 | $4.73\times10^{8}$ | $5.61\times10^{9}$ | $1.34\times10^{-4}$ | $1.60\times10^{-3}$ | | |
| | D3 | 11.099 | 0.75 | 0.21 | $3.29\times10^{8}$ | $5.95\times10^{7}$ | $7.88\times10^{-5}$ | $1.43\times10^{-5}$ | | |
| | D5 | 13.108 | 0.84 | 0.85 | $4.12\times10^{8}$ | $9.74\times10^{8}$ | $1.38\times10^{-4}$ | $2.34\times10^{-4}$ | | |

### 3.8. Linear and non-linear optical properties

The photoexcitation properties of the DCPP derivatives are analyzed form the UV-visible absorption spectra as presented in **Fig. 4**. Here, twenty states are considered to obtain the photoabsorption spectra using the aforementioned method. As evident from **Fig. 4**, all the studied compounds are optically active and show well-defined high intensity bands in the UV to visible region of the solar spectrum (206-877 nm). These absorption spectra can be roughly divided into three regions, namely, short wavelength (190-300 nm), medium wavelength (300-400 nm) and long wavelength (> 400 nm) regions. The notable absorption wavelengths ($\lambda_{abs}$), oscillator strengths ($f$), excitation energies, and major spectral compositions of the corresponding DCPPs and DCPP-TESs are listed in **Table S9** for reference. In DCPP-1 to 5 and their TES-counterparts, the most intense bands are observed in the short wavelength region, however, the intensities of TES-based compounds are found to be enhanced in this range (**Fig. 4(a, d)**), which is attributed to the extended π-electron delocalization in DCPP-TES structures.[63] On the other hand, the bands are found to be broadened in the medium to long wavelength region, indicating the presence of abundant vibrational and rotational sublevels. It is worth noting that, with indeno-substitutions and subsequent fluorination/imidation, the absorption spectra undergo bathochromic shifting, among which the imidized DCPPs (DCPP-5 and DCPP-TES-5) show better optical activity. In DCPP-5 and DCPP-TES-5, the absorption peaks found in the range of 520-530 nm are corresponding to $S_0 \rightarrow S_1$ transitions, originated from the H → L electronic transitions (**Table S9**). In contrast to the undoped DCPPs, the N-doped DCPPs (**Fig. 4(b)**) are found to possess high intensity bands in the range of 260-300 nm along with few broad peaks in the mid-wavelengths (< 400 nm). However, the TES-side chains help in shifting the spectra to the higher wavelengths. The distinctive bands could be found near 500 nm in DCPP-TES-6 to 8, which are majorly contributed by H-2 → L (~75 %) transitions. As seen in DCPP-9 to 12 and DCPP-TES-9 to 12, the absorption spectra are expanded towards the red region (~ 700 nm) upon structural modification, which is significantly observed in DCPP-TESs. In DCPP-TES-10, 11 and 12, the bands appearing beyond 650 nm are mainly contributed from the H-2→L and H-1→L transitions. As a whole, it can be concluded that the structural modification of DCPP-bowls influences the optical absorption property of the materials.

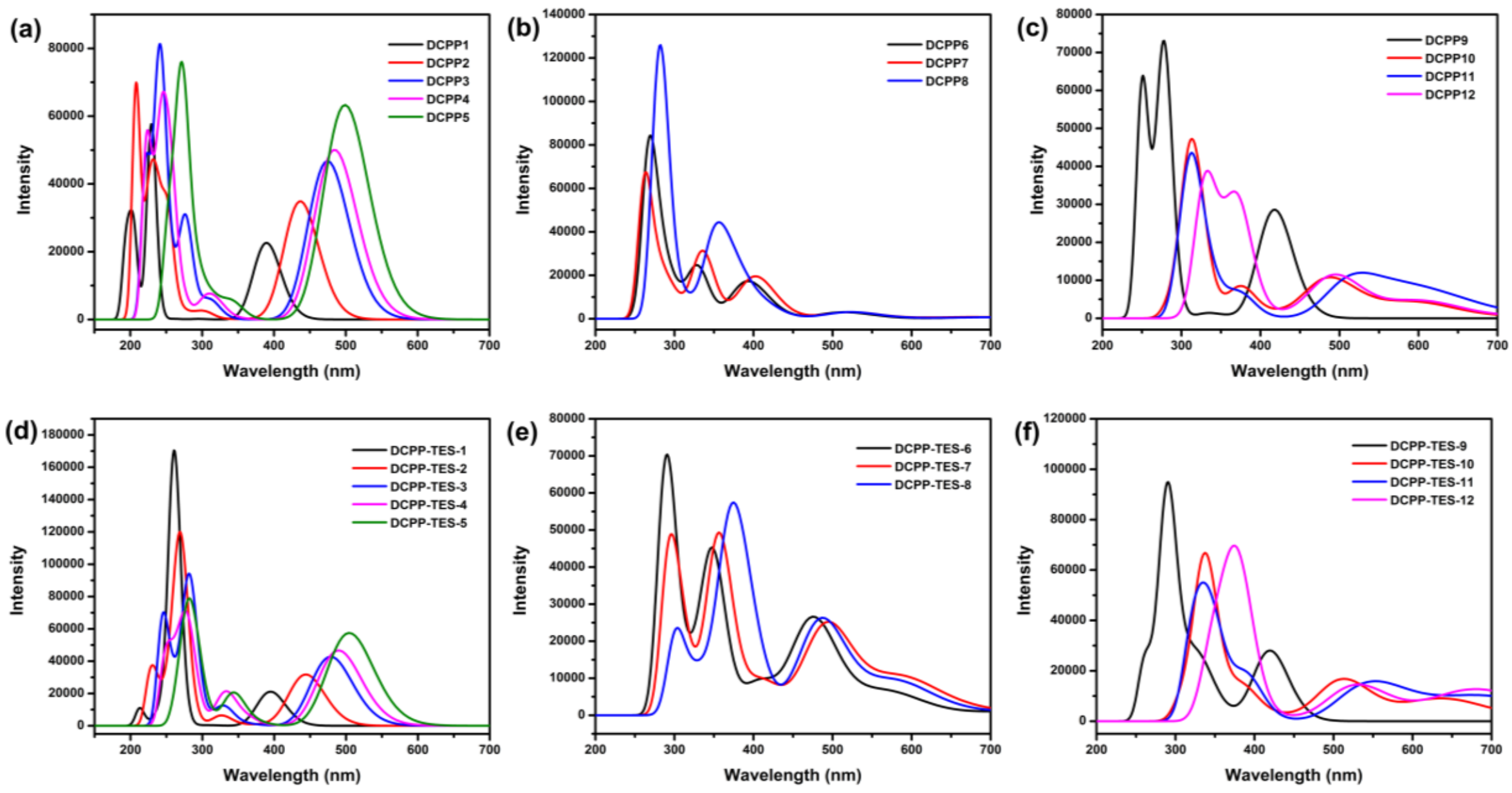

**Figure 5**. UV-visible absorption spectra of (a-c) DCPP and (d-f) DCPP-TES compounds in dichloromethane solvent simulated with PCM model.

## 4. Conclusion

To summarize and conclude, we have demonstrated the structure-property relationship of 24 DCPP-based organic compounds, all of which have bowl-shaped molecular features and mostly packed into bowl-in-bowl columnar stacking in the crystal. The electron injection criteria are satisfied by some of the studied compounds including all group-II derivatives; DCPP-10 and 12 in group-III and their TES-analogues. This is confirmed by their low-lying LUMO level (< -4 eV) and high electron affinity (> 3 eV) values, enabling the air-stability of these DCPP-derivatives under ambient operating condition. However, the fluorinated structures in group-III possess ultra-low LUMO levels and EAs > 4 eV, which may compromise their air-stability. Regarding the charge transport characteristics, the DCPP-derivatives are showing unipolar to ambipolar behavior upon structural modification. The highest $\mu_h$ and $\mu_e$ were found in DCPP-9 (6.296 $cm^2V^{-1}s^{-1}$) and DCPP-TES-6 (0.142 $cm^2V^{-1}s^{-1}$) respectively. On the other hand, DCPP-12 and DCPP-TES-12 showed ambipolar charge transport behavior. DCPP-TES-12 in particular enables the enhancement in both $\mu_h$ ($1.04\times10^{-2}$ $cm^2V^{-1}s^{-1}$) and $\mu_e$ ($1.22\times10^{-2}$ $cm^2V^{-1}s^{-1}$) by ten-times as compared to DCPP-12, making it suitable for ambipolar transistor applications. The DCPP-derivatives are also optically active in the UV-visible region, which is confirmed from the TD-DFT analysis. Inspired from their non-centrosymmetric molecular geometry and optical activity, we also investigated their non-linear optical (NLO) responses, which may pave their way towards applications in photonics and optoelectronics.

## Acknowledgement

SD and SS acknowledge Indian Institute of Technology (ISM), Dhanbad for research facilities and financial support. We also acknowledge Prof. B. R. K. Nanda, Department of Physics, IIT Madras and HPCE, IIT Madras for providing the computational facilities.

## Supporting Information Available

It contains (a) comparison of DFT methods; (b) optimized geometry of DCPP derivatives; (c) bowl-depth and cone angle of DCPP derivatives; (d) FMO distribution; (e) ESP map of DCPP derivatives; (f) molecular packing motifs; (g) crystallographic information; and (h) Optical absorption details.

# Supporting Information

**Table S1**. Comparison of the frontier molecular orbital (FMO) energies of DCPP-TFMP-1, 2, and 3, calculated at different density functional theory (DFT) methods using 6-311G(d,p) basis set, with the experimental values. (B3LYP: 20% HF, PBE0: 25% HF).

| **Compound** | **Method** | **HOMO (eV)** | **LUMO (eV)** | **$E_g$ (eV)** |
|---|---|---|---|---|
| | Expt. | -5.48 | -3.08 | 2.40 |
| DCPP-1 | B3LYP | -5.666 | -2.664 | 3.002 |
| | PBE0 | -5.851 | -2.552 | 3.299 |
| | B3LYP-D3 | -5.665 | -2.663 | 3.002 |
| | B3LYP-D3(BJ) | -5.667 | -2.665 | 3.002 |
| | Expt. | -5.50 | -3.29 | 2.21 |
| DCPP-2 | B3LYP | -5.662 | -2.967 | 2.695 |
| | PBE0 | -5.841 | -2.869 | 2.972 |
| | B3LYP-D3 | -5.668 | -2.951 | 2.717 |
| | B3LYP-D3(BJ) | -5.671 | -2.949 | 2.722 |
| | Expt. | -5.46 | -2.70 | 2.76 |
| DCPP-3 | B3LYP | -5.673 | -3.216 | 2.457 |
| | PBE0 | -5.854 | -3.134 | 2.720 |
| | B3LYP-D3 | -5.685 | -3.201 | 2.484 |
| | B3LYP-D3(BJ) | -5.687 | -3.198 | 2.489 |

**Table S2:** Front and side views of B3LYP/6-311G(d,p) optimized molecular geometries of the experimentally available [S1] DCPP-TFMP-1, 2, 3 molecules (TFMP: p-trifluoromethylphenyl groups).

| Compound | Front view | Side view |
|---|---|---|
| DCPP-TFMP-1 | 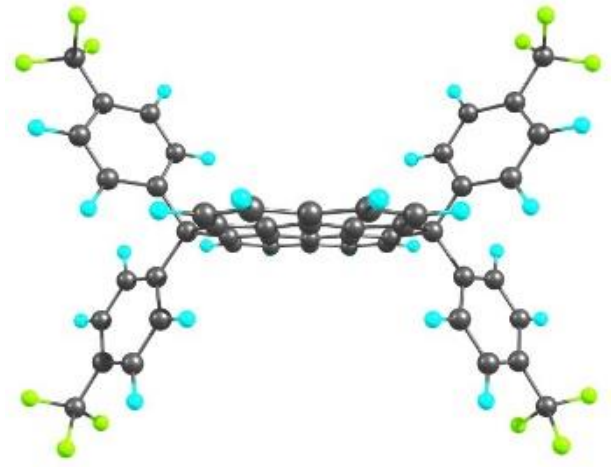 | 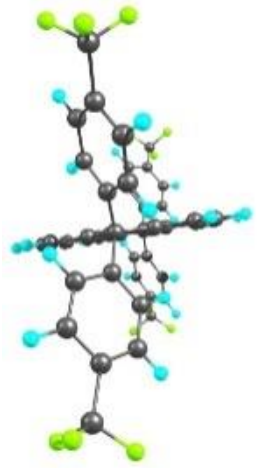 |
| DCPP-TFMP-2 | 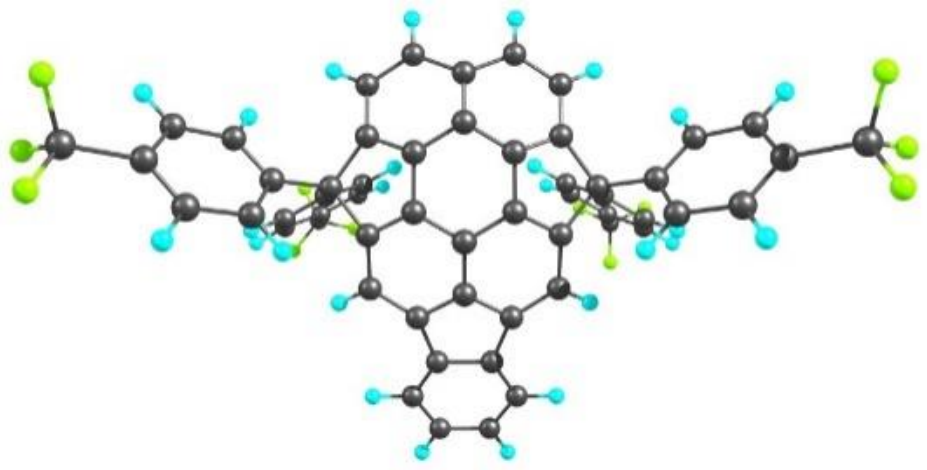 | 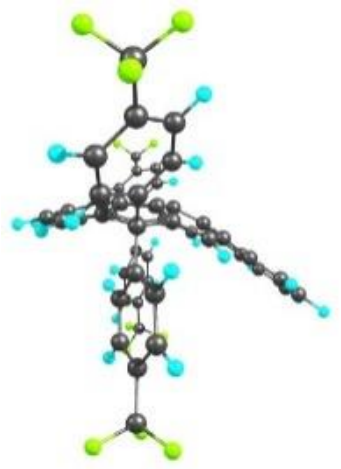 |
| DCPP-TFMP-3 | 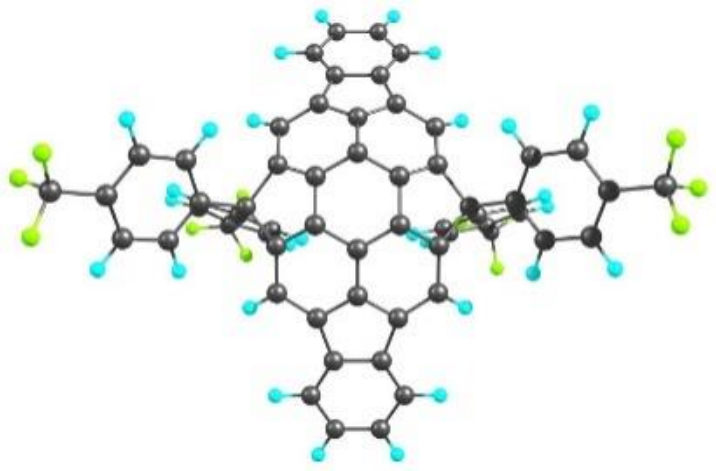 | 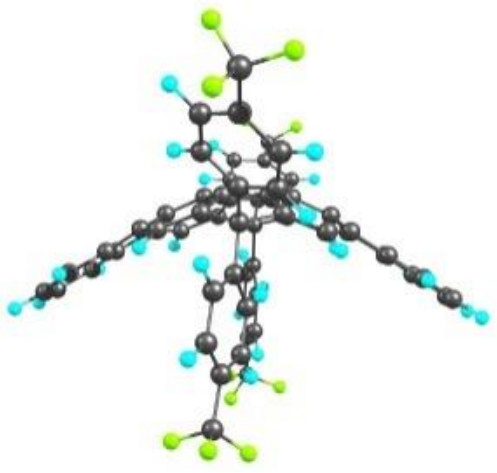 |

**Table S3**. Front and side views of B3LYP/6-311G(d,p) optimized molecular geometries of DCPP compounds.

| Compound | Front view | | Side view |
|---|---|---|---|
| | Concave surface | Convex surface | |
| DCPP-1 | 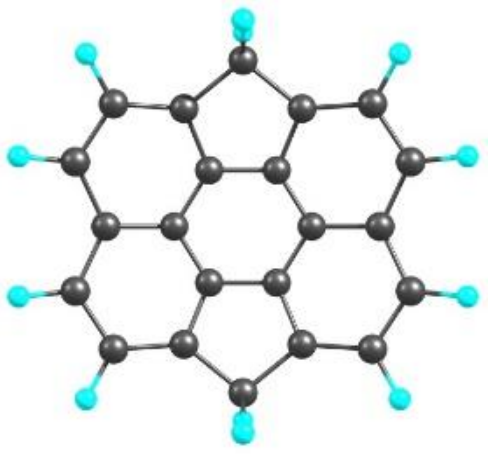 | 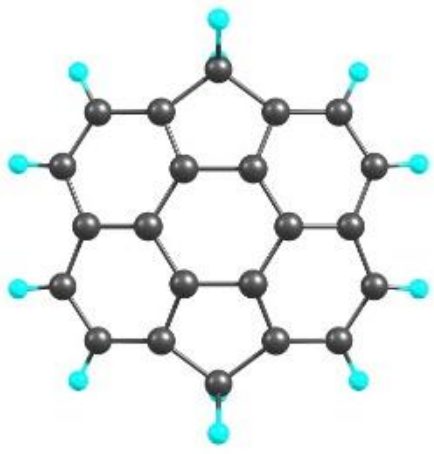 | 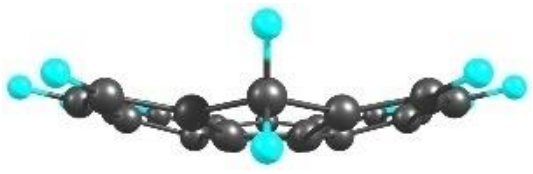 |
| DCPP-2 | 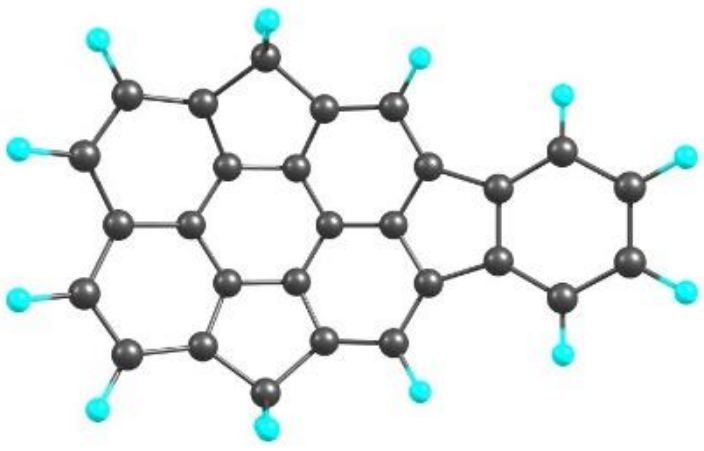 | 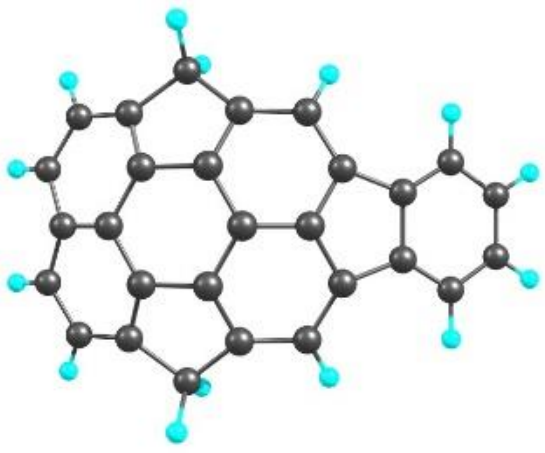 | 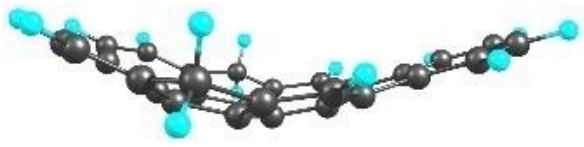 |
| DCPP-3 | 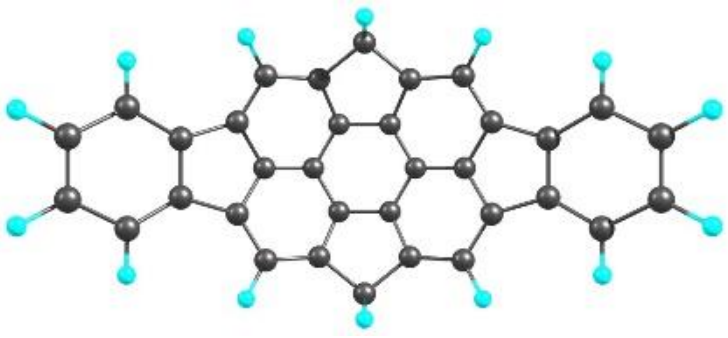 | 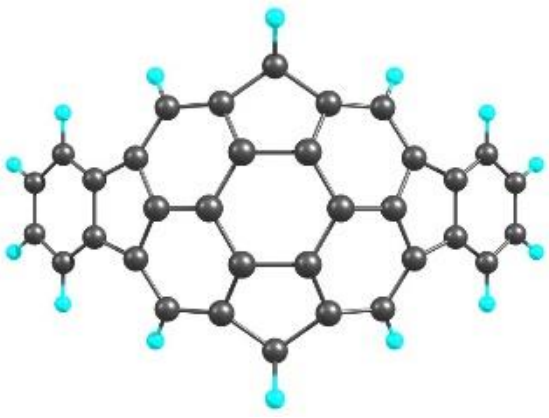 | 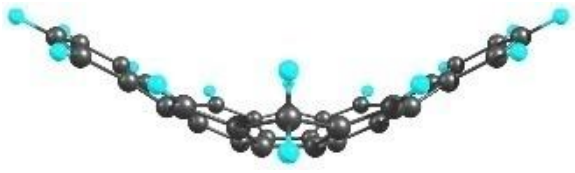 |
| DCPP-4 | 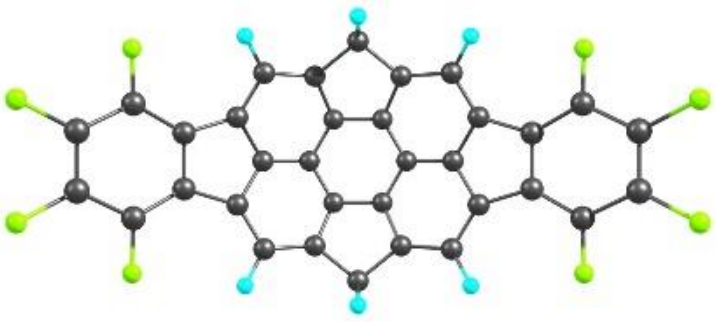 | 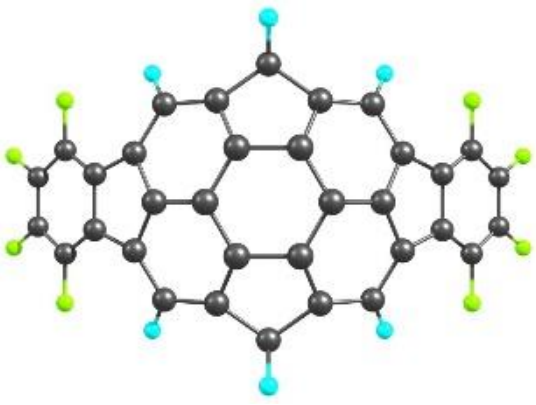 | 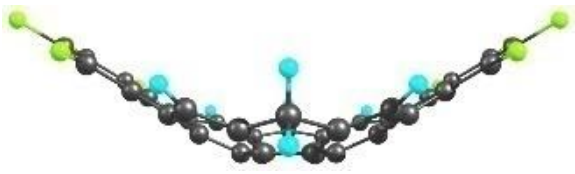 |
| DCPP-5 | 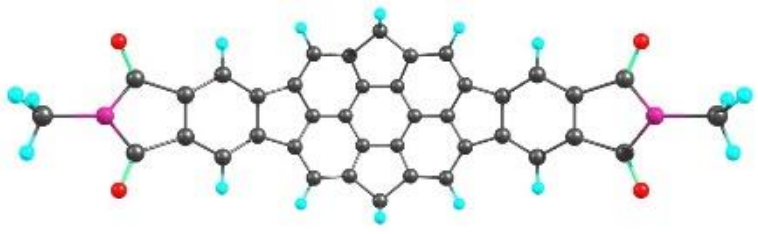 | 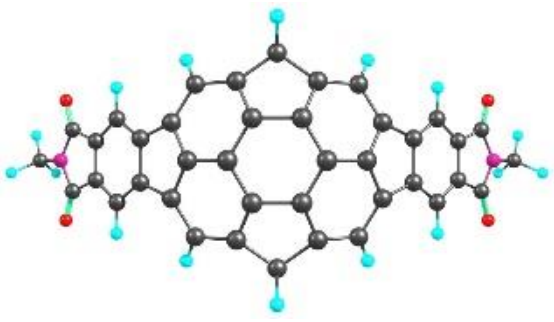 | 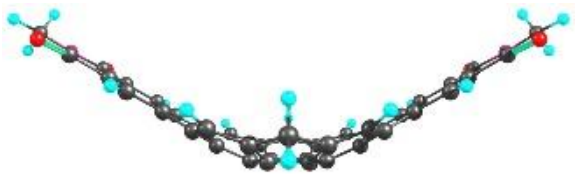 |
| DCPP-6 | 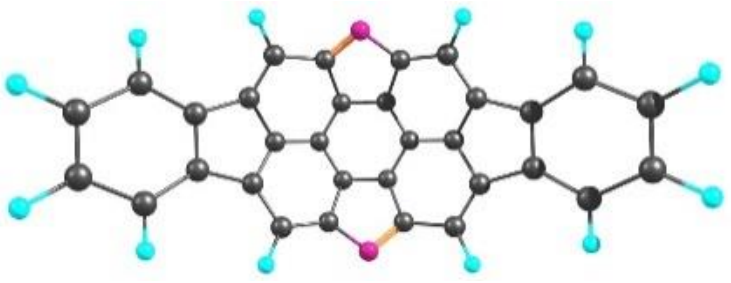 | 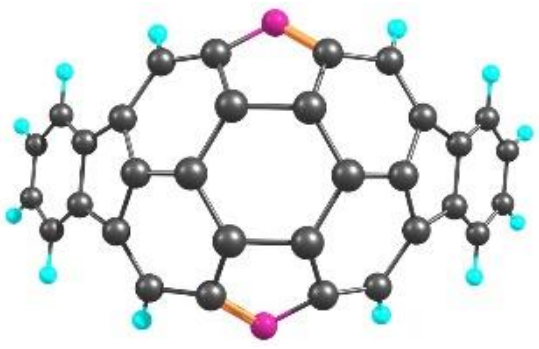 | 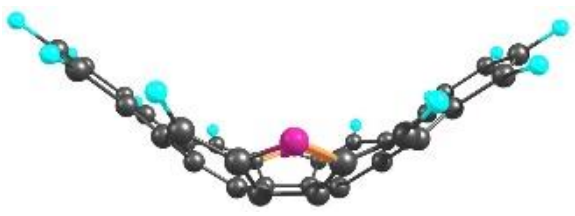 |
| DCPP-7 | 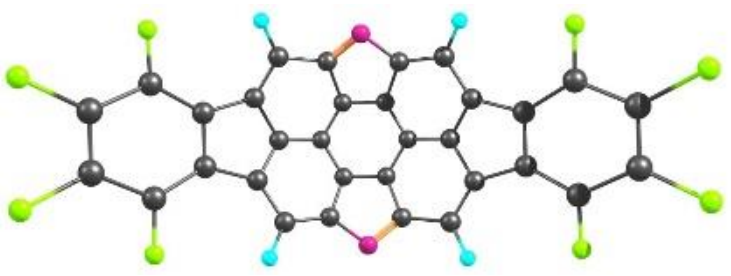 | 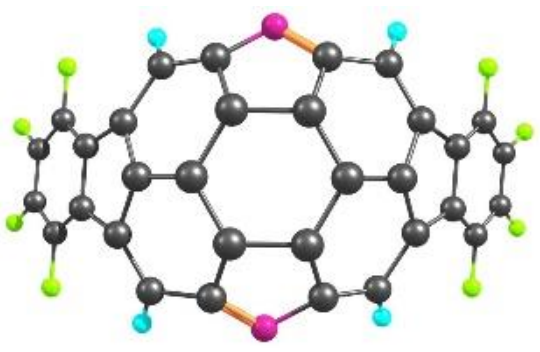 | 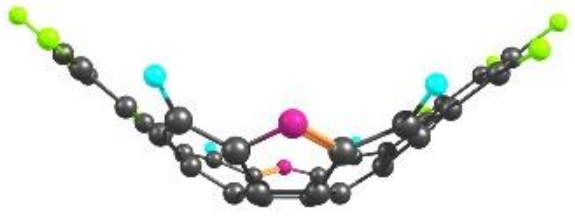 |
| DCPP-8 | 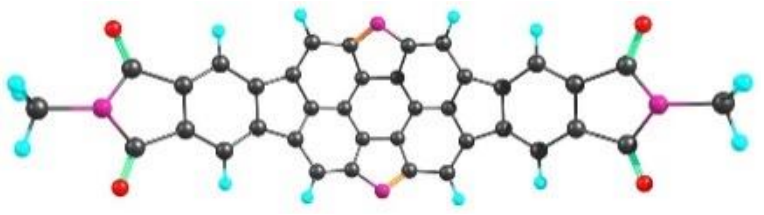 | 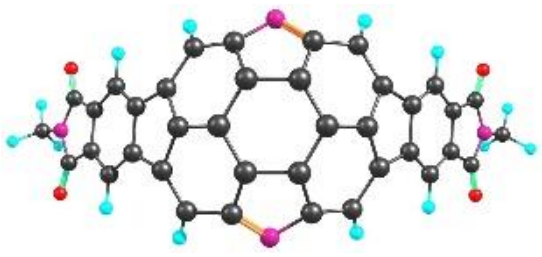 | 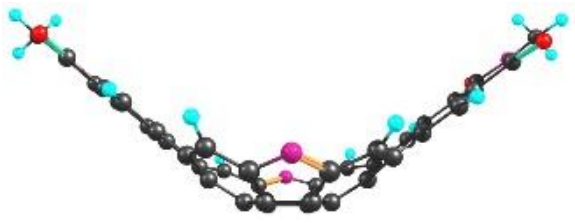 |
| DCPP-9 | 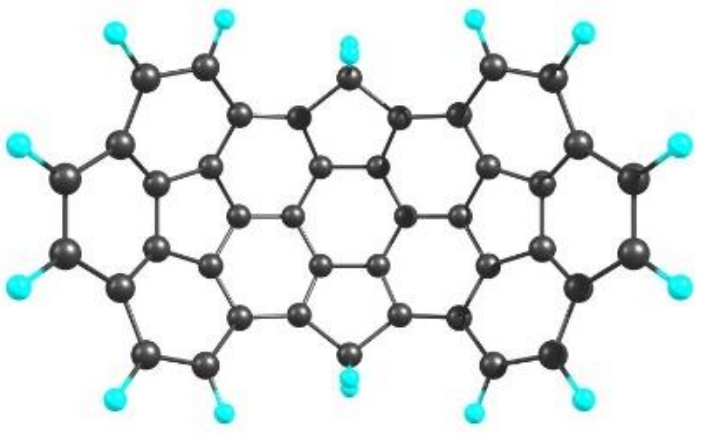 | 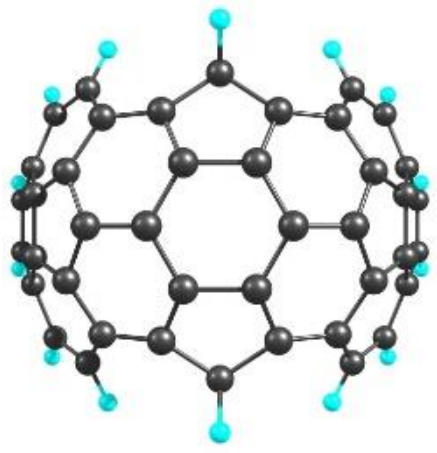 | 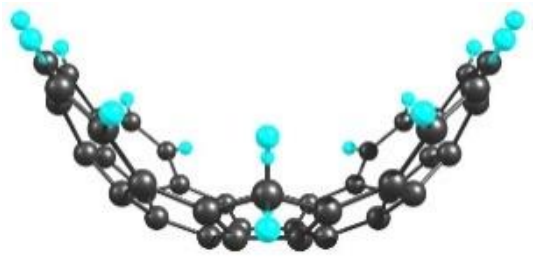 |
| DCPP-10 | 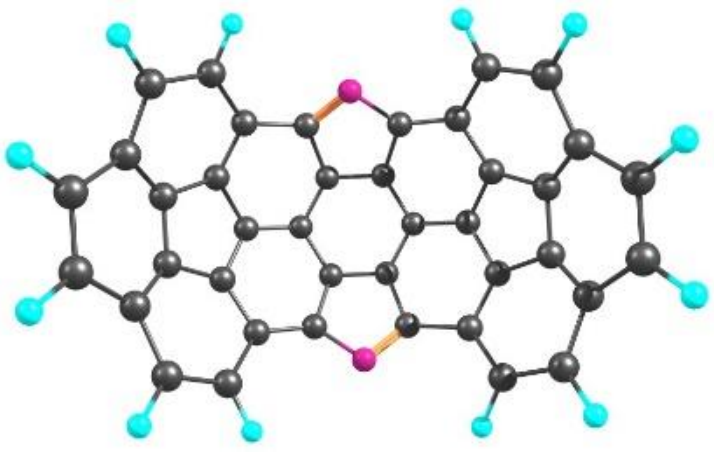 | 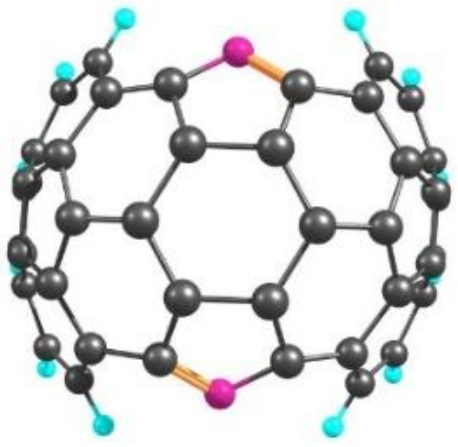 | 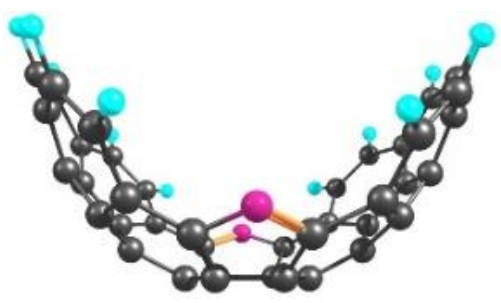 |
| DCPP-11 | 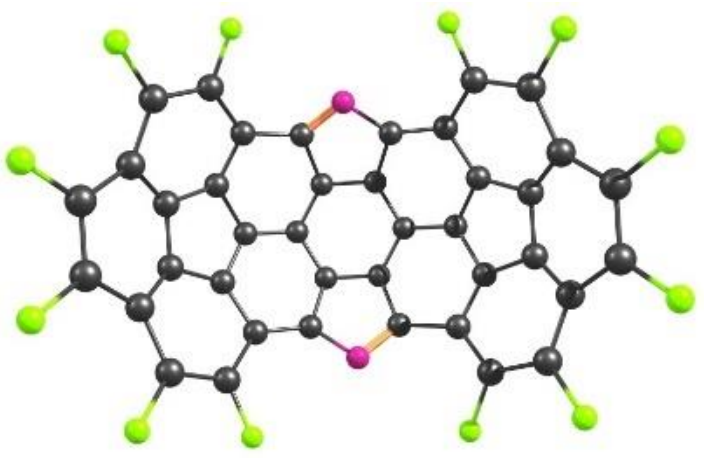 | 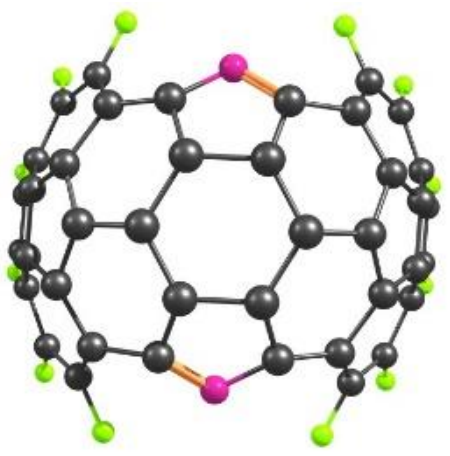 | 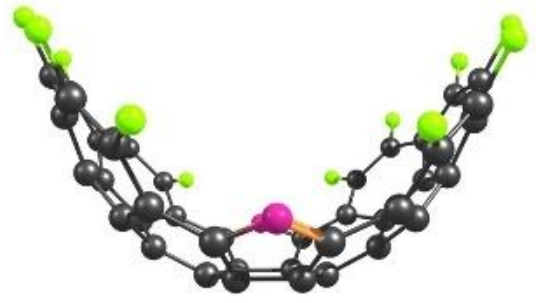 |
| DCPP-12 | 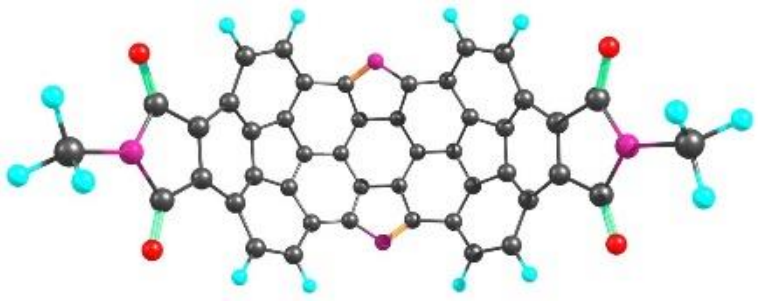 | 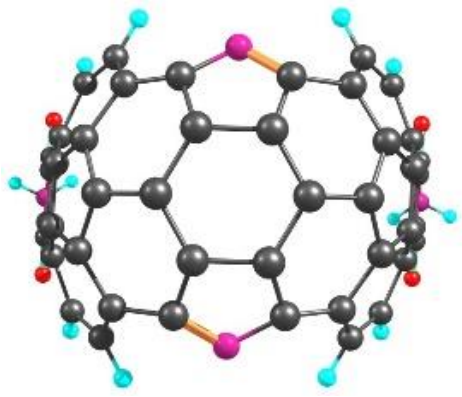 | 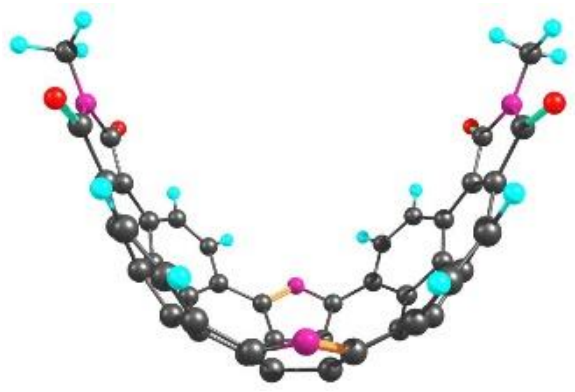 |

**C:● ; N: ● ; O: ● F: ● ; H: ●**

**Table S4**. Front and side views of B3LYP/6-311G(d,p) optimized molecular geometries of DCPP compounds with TES side chains.

| Compound | Front view | Side view |
|---|---|---|
| DCPP-TES-1 | 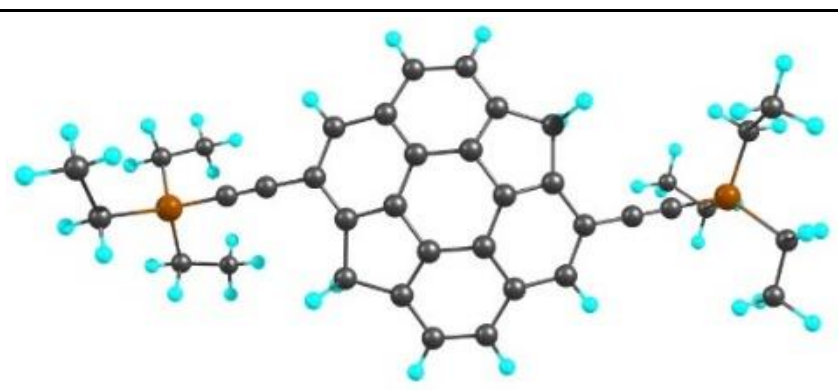 | 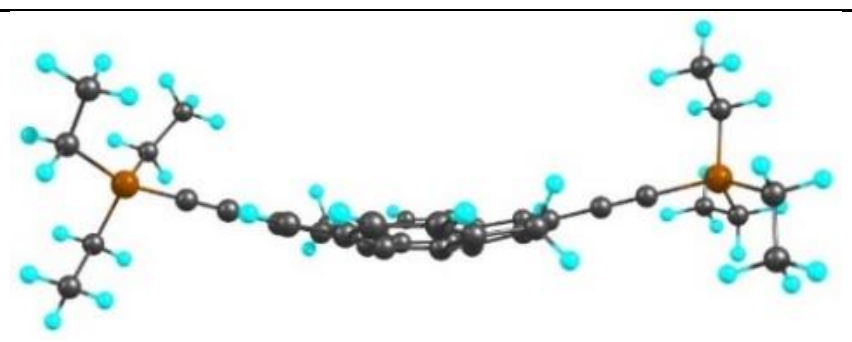 |
| DCPP-TES-2 | 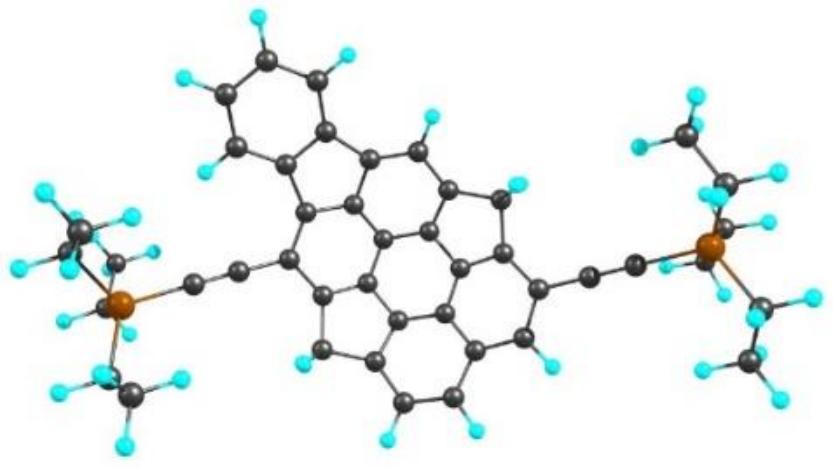 | 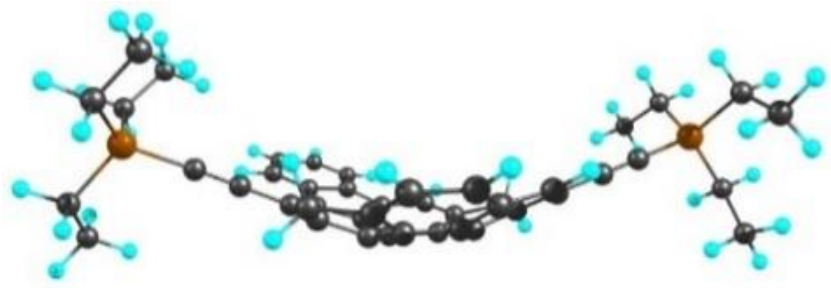 |
| DCPP-TES-3 | 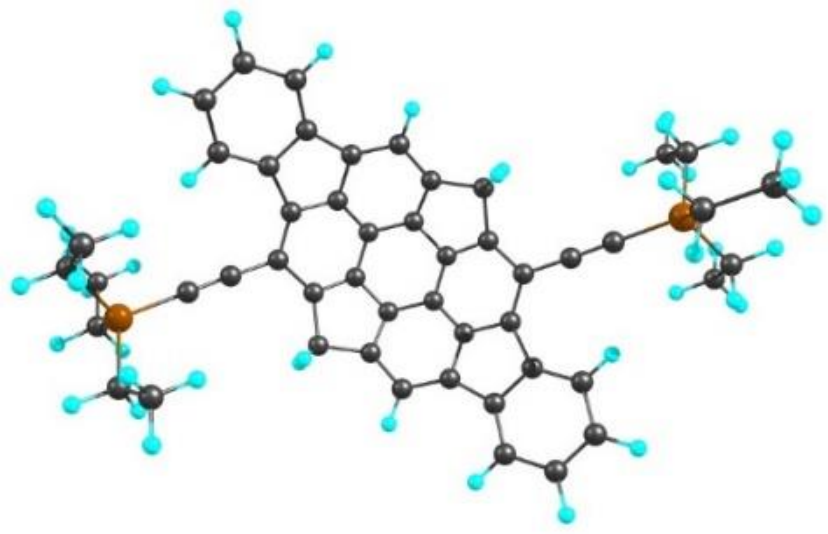 | 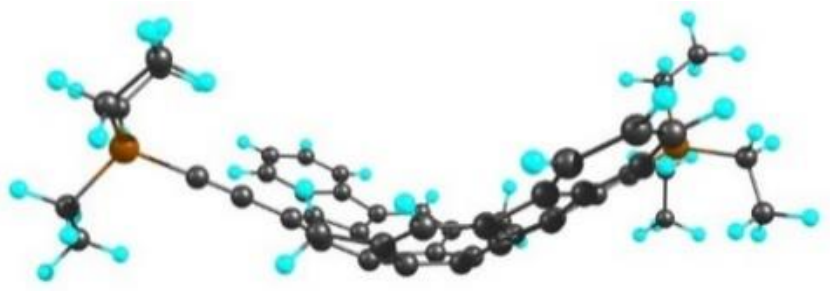 |
| DCPP-TES-4 | 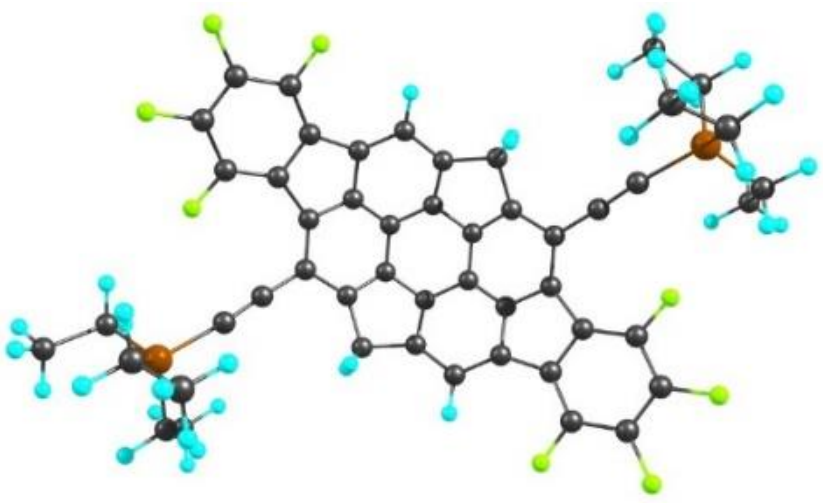 | 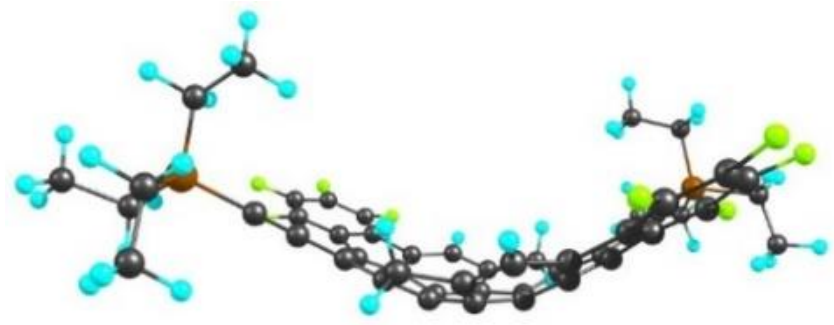 |
| DCPP-TES-5 | 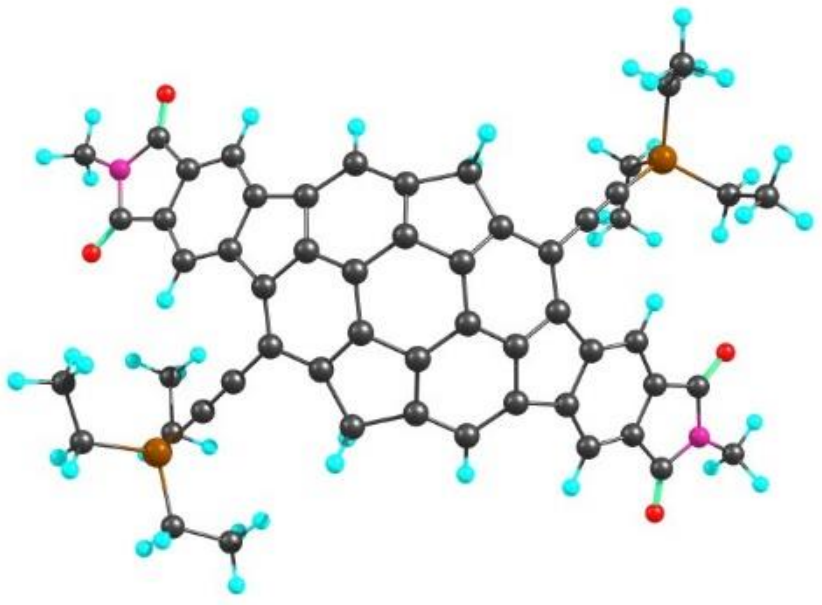 | 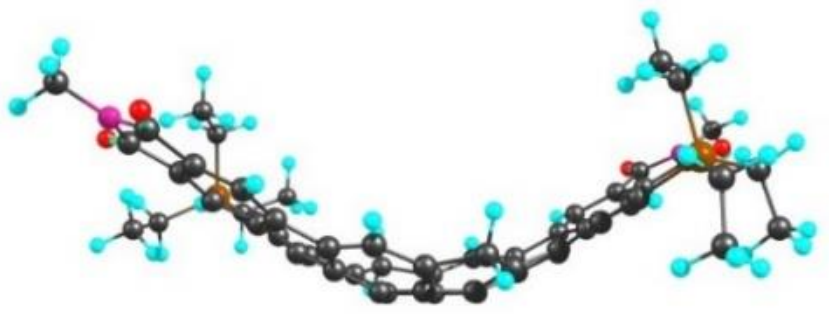 |
| DCPP-TES-6 | 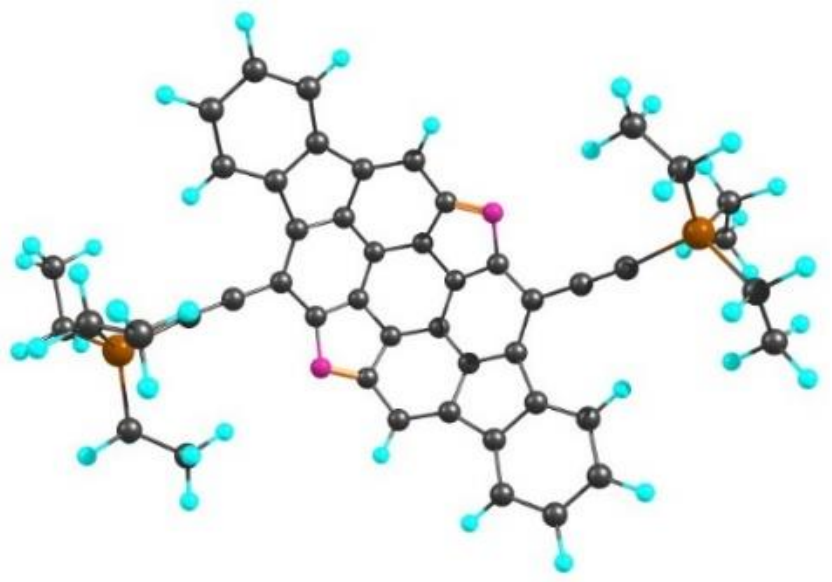 | 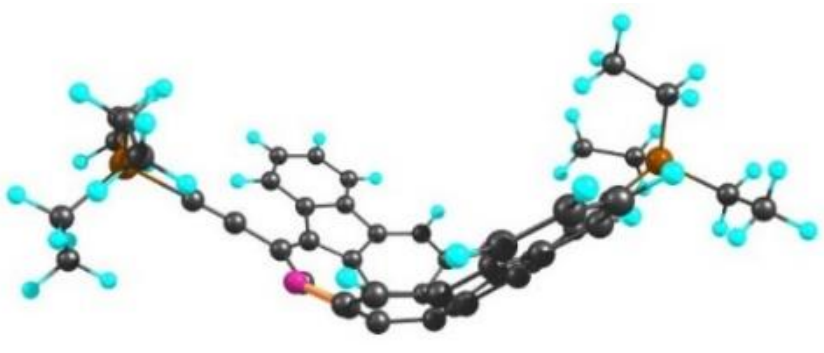 |
| DCPP-TES-7 | 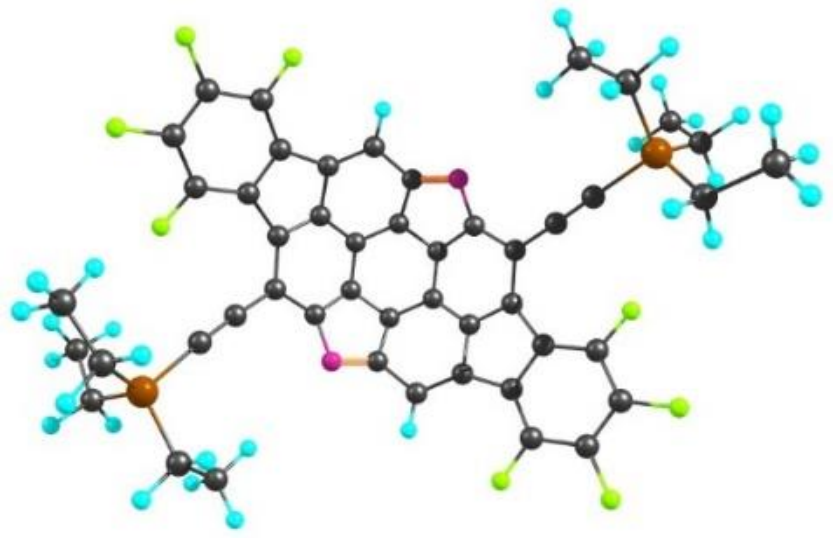 | 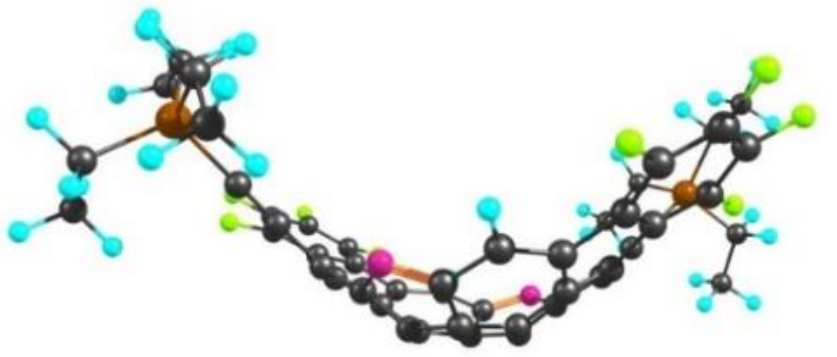 |
| DCPP-TES-8 | 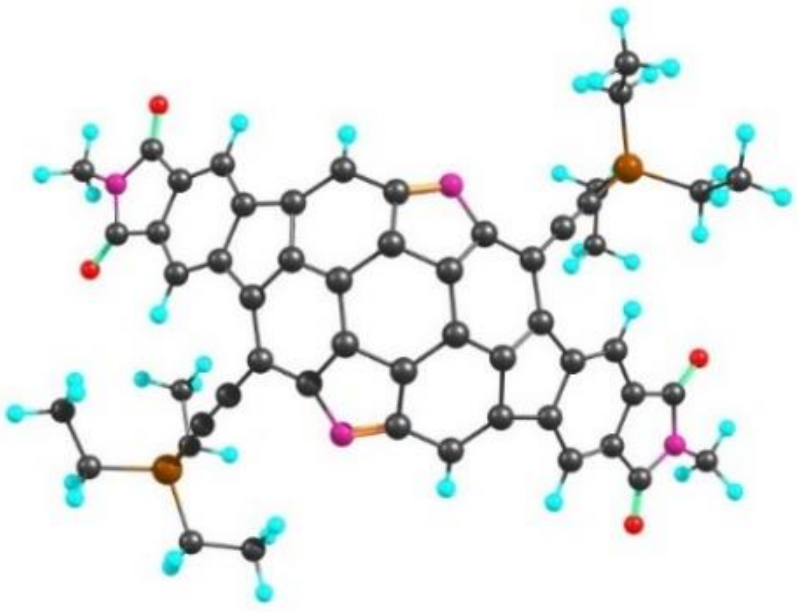 | 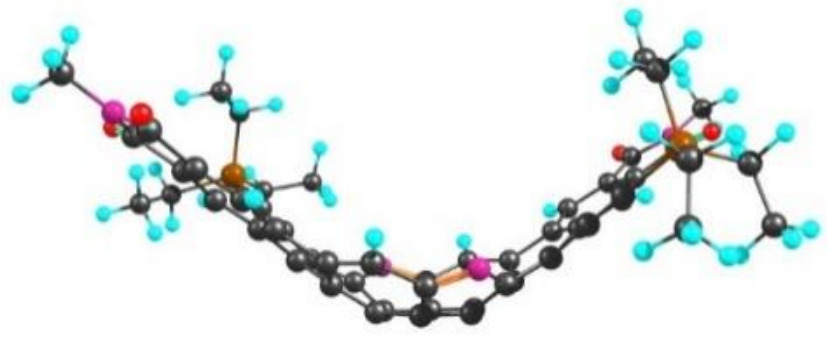 |
| DCPP-TES-9 | 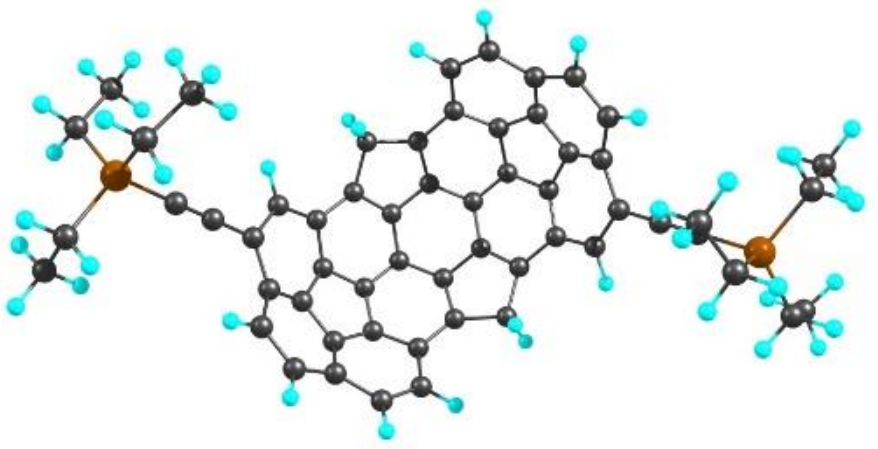 | 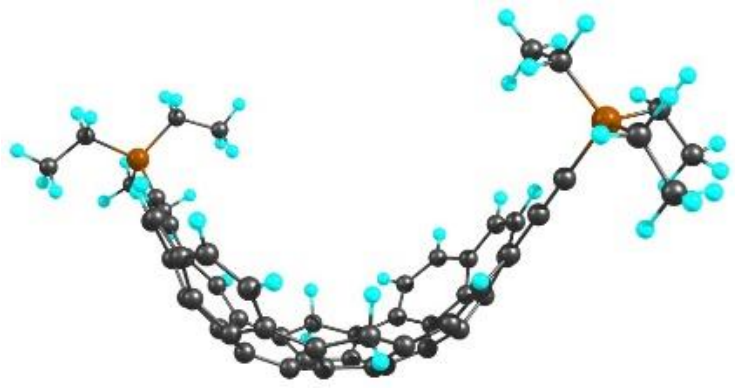 |
| DCPP-TES-10 | 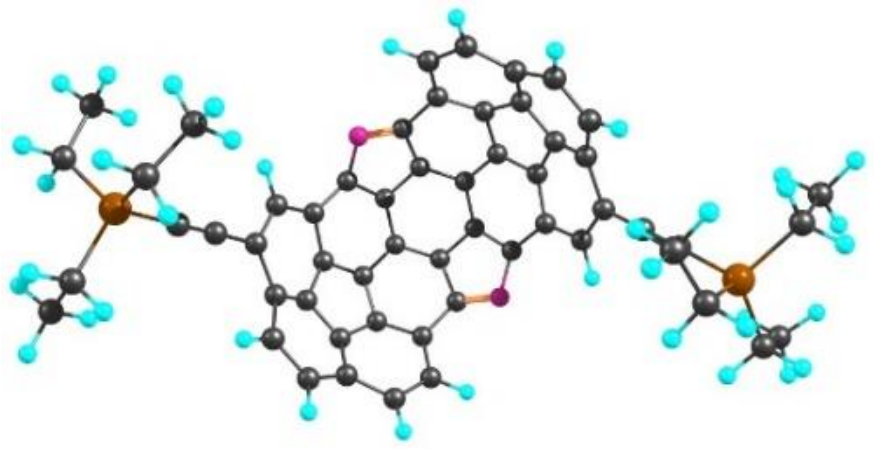 | 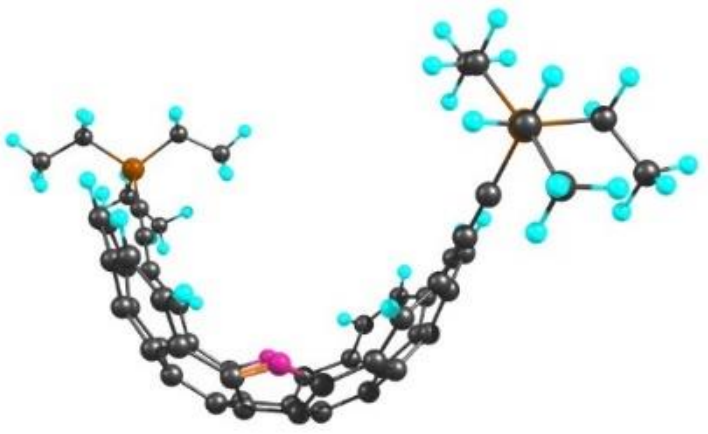 |
| DCPP-TES-11 | 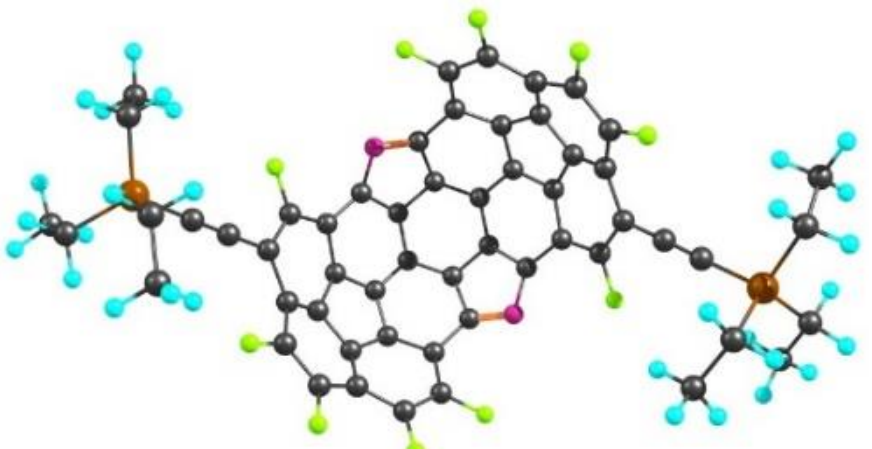 | 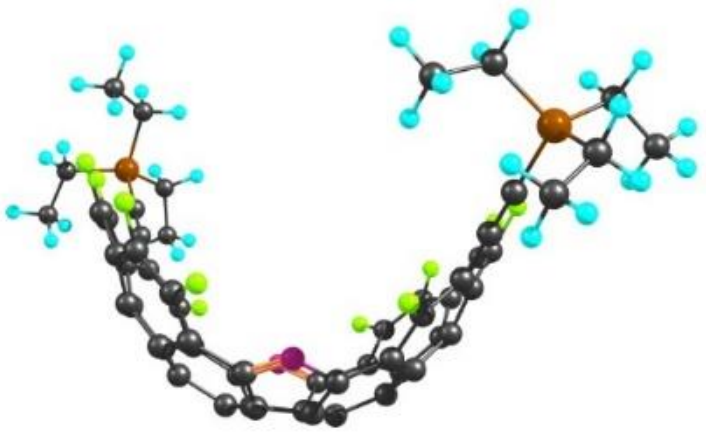 |
| DCPP-TES-12 | 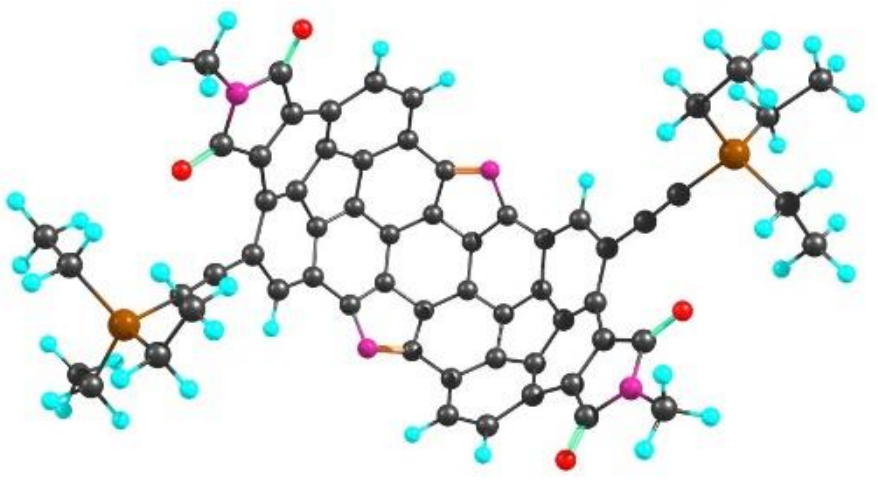 | 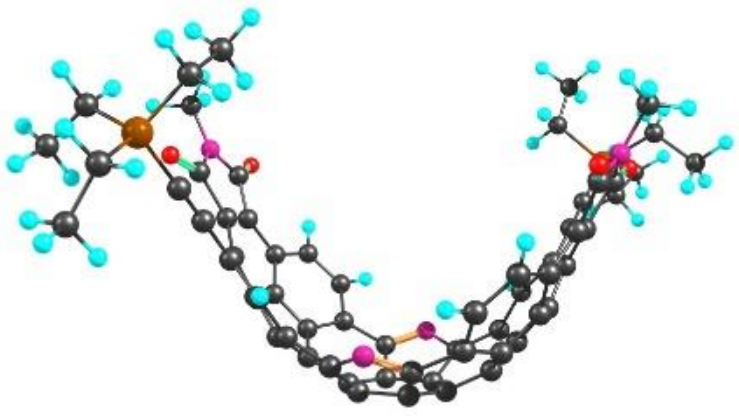 |

C: ● ; N: ● ; O: ● ; F: ● ; Si: ● ; H: ●

**Table S5.** Bowl depths, cone angles, dipole moments (μ), bowl-to-bowl inversion barriers, HOMOs, LUMOs and HOMO-LUMO gaps ($E_g$) of the DCPP and DCPP-TES derivatives.

| Compound | Bowl depth (Å) | Cone angle (°) | BWBI | HOMO (eV) | LUMO (eV) | Eg (eV) |
|---|---|---|---|---|---|---|
| DCPP-1 | 0.595 | 147.06 | | -5.048 | -1.919 | 3.129 |
| DCPP-2 | ---- | ---- | | -5.133 | -2.349 | 2.784 |
| DCPP-3 | 3.007 | 119.64 | | -5.205 | -2.661 | 2.544 |
| DCPP-4 | 3.035 | 118.89 | | -5.739 | -3.267 | 2.472 |
| DCPP-5 | 4.144 | 118.53 | | -5.8 | -3.408 | 2.392 |
| DCPP-6 | 3.640 | 106.13 | | -5.778 | -4.113 | 1.665 |
| DCPP-7 | 3.677 | 105.13 | | -6.285 | -4.644 | 1.641 |
| DCPP-8 | 4.967 | 105.05 | | -6.309 | -4.646 | 1.663 |
| DCPP-9 | 4.104 | 86.77 | | -5.421 | -2.47 | 2.951 |
| DCPP-10 | 4.542 | 75.70 | | -5.705 | -4.264 | 1.441 |
| DCPP-11 | 4.479 | 77.73 | | -6.57 | -5.157 | 1.413 |
| DCPP-12 | 6.584 | 65.21 | | -6.139 | -4.704 | 1.435 |
| DCPP-TES-1 | 0.590 | 146.85 | | -5.225 | -2.174 | 3.051 |
| DCPP-TES-2 | ---- | ---- | | -5.258 | -2.54 | 2.718 |
| DCPP-TES-3 | 2.938 | 121.20 | | -5.285 | -2.782 | 2.503 |
| DCPP-TES-4 | 3.035 | 123.46 | | -5.746 | -3.317 | 2.429 |
| DCPP-TES-5 | 4.164 | 118.21 | | -5.8 | -3.443 | 2.357 |
| DCPP-TES-6 | 3.520 | 108.90 | | -5.692 | -4.173 | 1.519 |
| DCPP-TES-7 | 3.398 | 111.43 | | -6.102 | -4.598 | 1.504 |
| DCPP-TES-8 | 4.843 | 107.23 | | -6.124 | -4.597 | 1.527 |
| DCPP-TES-9 | 4.109 | 53.67 | | -5.505 | -2.578 | 2.927 |
| DCPP-TES-10 | 4.539 | 75.86 | | -5.692 | -4.3 | 1.392 |
| DCPP-TES-11 | 4.471 | 77.89 | | -6.315 | -4.962 | 1.353 |
| DCPP-TES-12 | 6.466 | 69.72 | | -6.025 | -4.679 | 1.346 |

**Table S6.** Distribution of frontier molecular orbitals (HOMO and LUMO) of the DCPP compounds computed at B3LYP/6-311G(d,p) level.

| Compound | HOMO | LUMO |
|---|---|---|
| DCPP-1 | 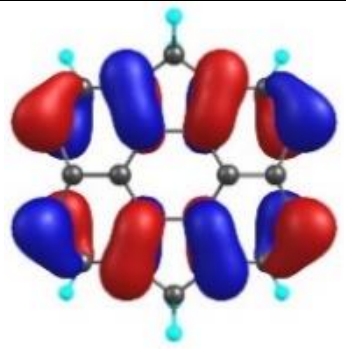 | 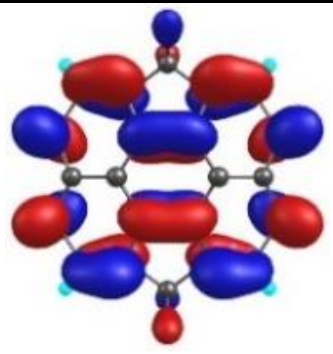 |
| DCPP-2 | 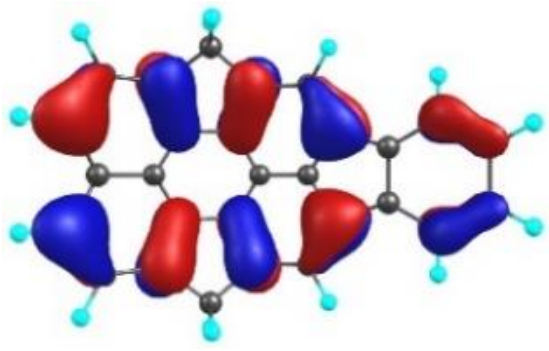 | 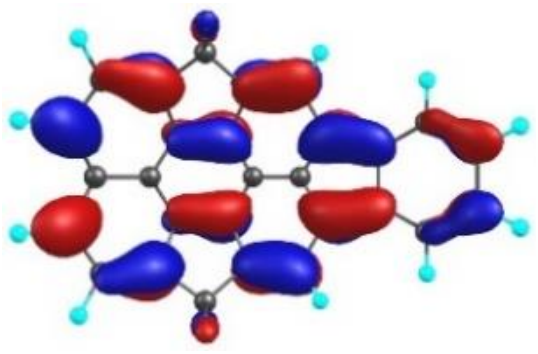 |
| DCPP-3 | 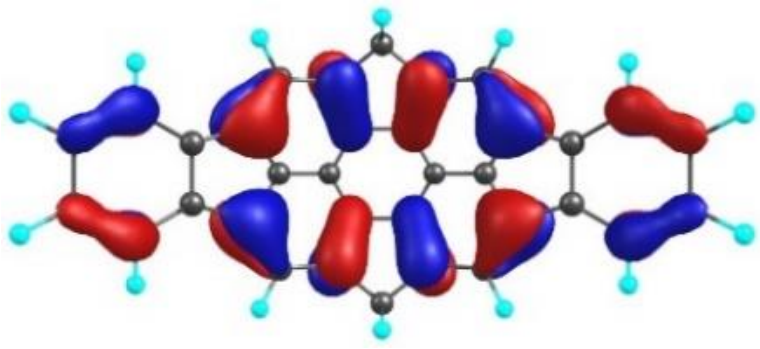 | 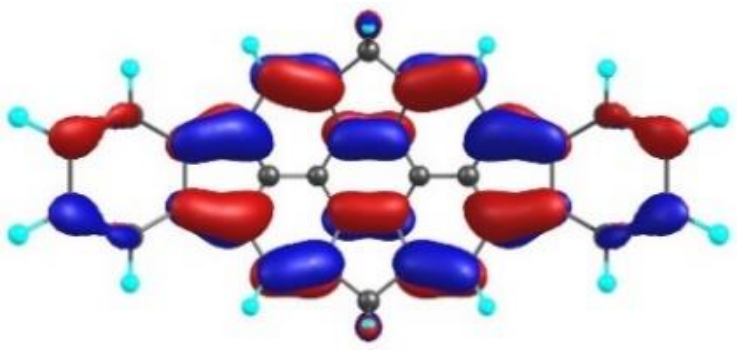 |
| DCPP-4 | 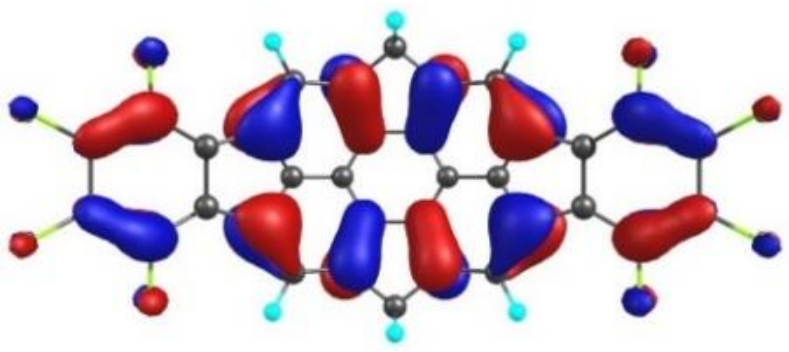 | 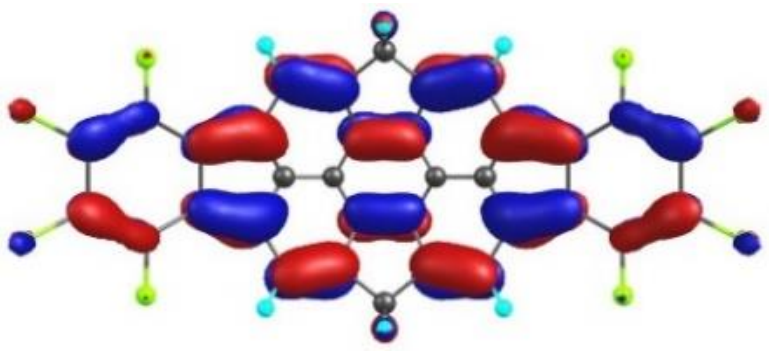 |
| DCPP-5 | 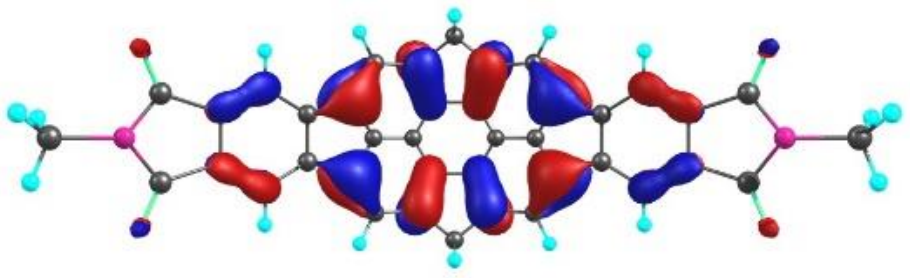 | 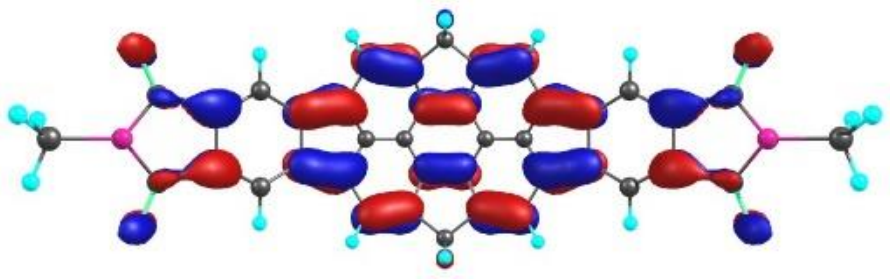 |
| DCPP-6 | 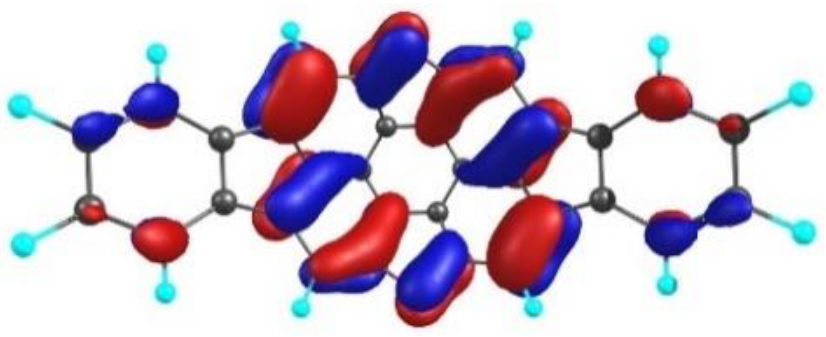 | 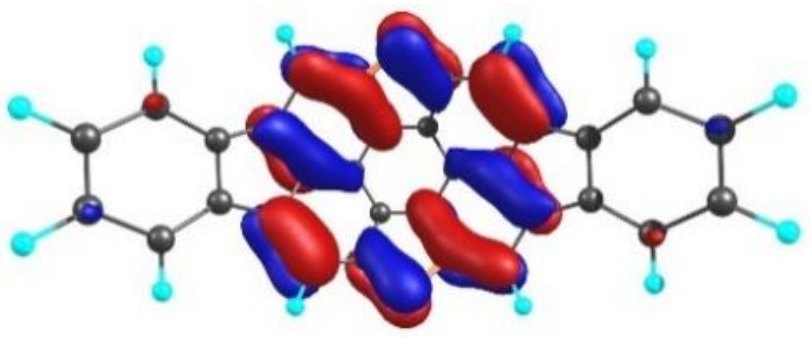 |
| DCPP-7 | 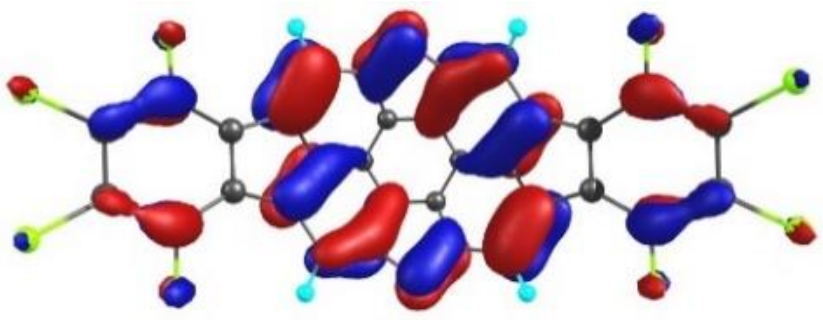 | 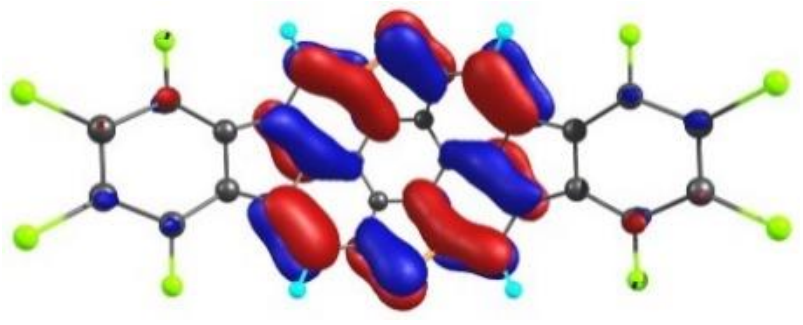 |
| DCPP-8 | 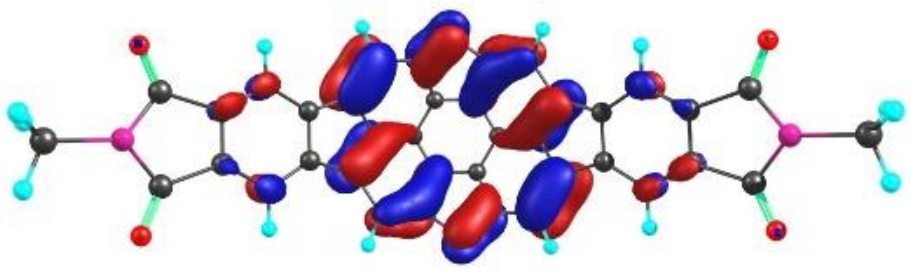 | 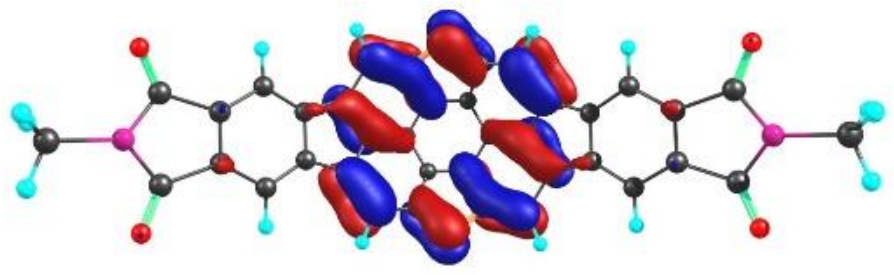 |
| DCPP-9 | 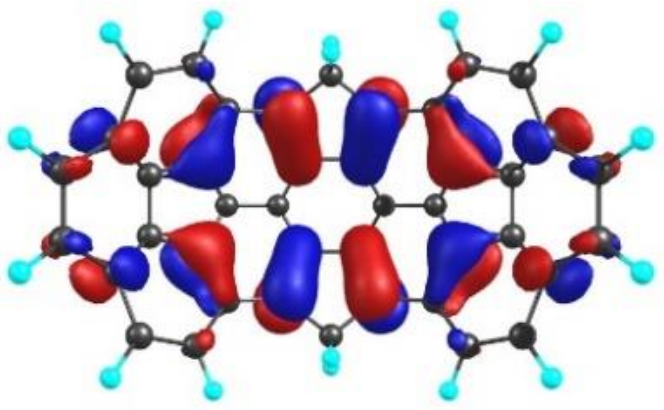 | 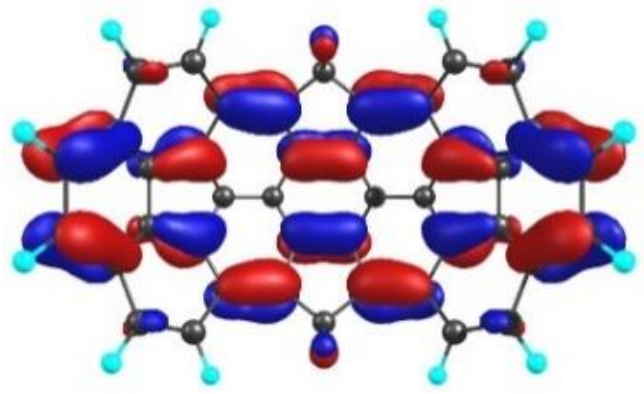 |
| DCPP-10 | 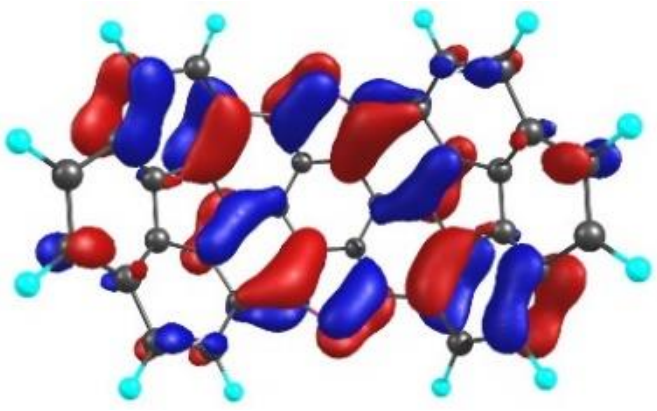 | 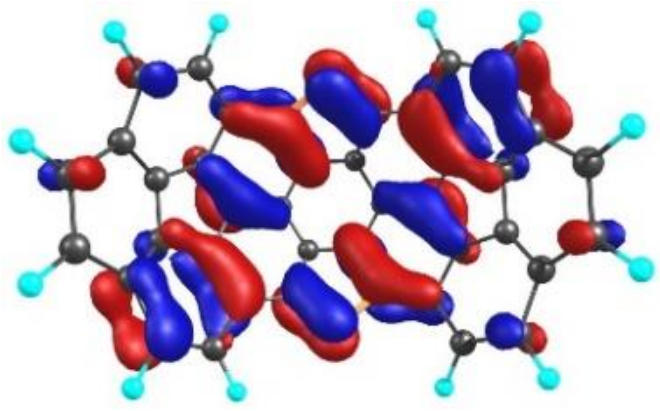 |
| DCPP-11 | 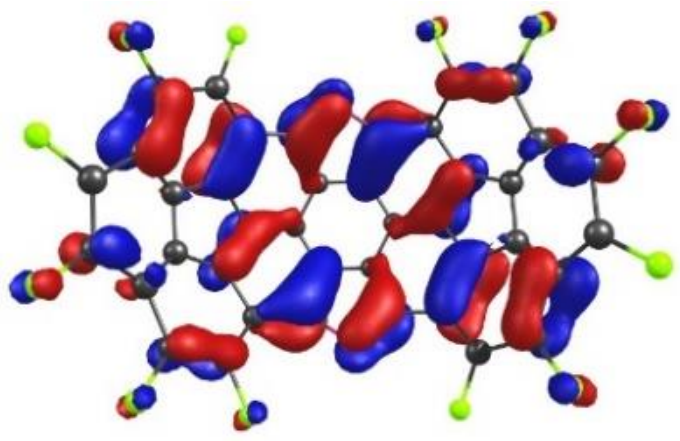 | 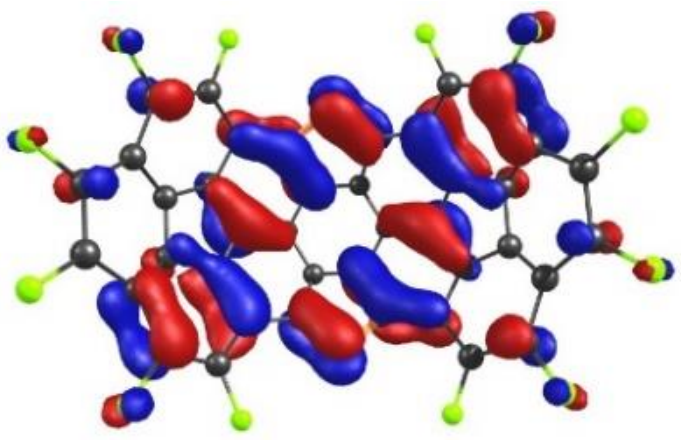 |
| DCPP-12 | 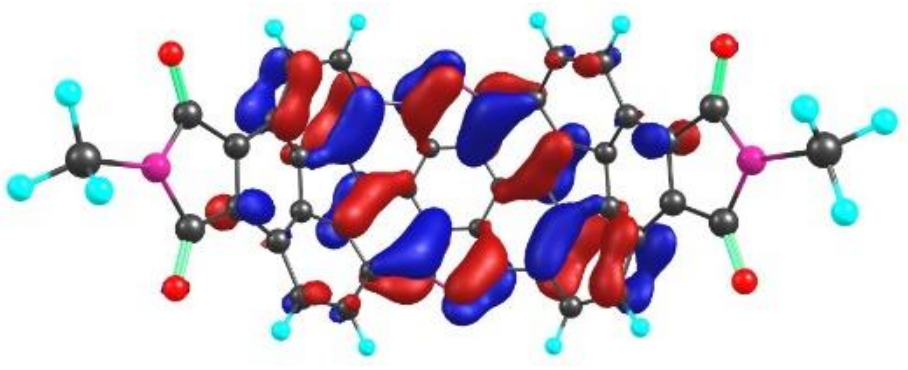 | 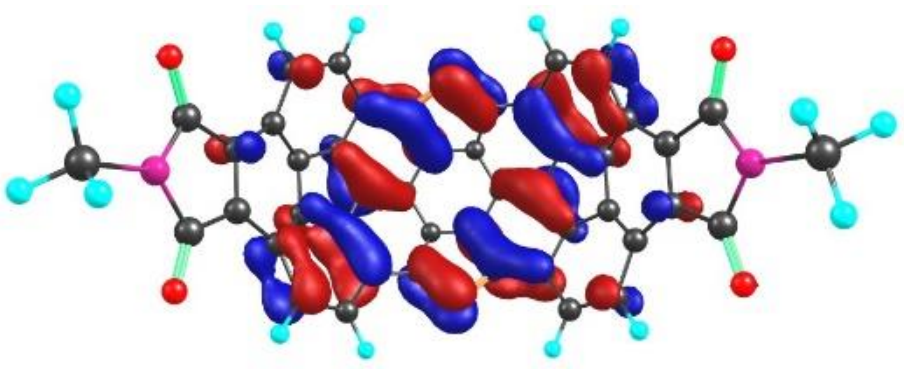 |

**Table S7.** Distribution of frontier molecular orbitals (HOMO and LUMO) of the DCPP-TES compounds computed at B3LYP/6-311G(d,p) level.

| Compound | HOMO | LUMO |
|---|---|---|
| DCPP-TES-1 | 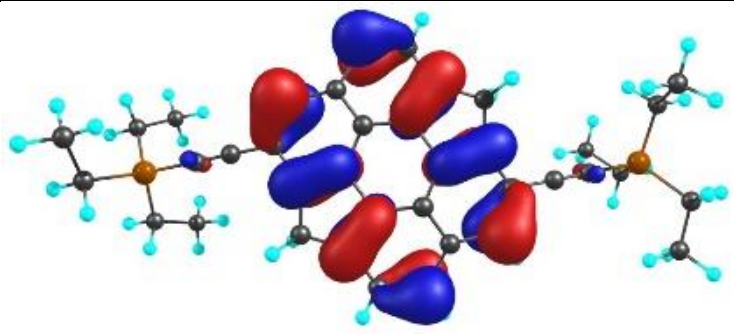 | 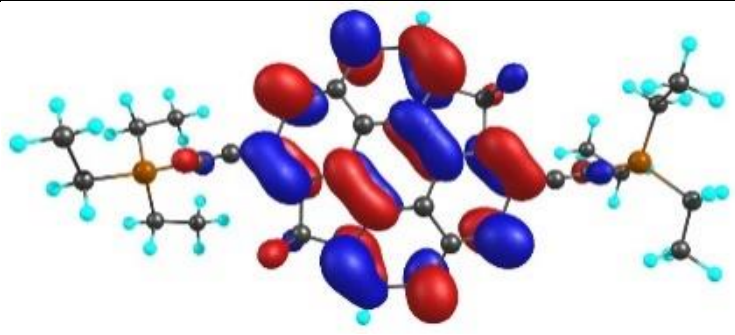 |
| DCPP-TES-2 | 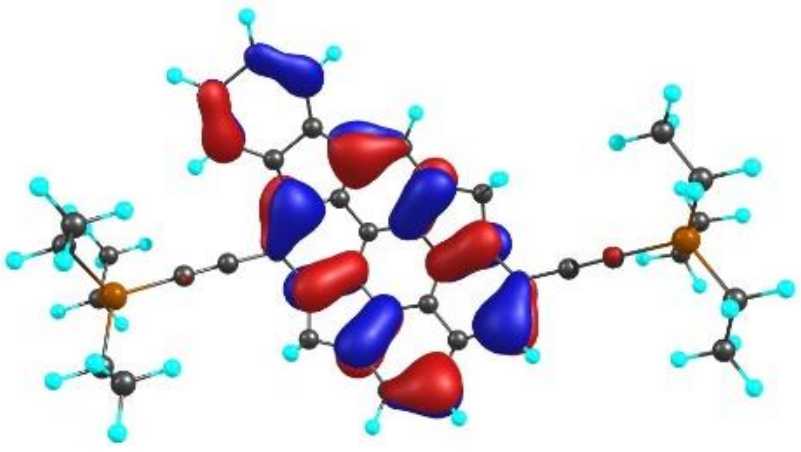 | 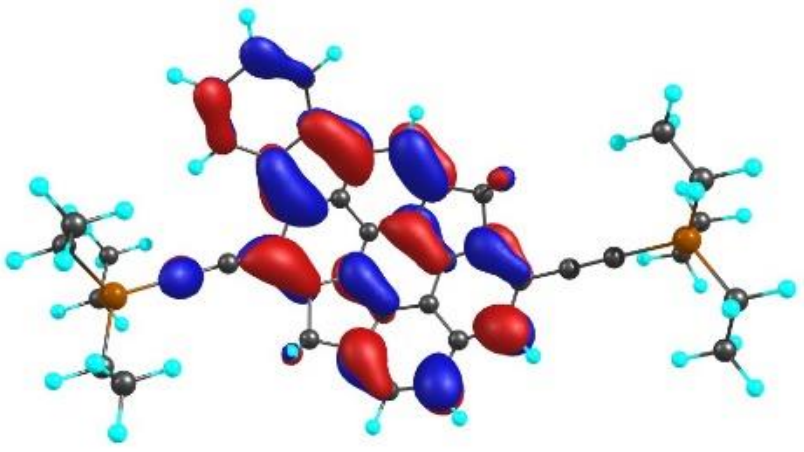 |
| DCPP-TES-3 | 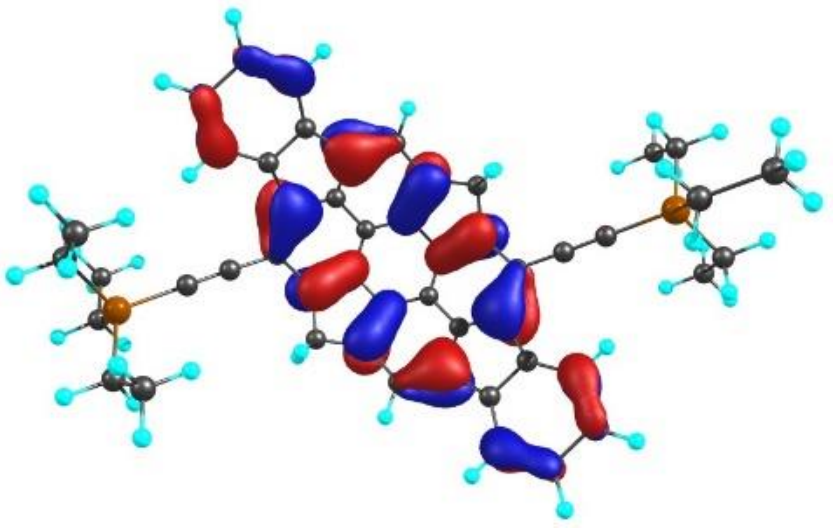 | 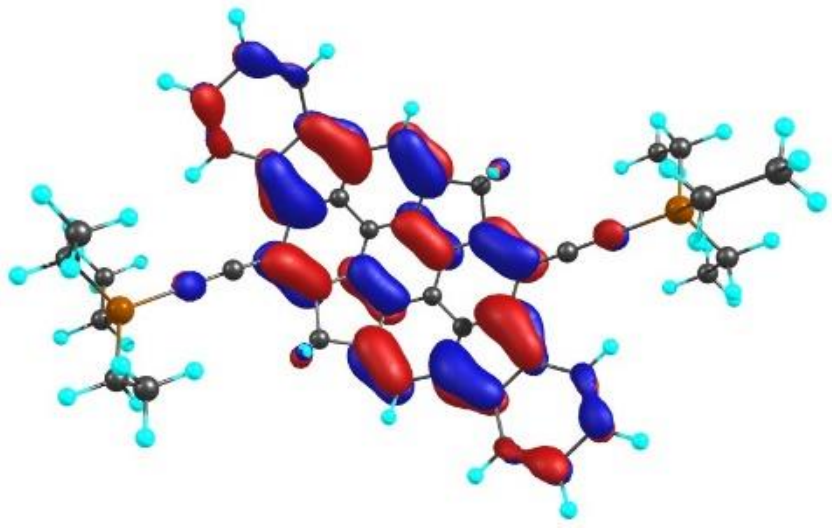 |
| DCPP-TES-4 | 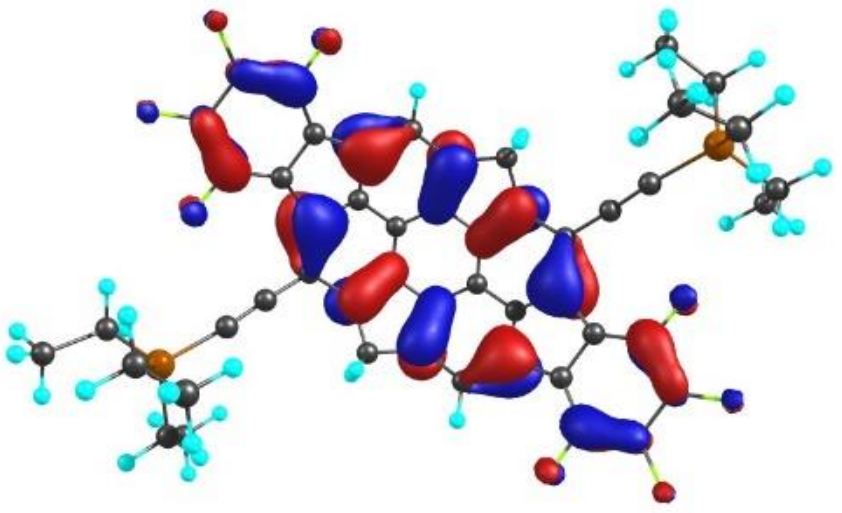 | 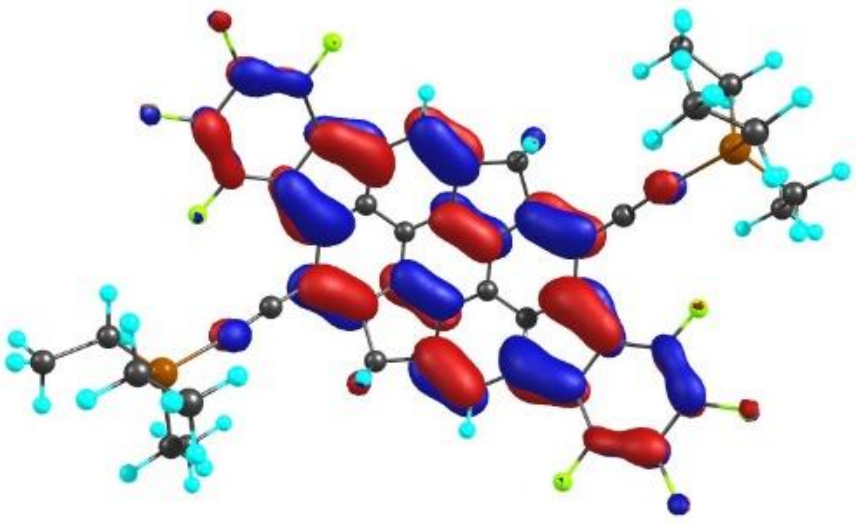 |
| DCPP-TES-5 | 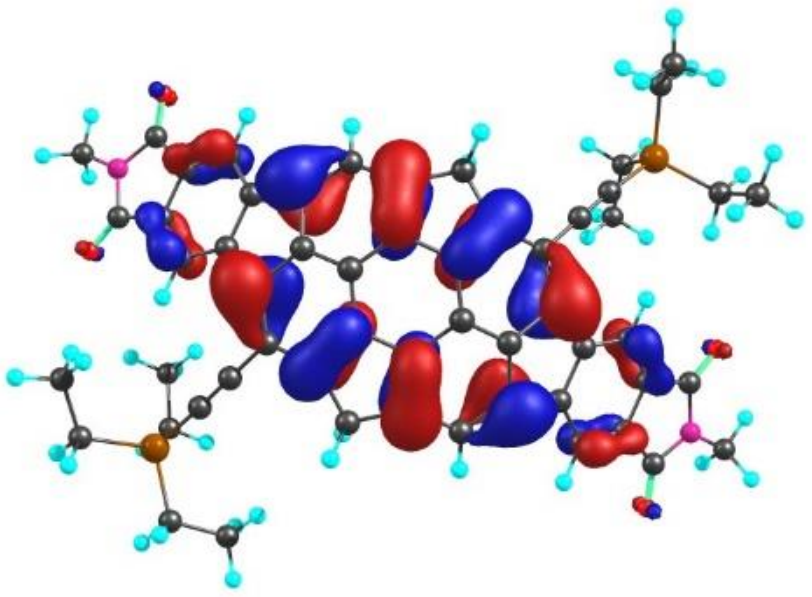 | 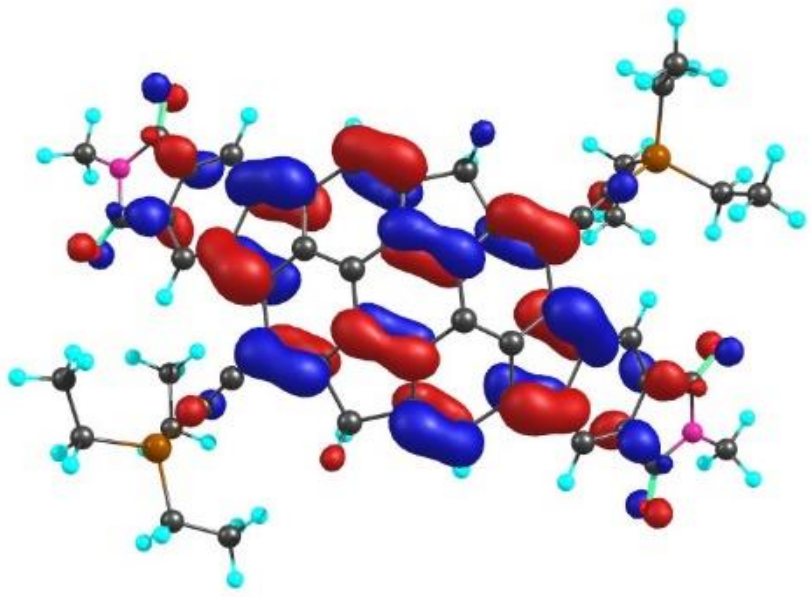 |
| DCPP-TES-6 | 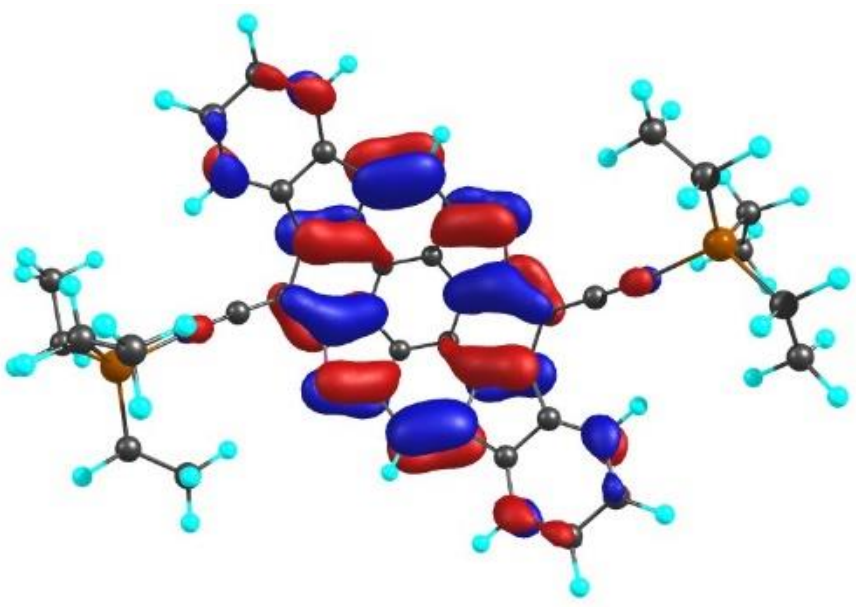 | 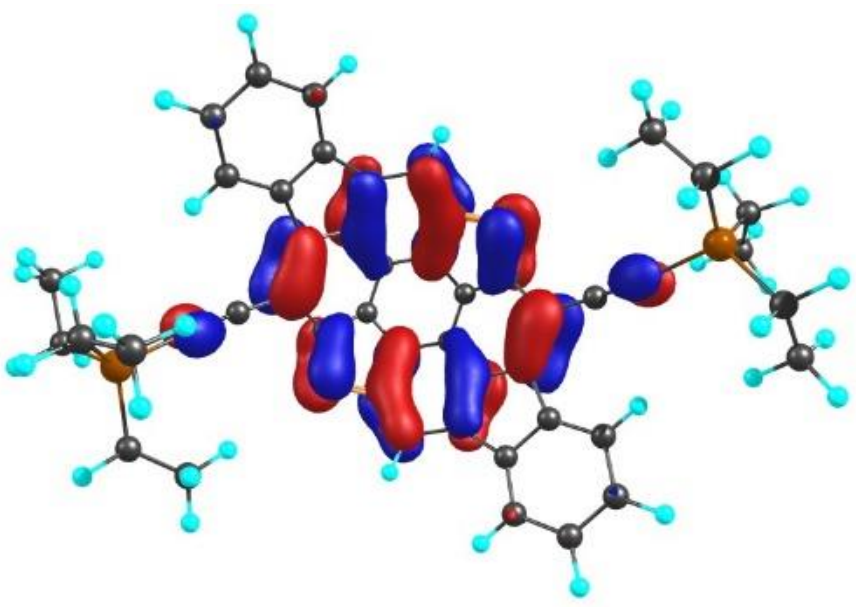 |
| DCPP-TES-7 | 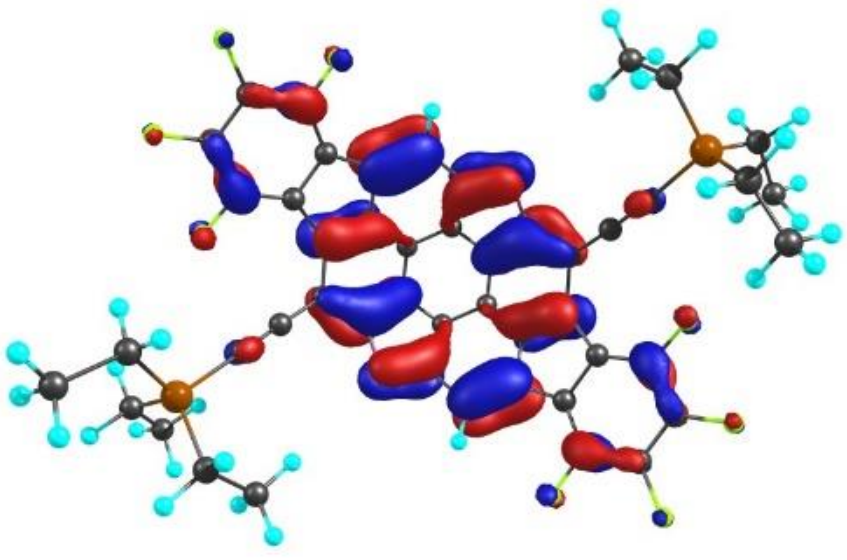 | 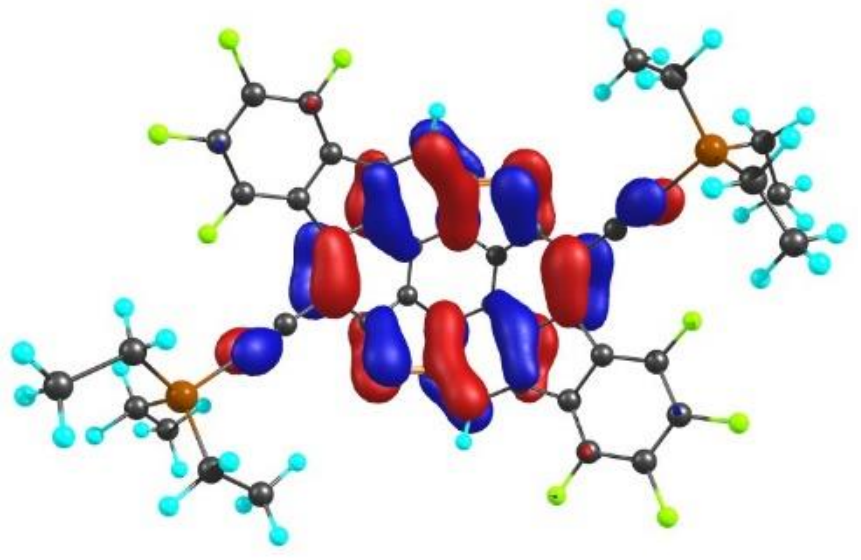 |
| DCPP-TES-8 | 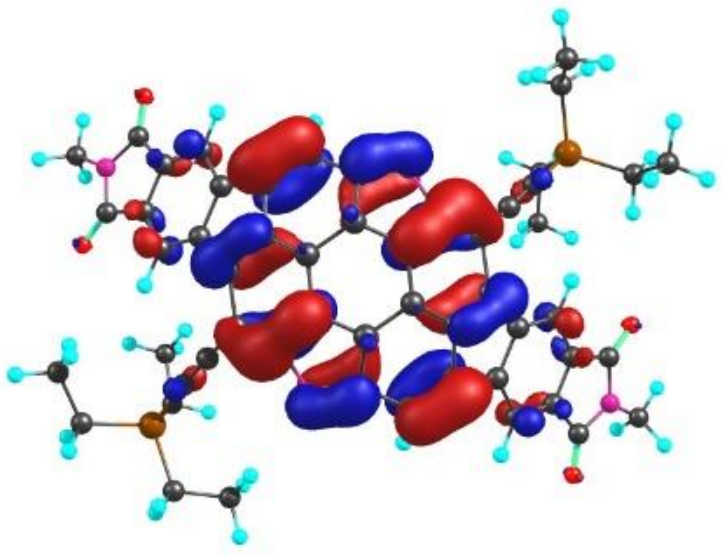 | 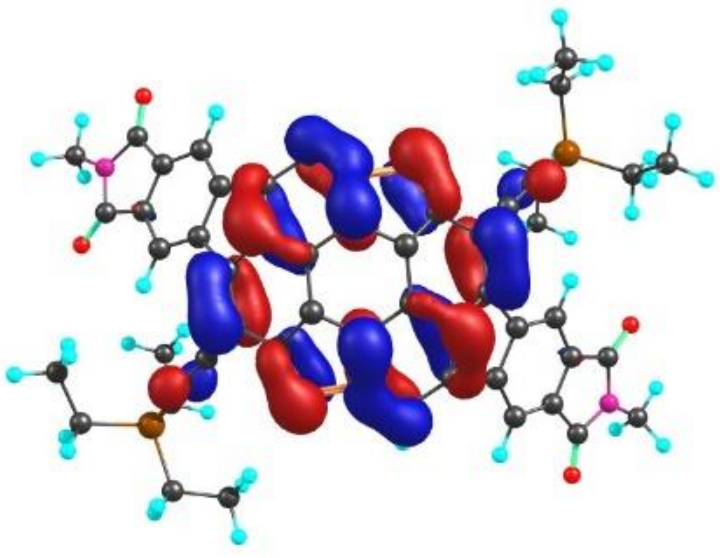 |
| DCPP-TES-9 | 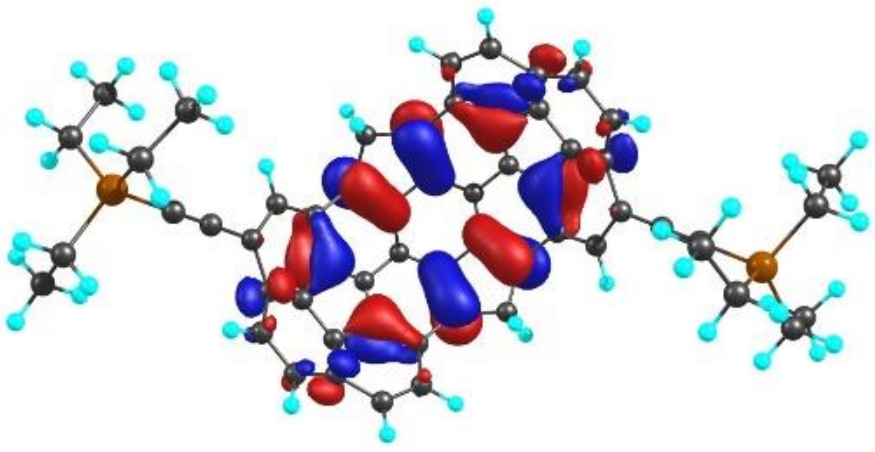 | 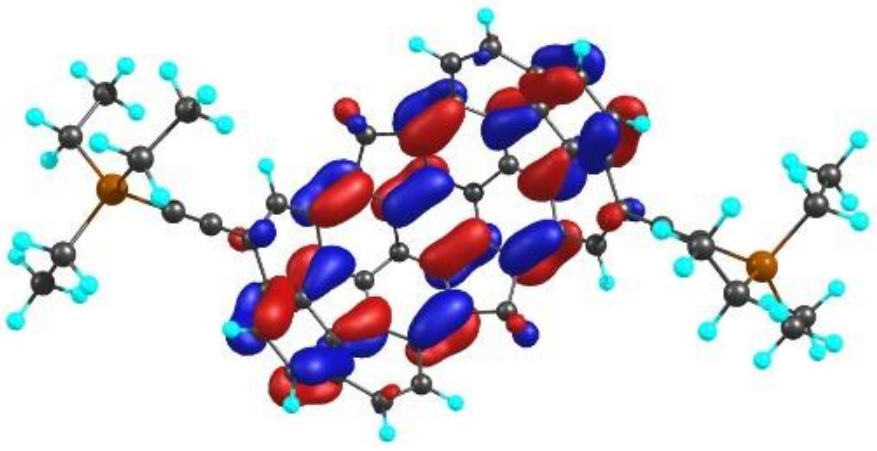 |
| DCPP-TES-10 | 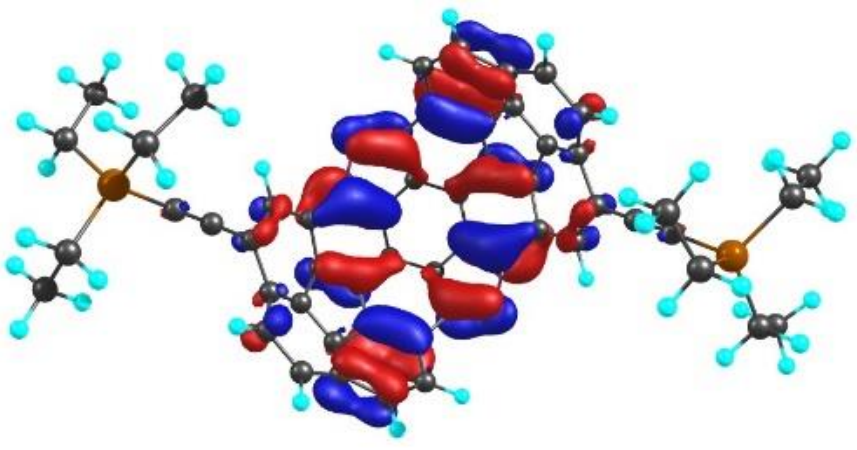 | 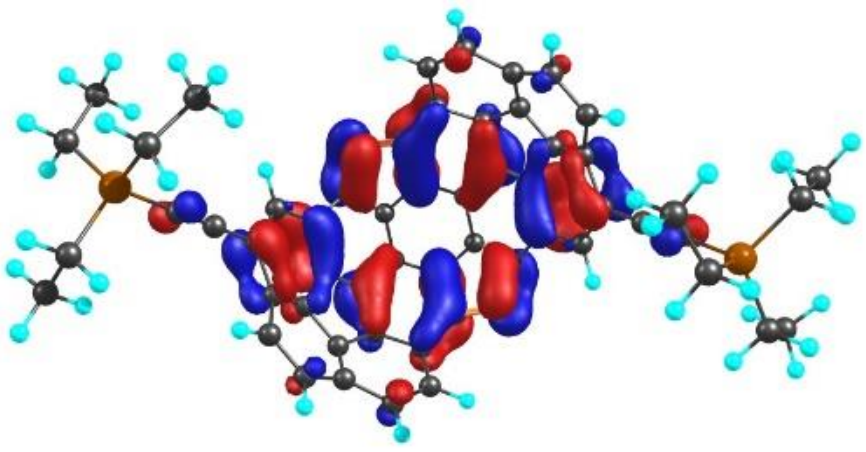 |
| DCPP-TES-11 | 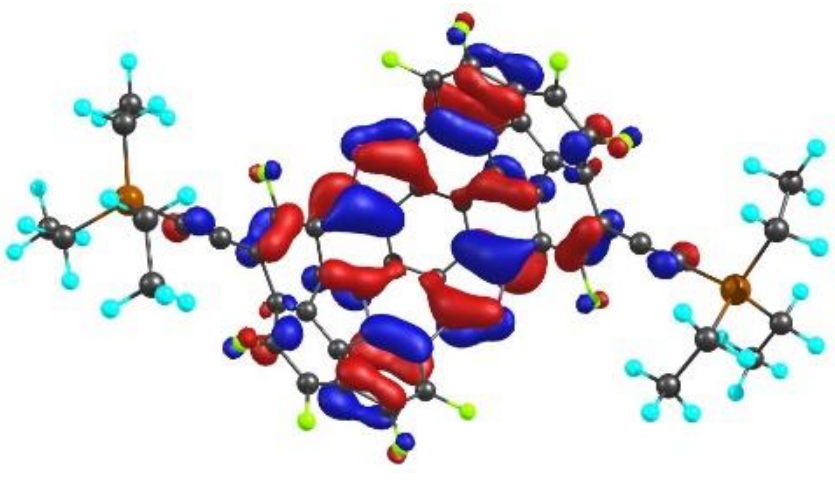 | 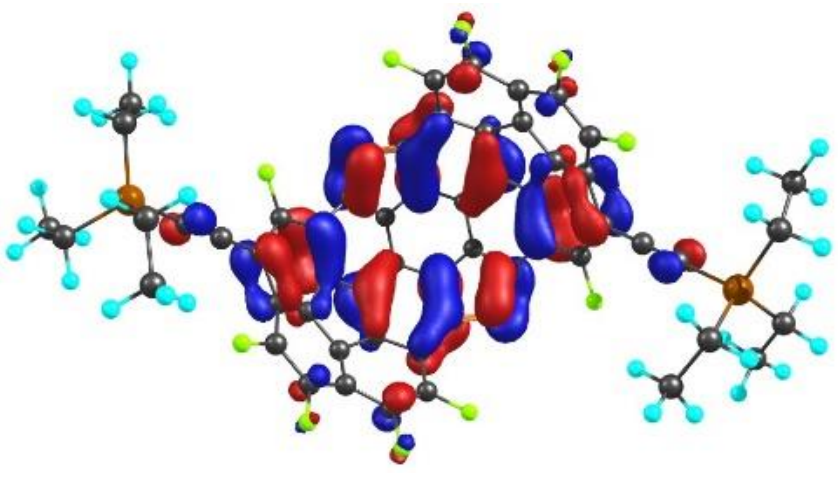 |
| DCPP-TES-12 | 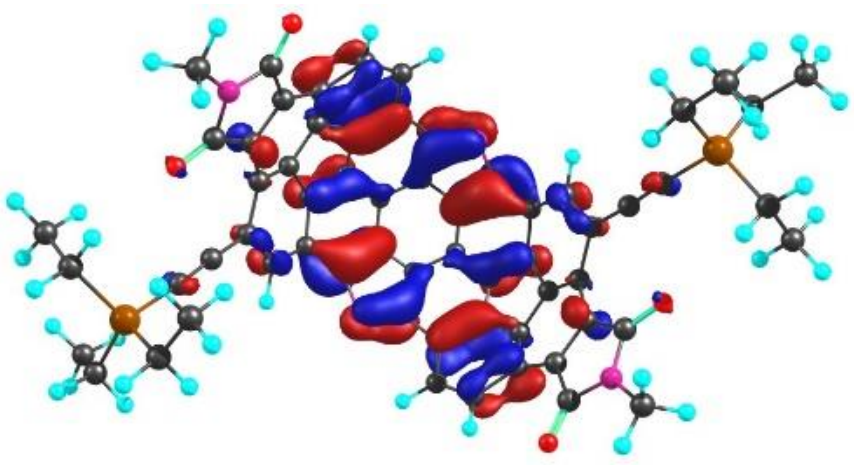 | 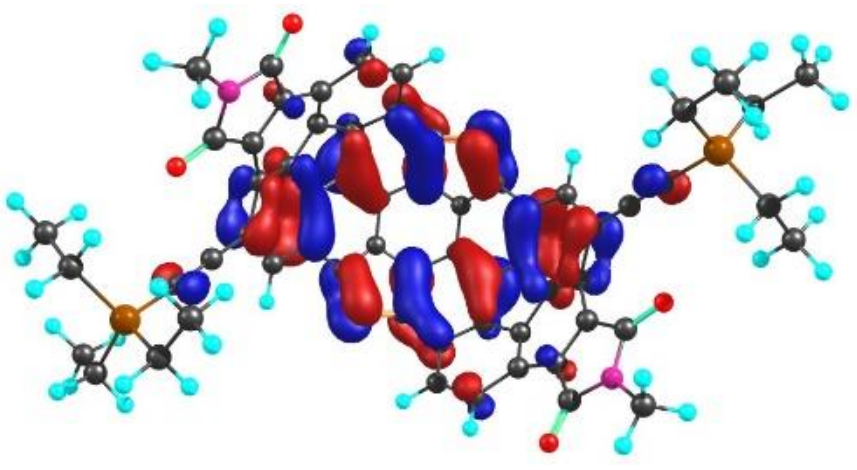 |

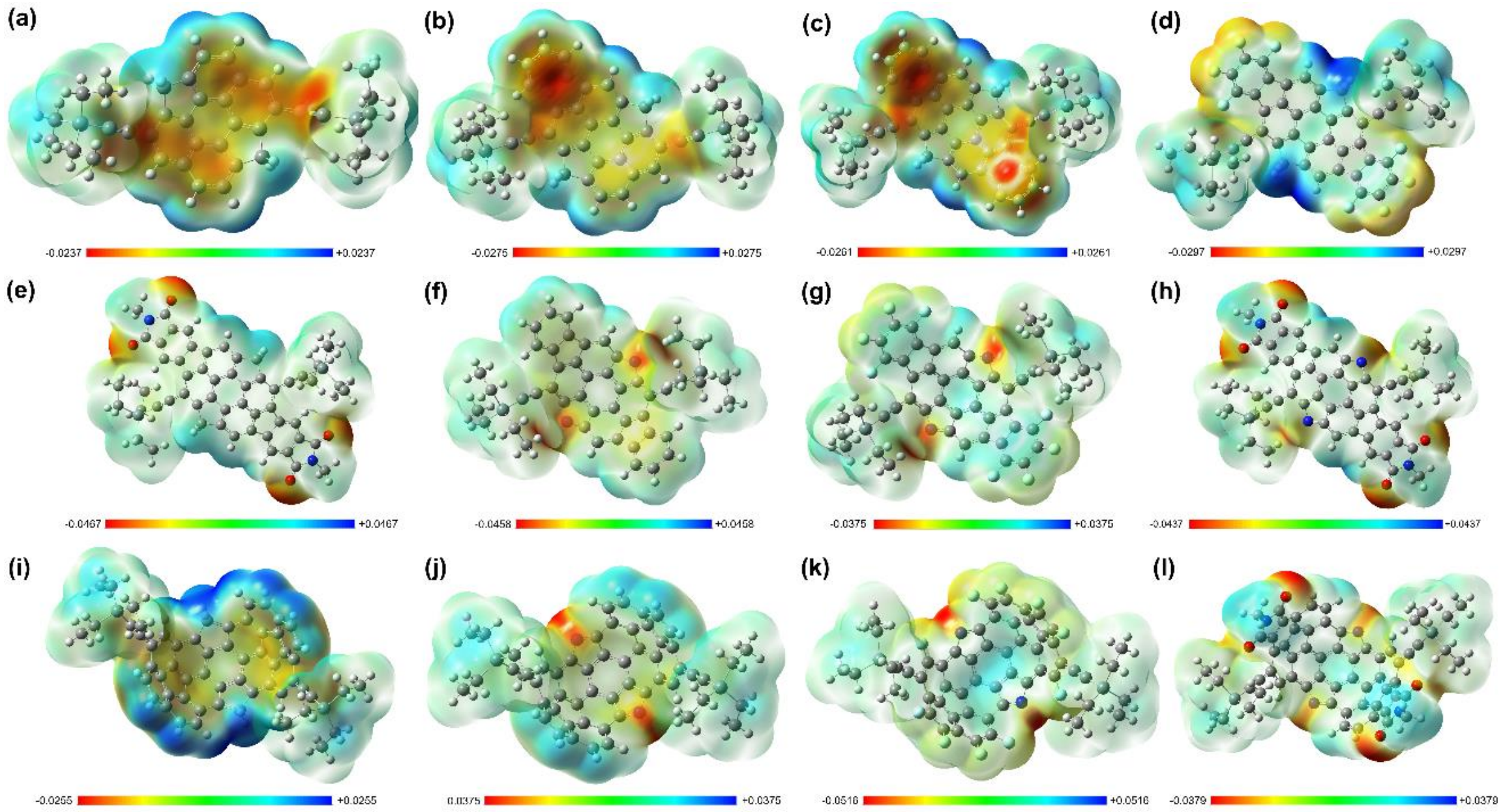

**Figure S1.** Electrostatic potential (ESP) mapping of DCPP-TES compounds.

**Section S1. Predicting the crystal structures of DCPP-derivatives.**

Crystal structure predictions are conducted over the Polymorph Predictor module in BIOVIA Materials Studio program (17.1.0.48) [S2], using the previously optimized molecular geometries. All the crystal structures of DCPP-derivatives are obtained with the fine quality simulation, using Dreiding forcefield, as discussed in our previous work [S3]. However, we chose Gasteiger and ESP charges separately for generating the molecular packing motifs of DCPPs and DCPP-TESs, respectively. Each molecule is allowed to pack into the following space groups; $P2_1/c$, $P1$, $P2_12_12_1$, $C2/c$, $P2_1$, $Pbca$, $Pna2_1$, $Pbcn$, $Cc$, and $C2$, out of which the crystal with lowest lattice energy is selected for the further studies [S4, S5].

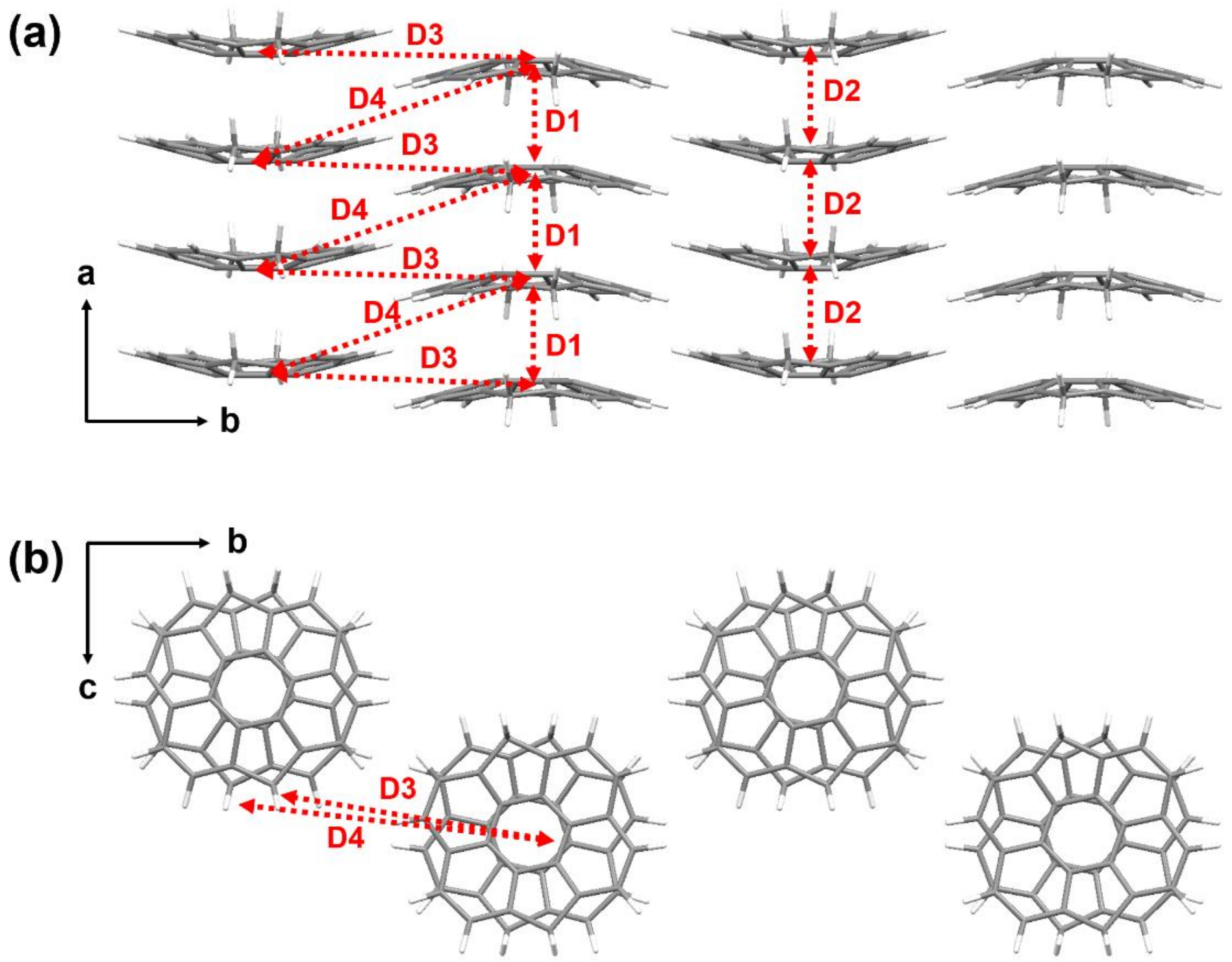

**Figure S2.** Crystal structure of DCPP1 showing different conducting channels: (a) side view and (b) top view.

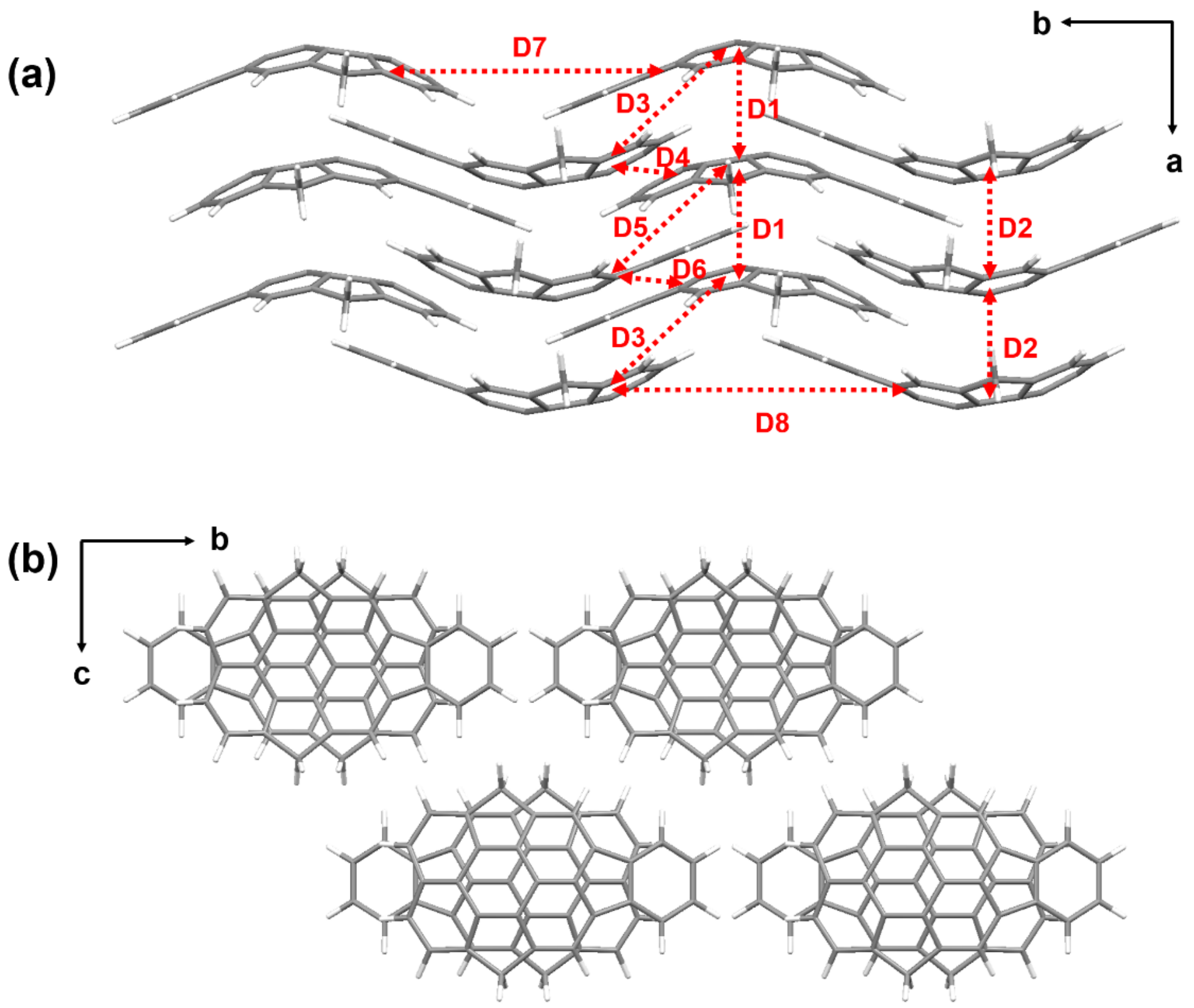

**Figure S3.** Crystal structure of DCPP2 showing different conducting channels: (a) side view and (b) top view.

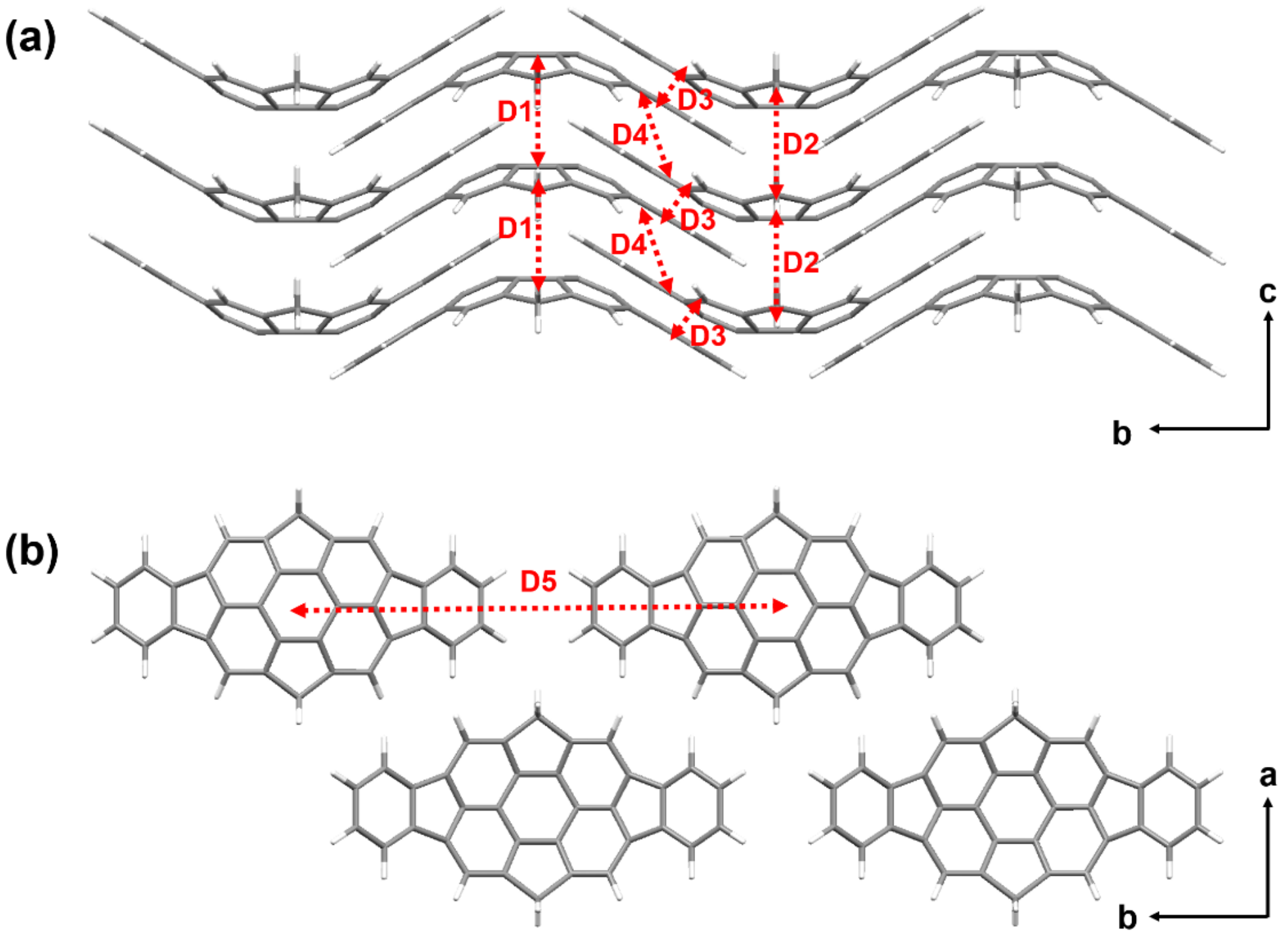

**Figure S4.** Crystal structure of DCPP3 showing different conducting channels: (a) side view and (b) top view.

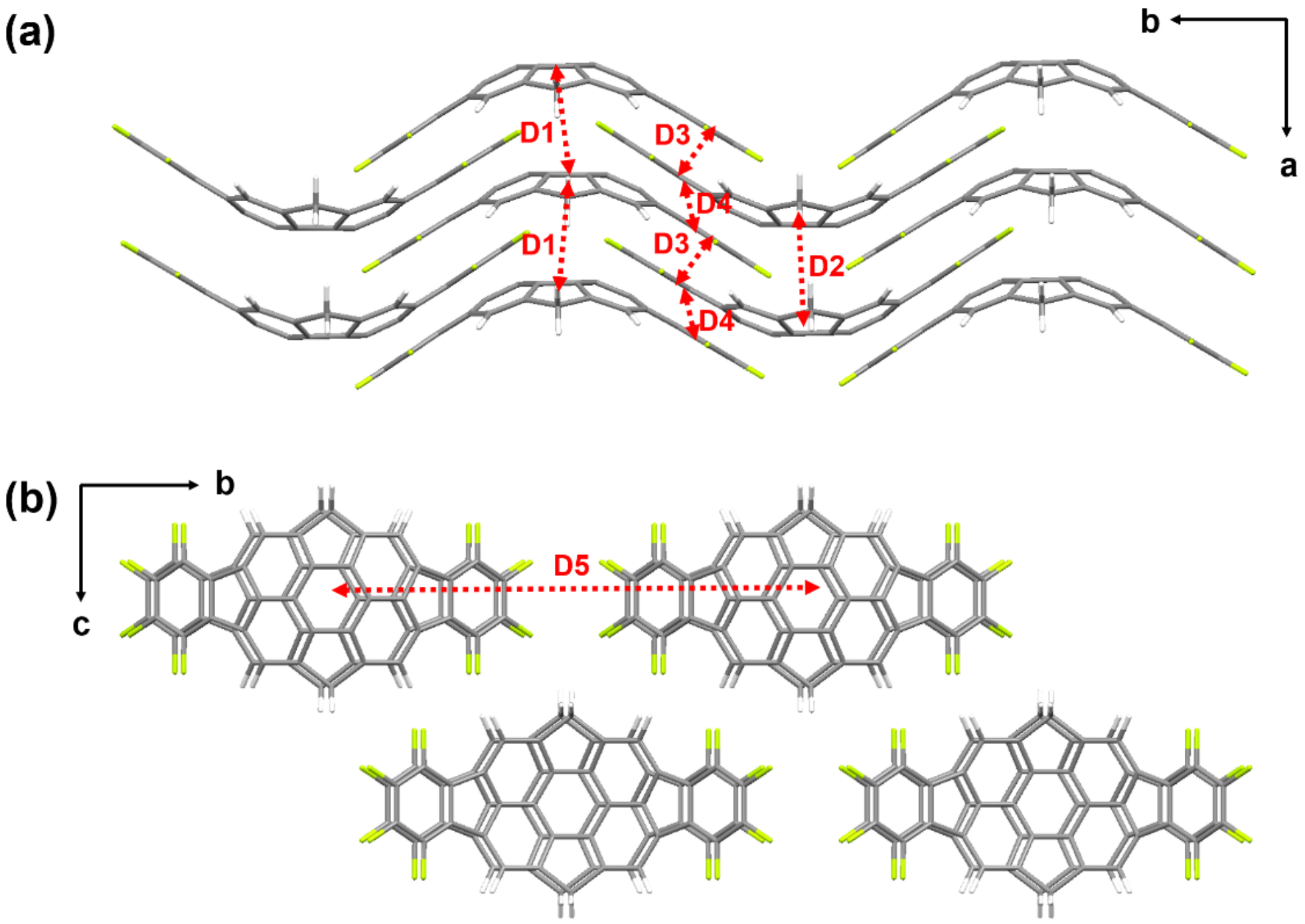

**Figure S5.** Crystal structure of DCPP4 showing different conducting channels: (a) side view and (b) top view.

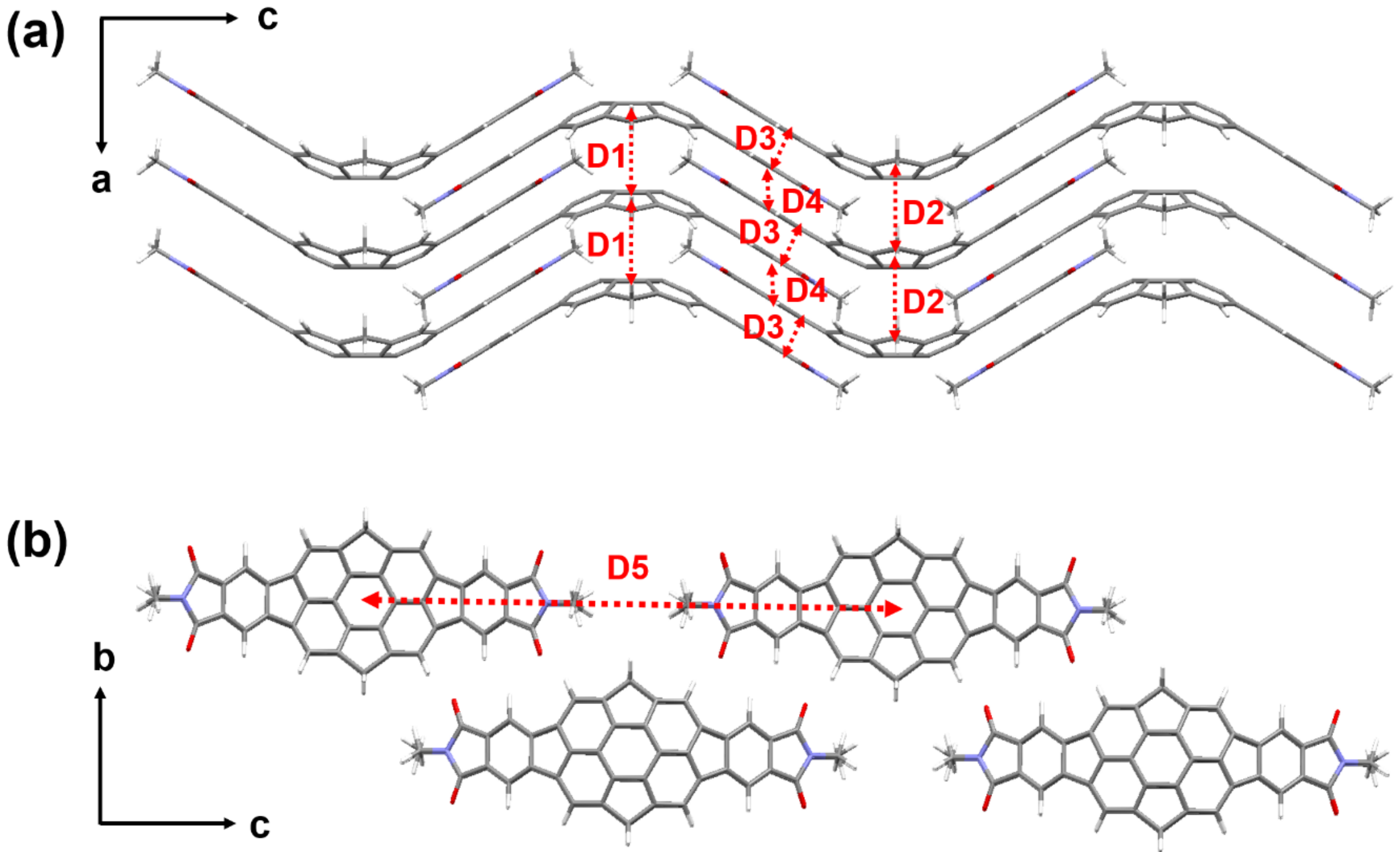

**Figure S6.** Crystal structure of DCPP5 showing different conducting channels: (a) side view and (b) top view.

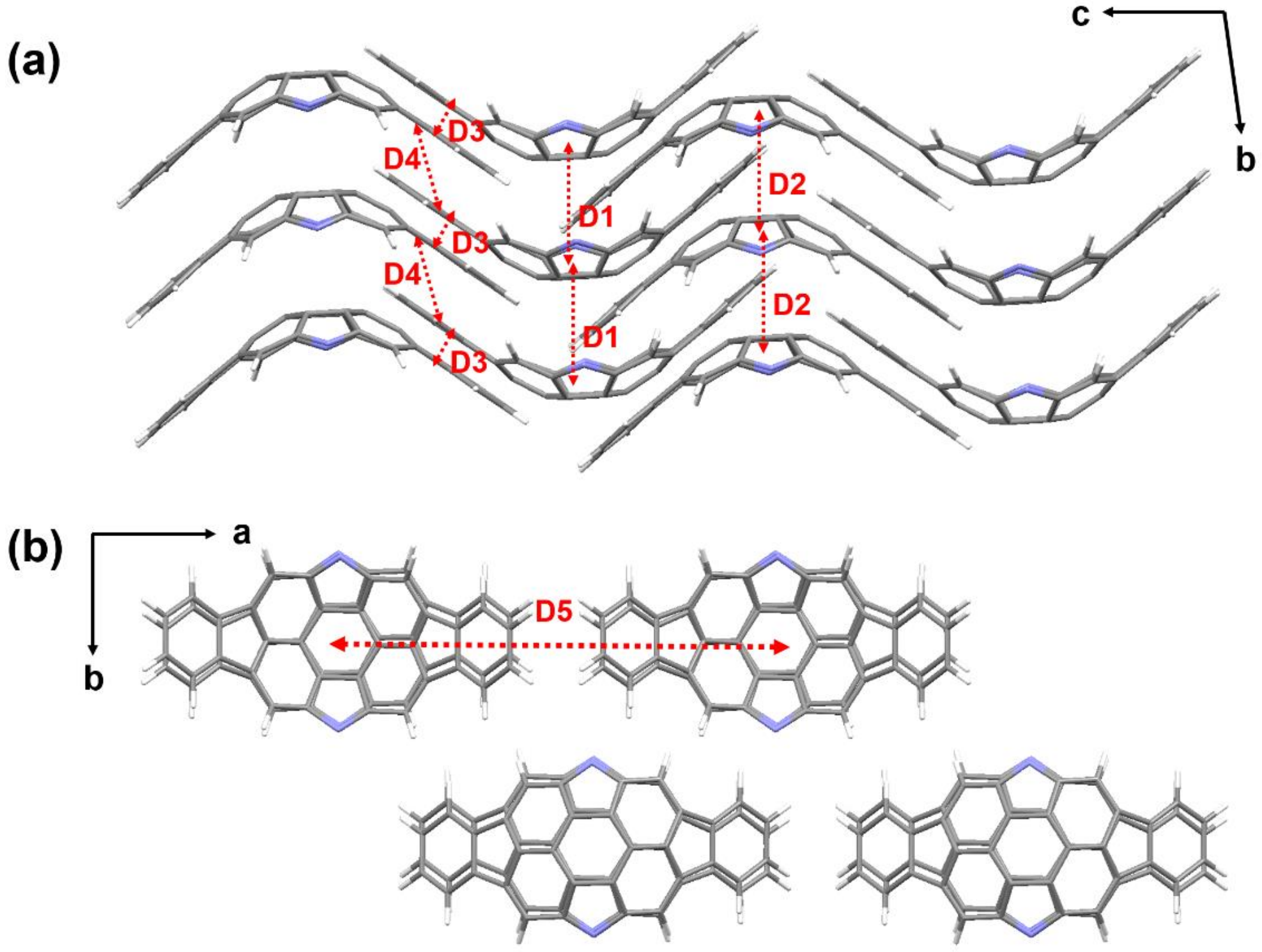

**Figure S7.** Crystal structure of DCPP6 showing different conducting channels: (a) side view and (b) top view.

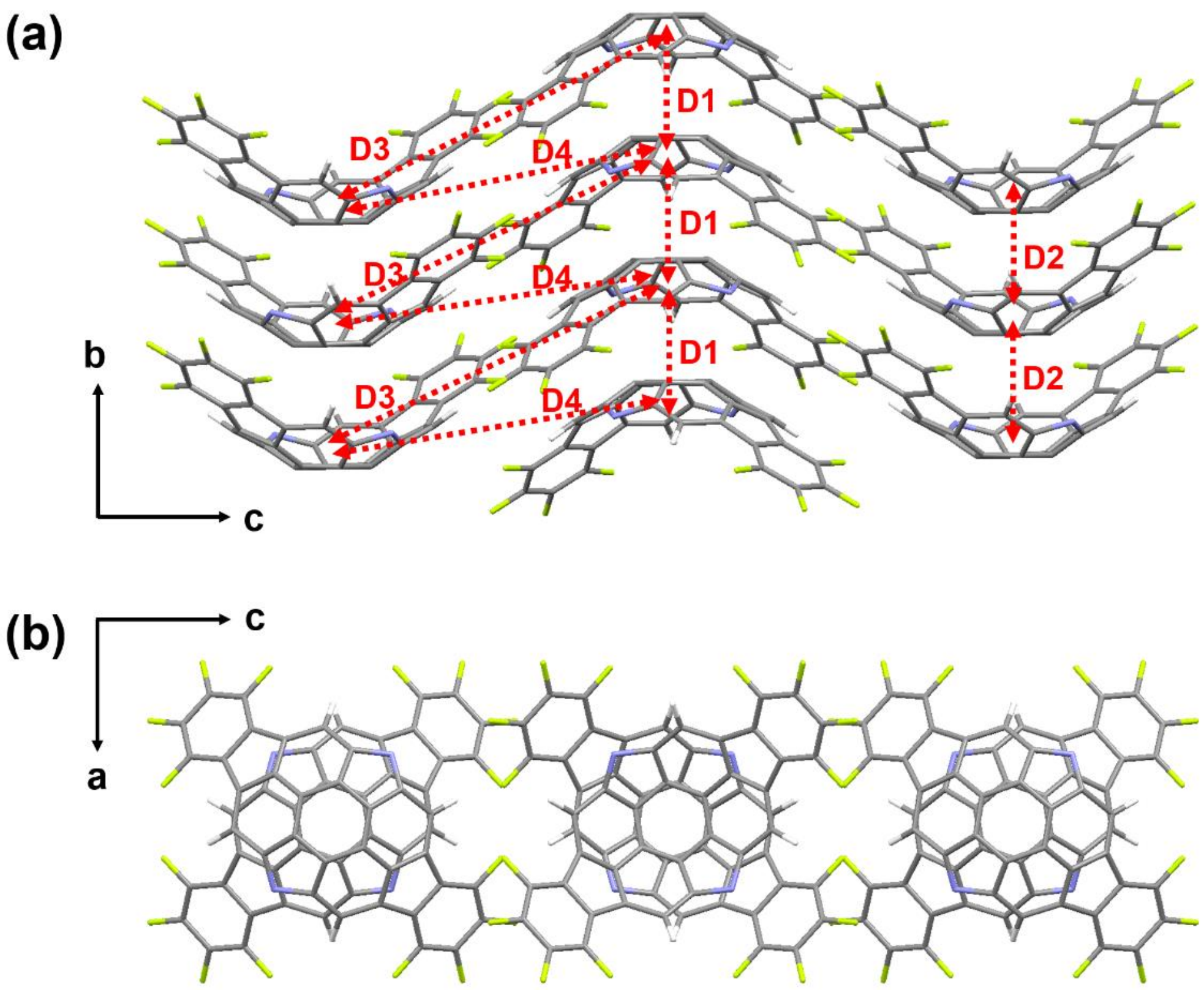

**Figure S8.** Crystal structure of DCPP7 showing different conducting channels: (a) side view and (b) top view.

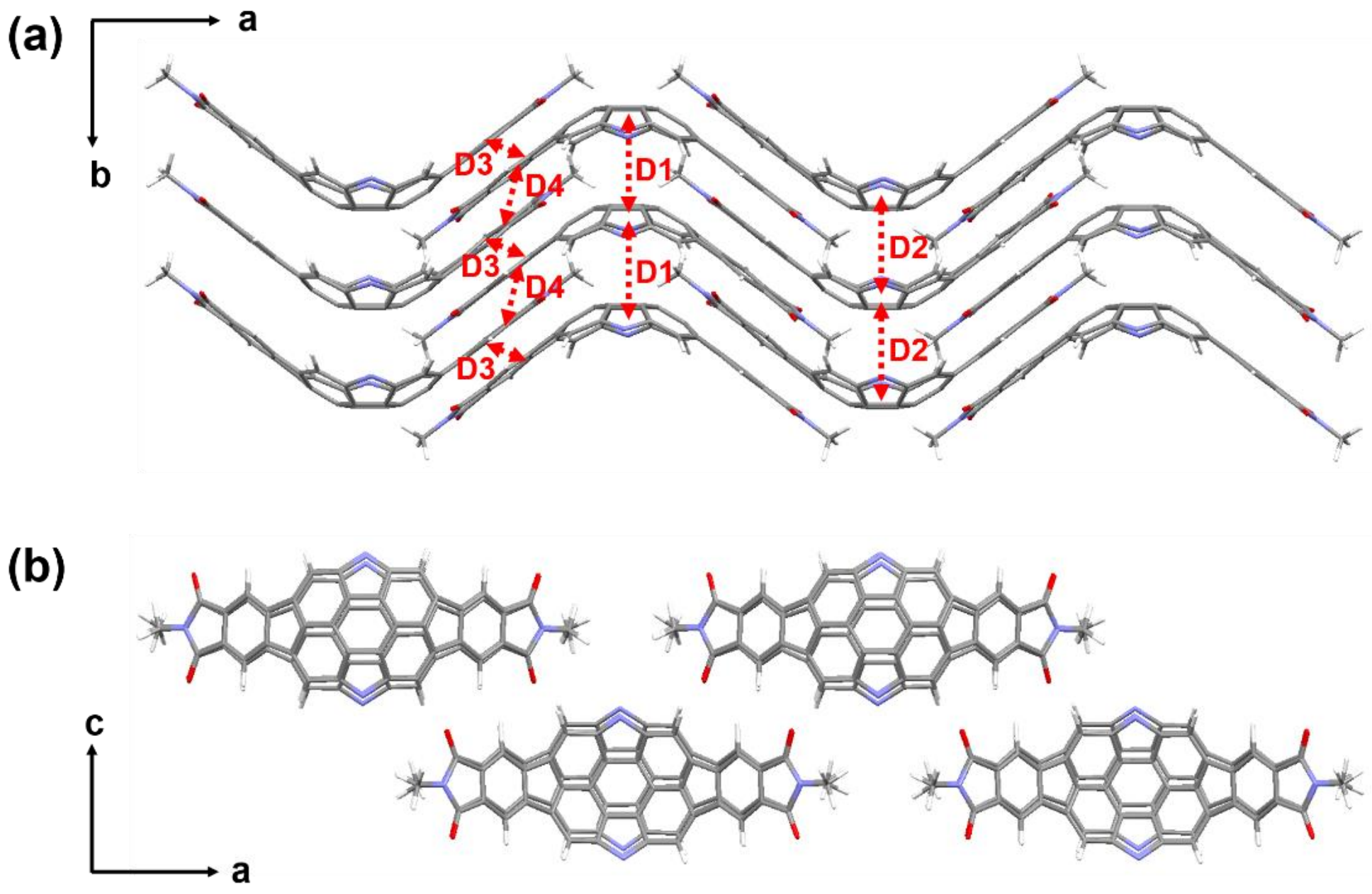

**Figure S9.** Crystal structure of DCPP8 showing different conducting channels: (a) side view and (b) top view.

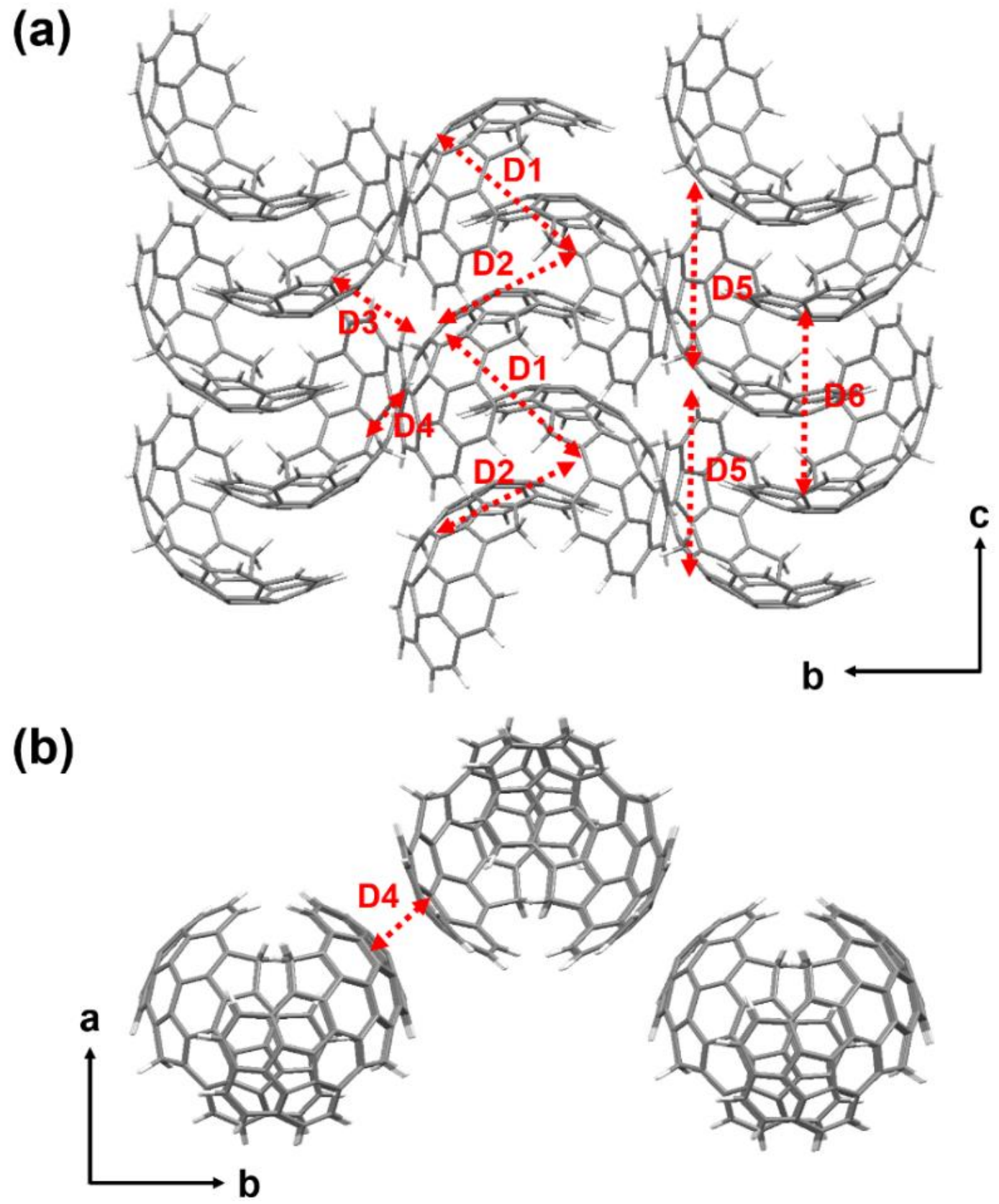

**Figure S10.** Crystal structure of DCPP9 showing different conducting channels: (a) side view and (b) top view.

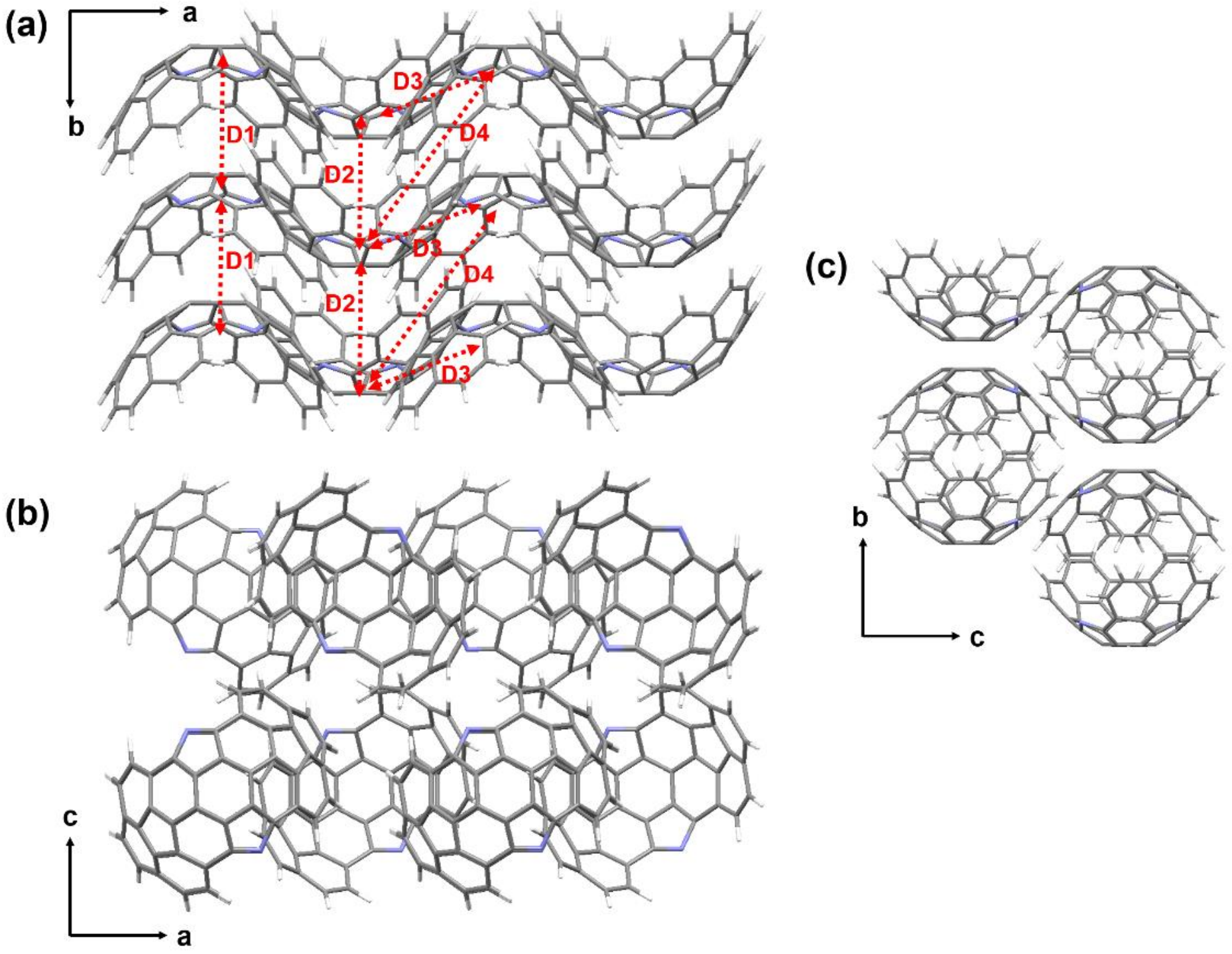

**Figure S11.** Crystal structure of DCPP10 showing different conducting channels: (a) side view and (b) top view.

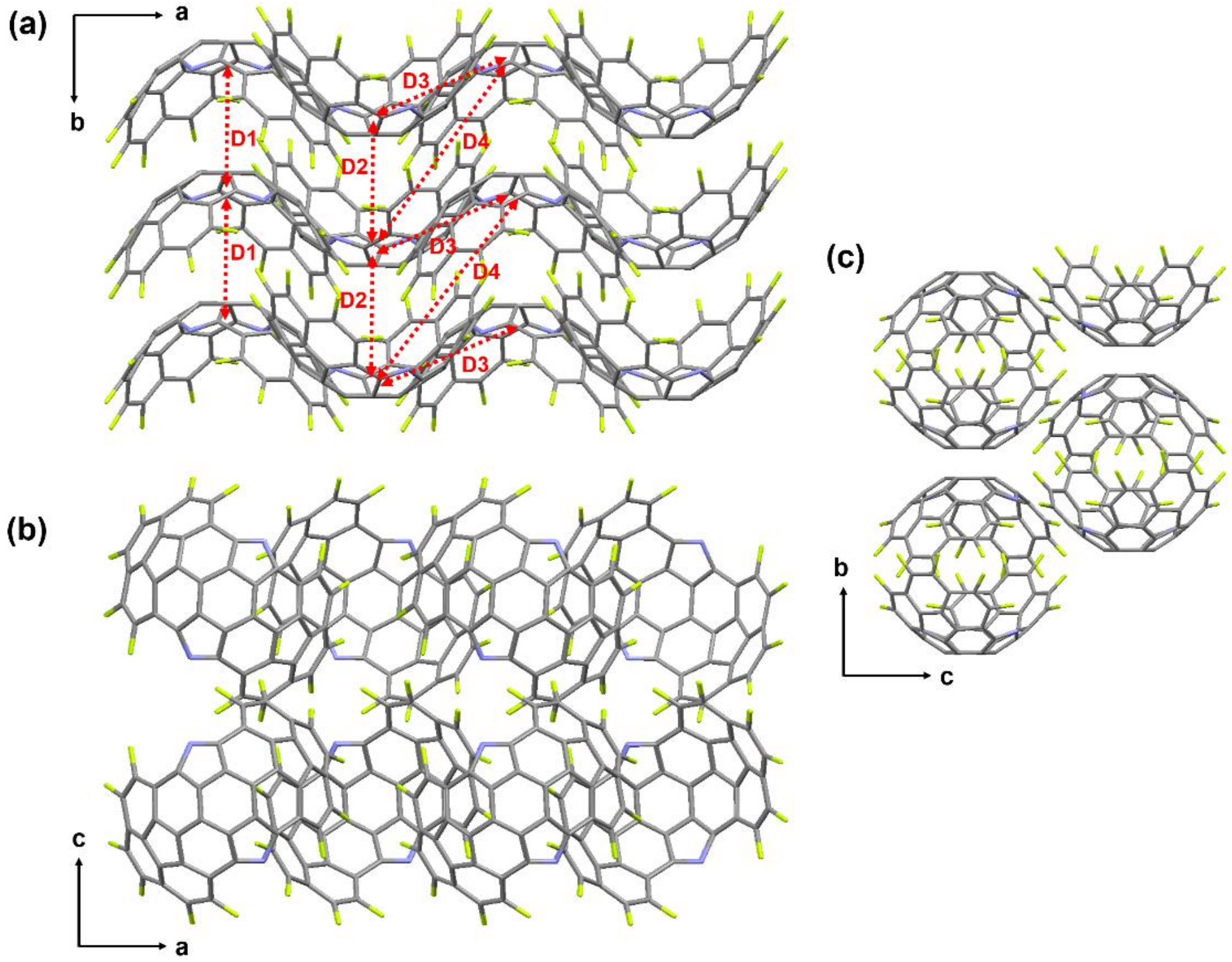

**Figure S12.** Crystal structure of DCPP11 showing different conducting channels: (a) side view and (b) top view.

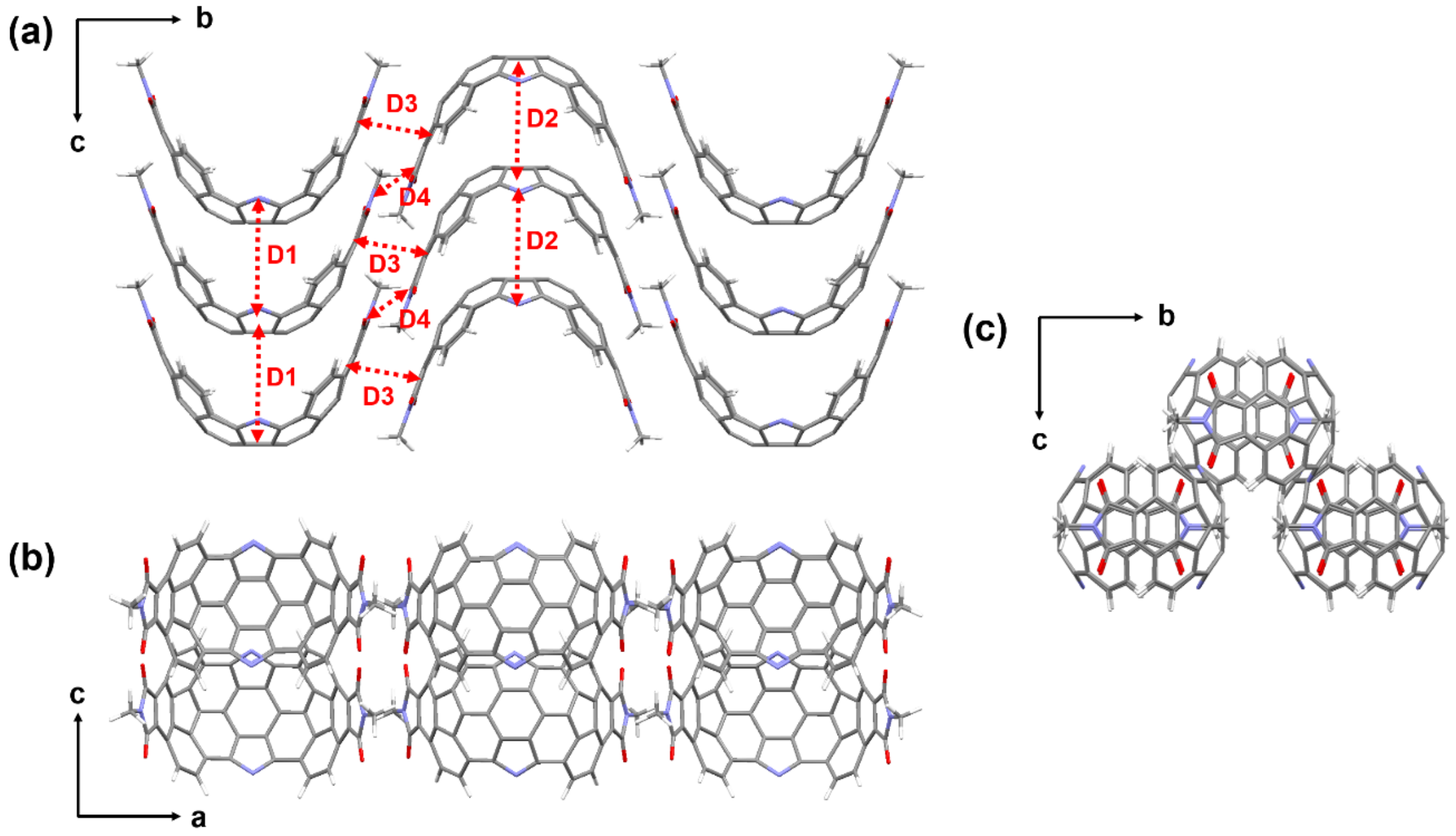

**Figure S13.** Crystal structure of DCPP12 showing different conducting channels: (a) side view and (b) top view.

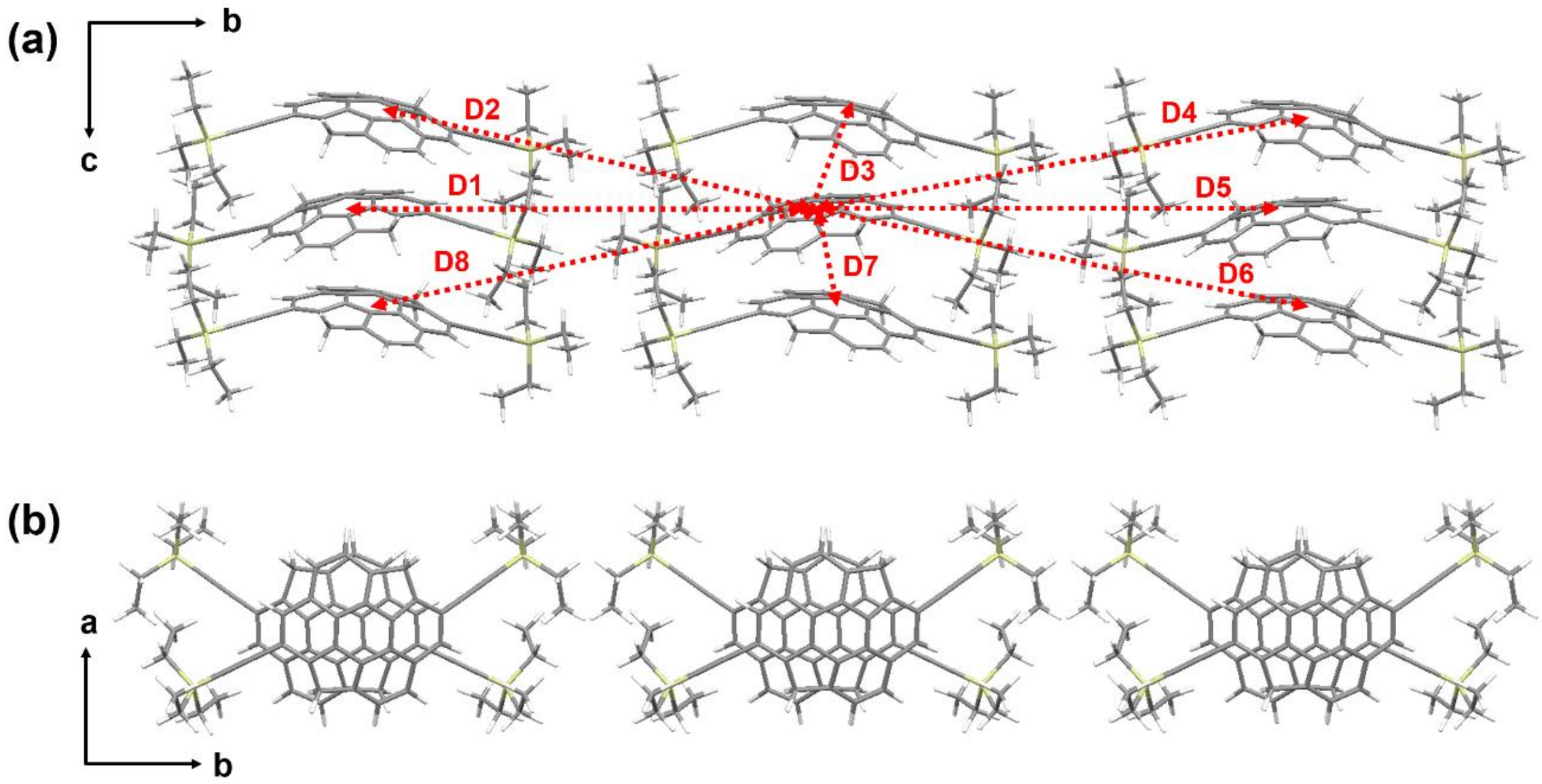

**Figure S14.** Crystal structure of DCPP1-TES showing different conducting channels: (a) side view and (b) top view.

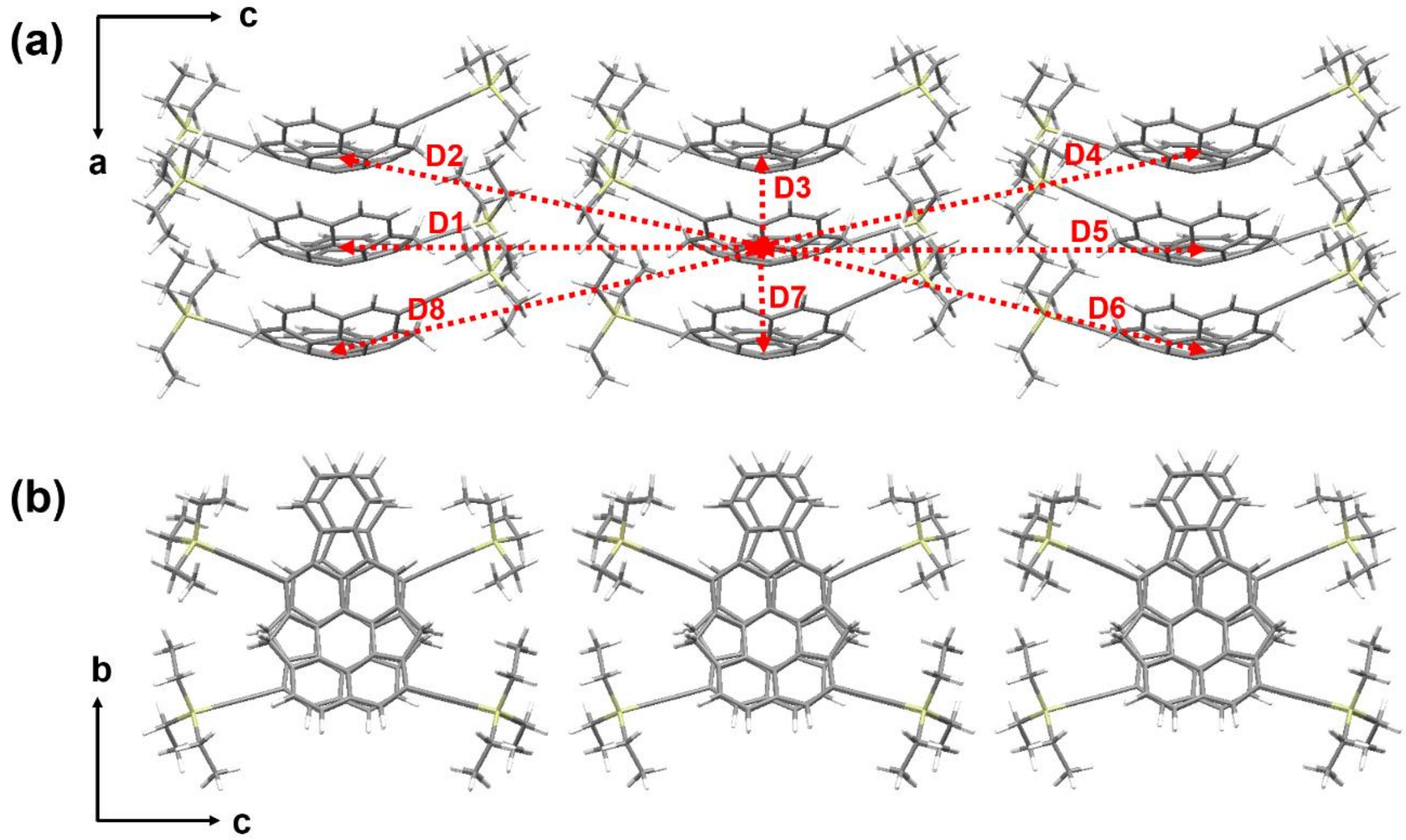

**Figure S15.** Crystal structure of DCPP2-TES showing different conducting channels: (a) side view and (b) top view.

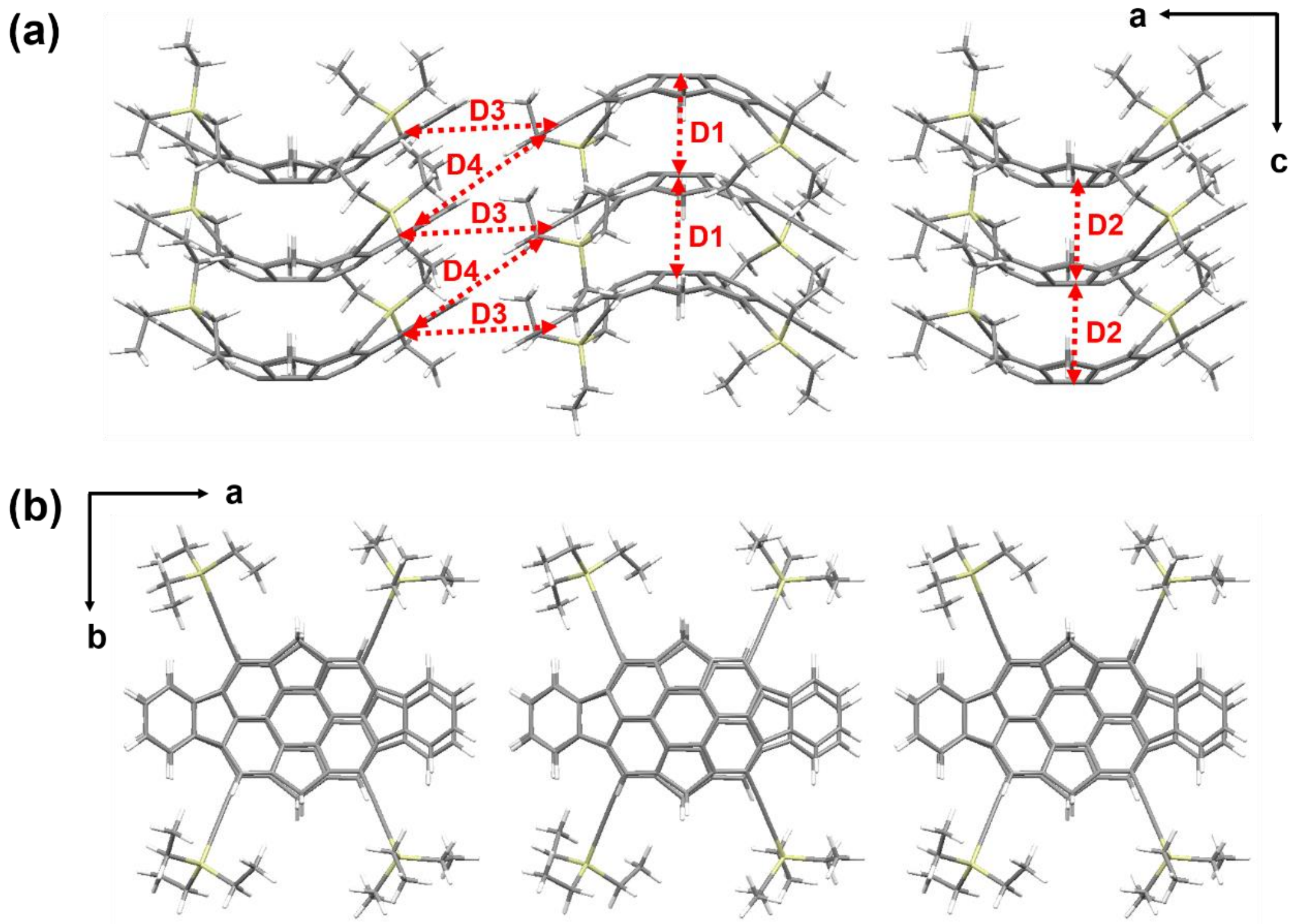

**Figure S16.** Crystal structure of DCPP3-TES showing different conducting channels: (a) side view and (b) top view.

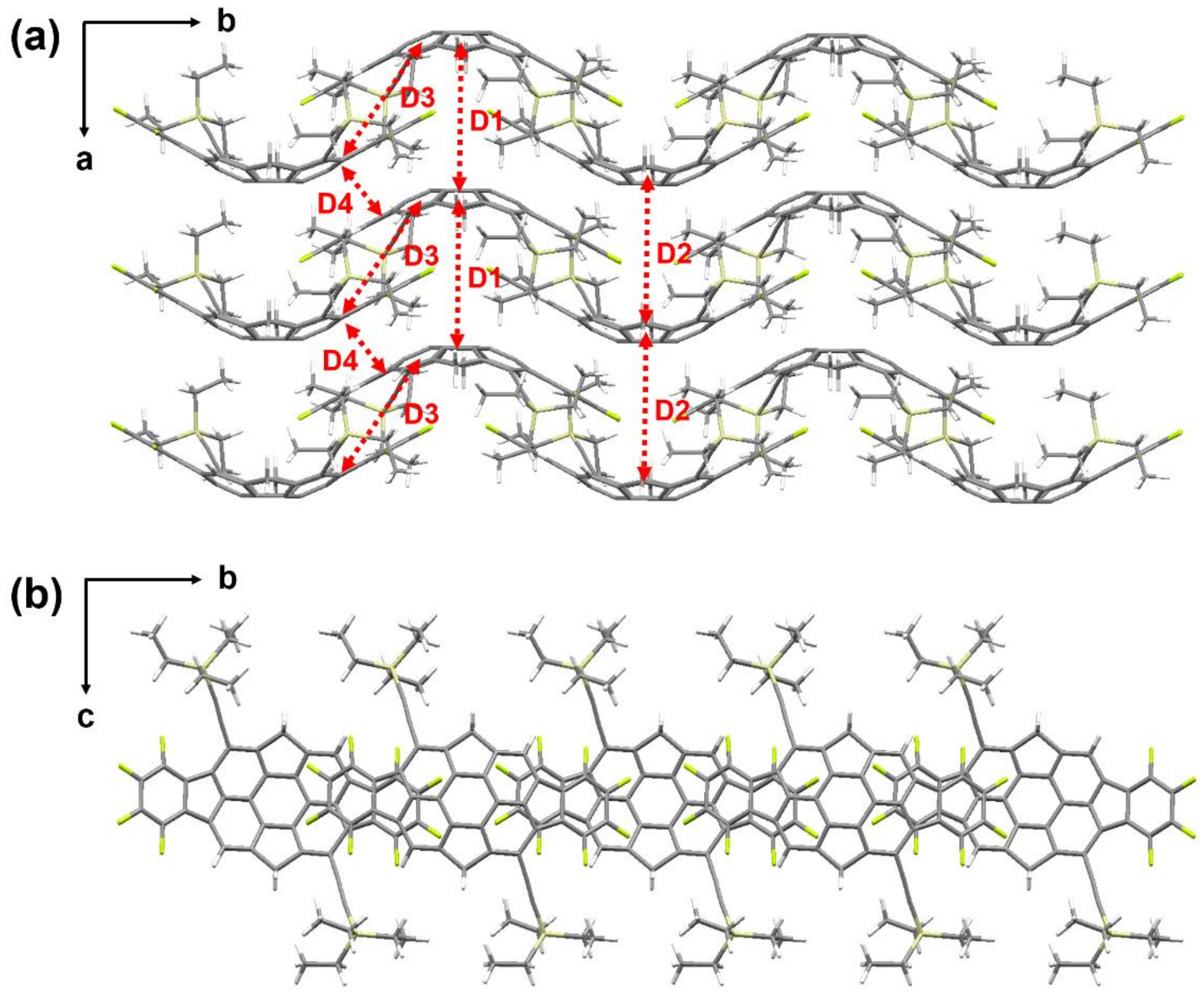

**Figure S17.** Crystal structure of DCPP4-TES showing different conducting channels: (a) side view and (b) top view.

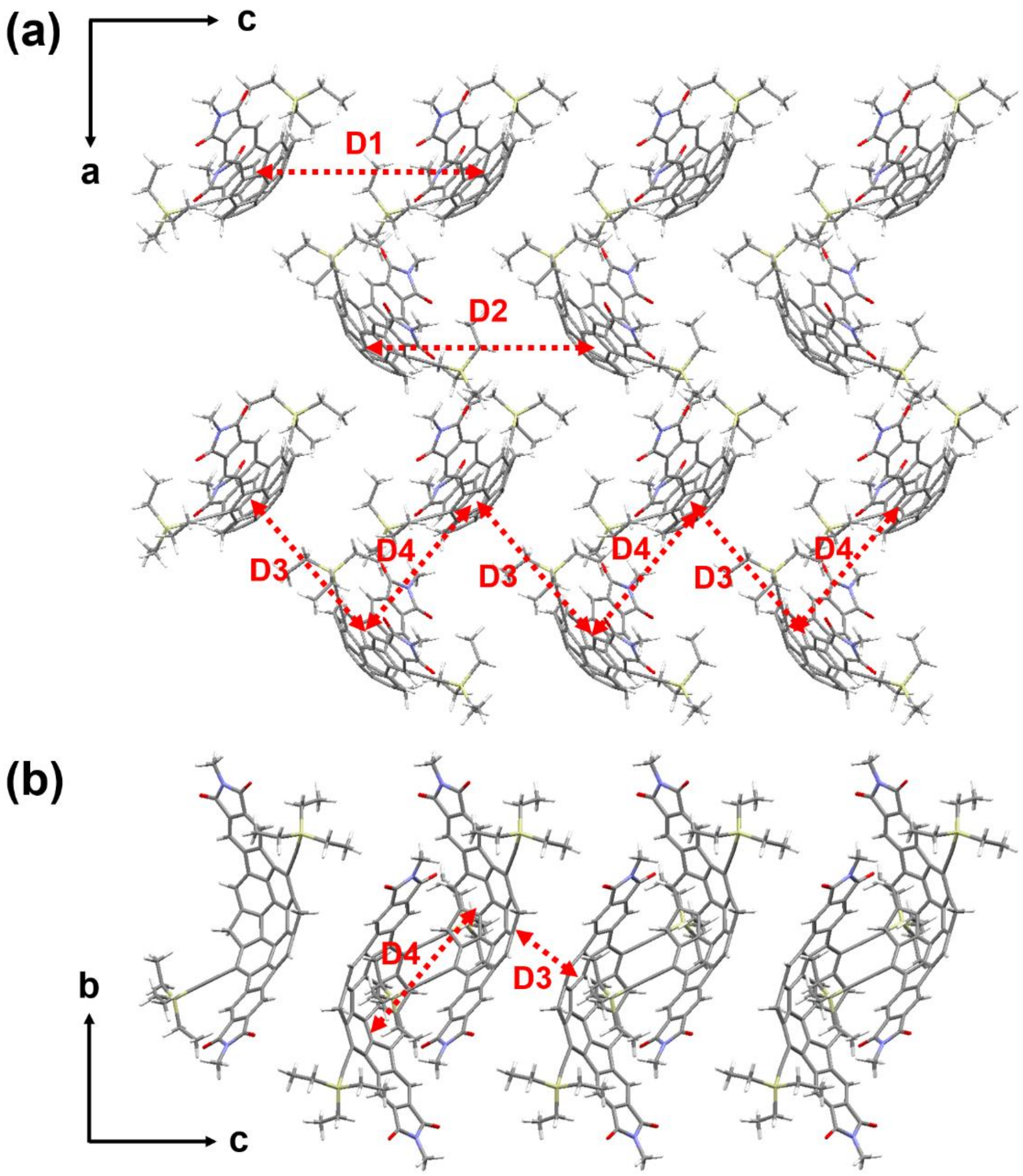

**Figure S18.** Crystal structure of DCPP5-TES showing different conducting channels: (a) side view and (b) top view.

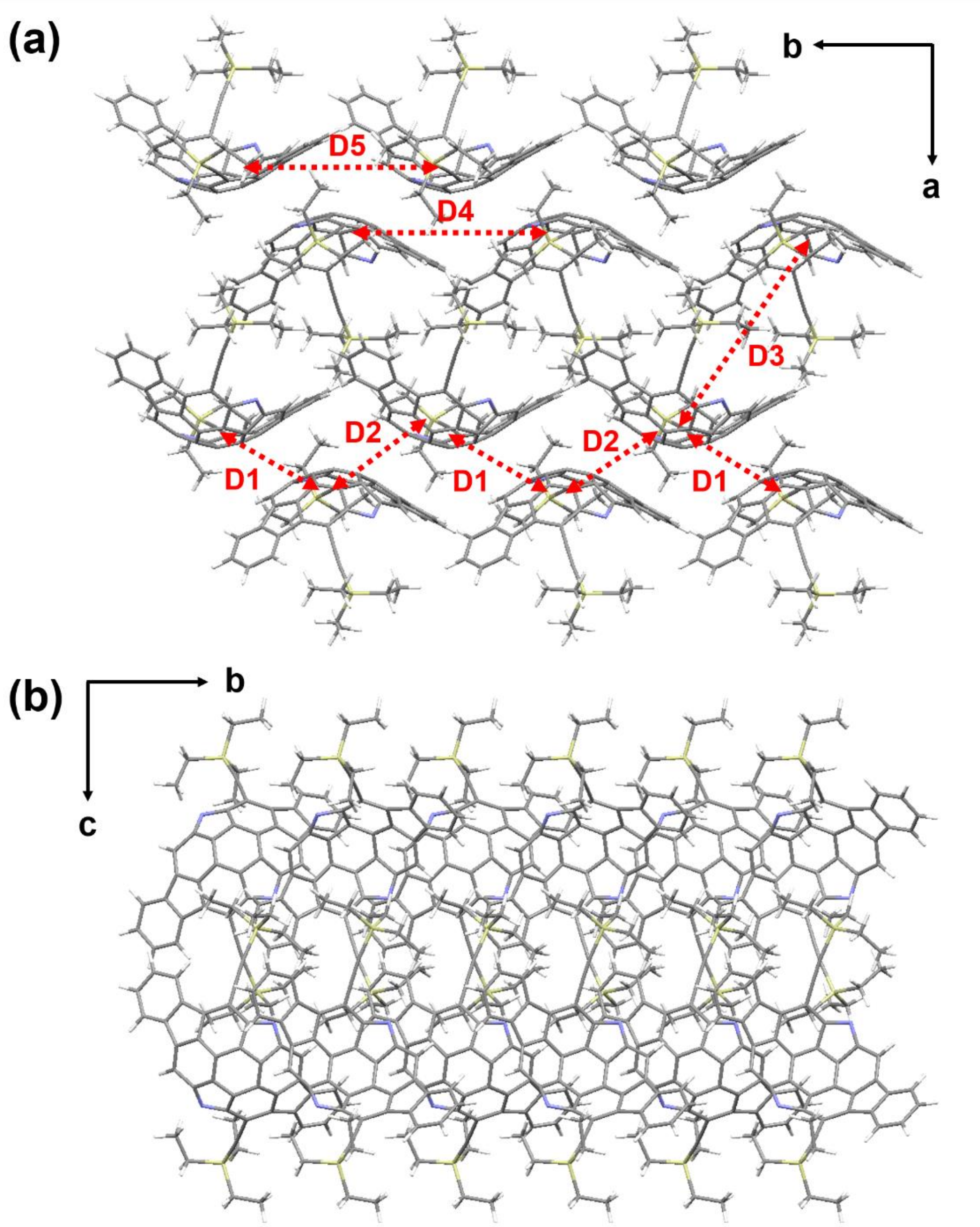

**Figure S19.** Crystal structure of DCPP6-TES showing different conducting channels: (a) side view and (b) top view.

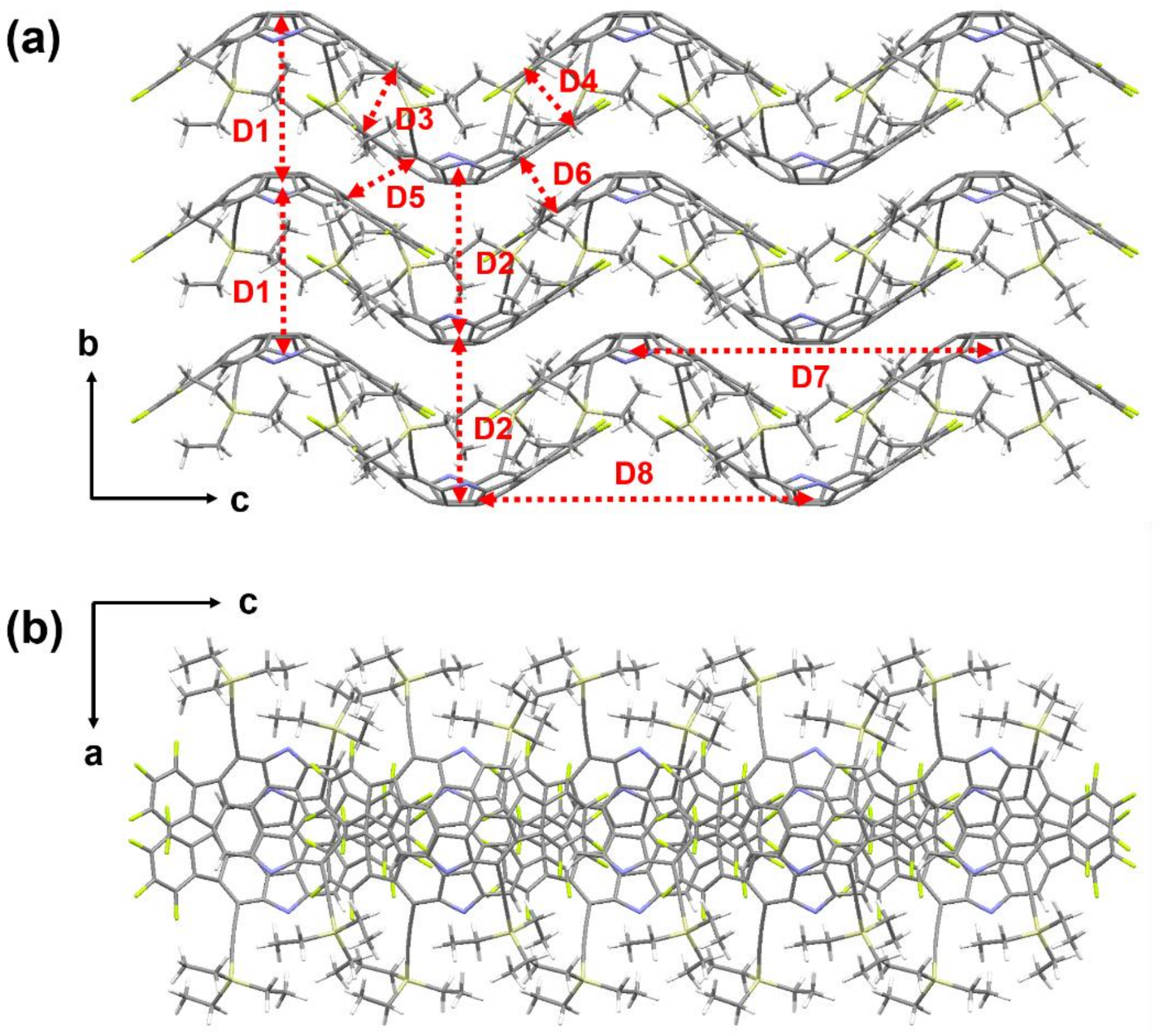

**Figure S20.** Crystal structure of DCPP7-TES showing different conducting channels: (a) side view and (b) top view.

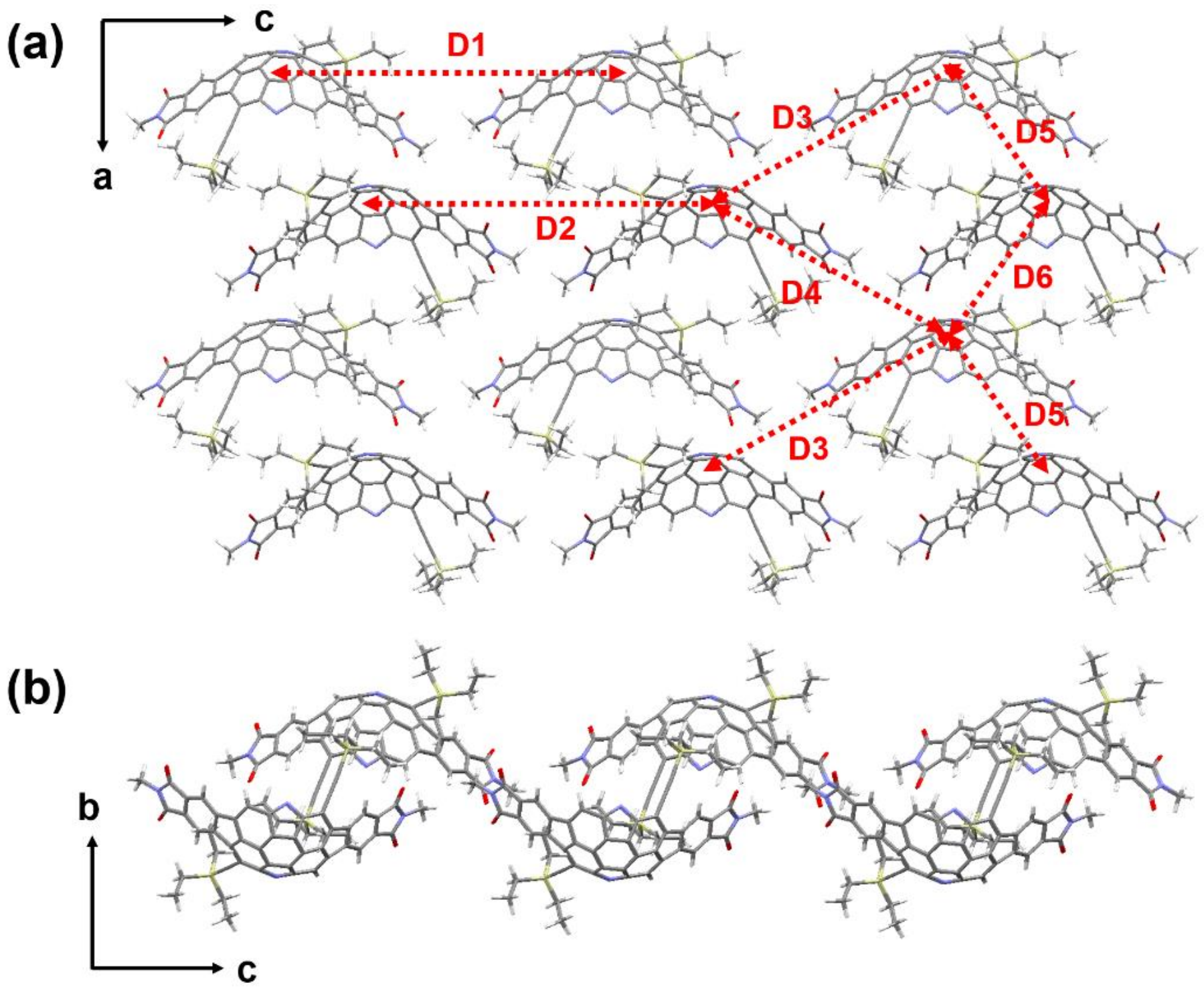

**Figure S21.** Crystal structure of DCPP8-TES showing different conducting channels: (a) side view and (b) top view.

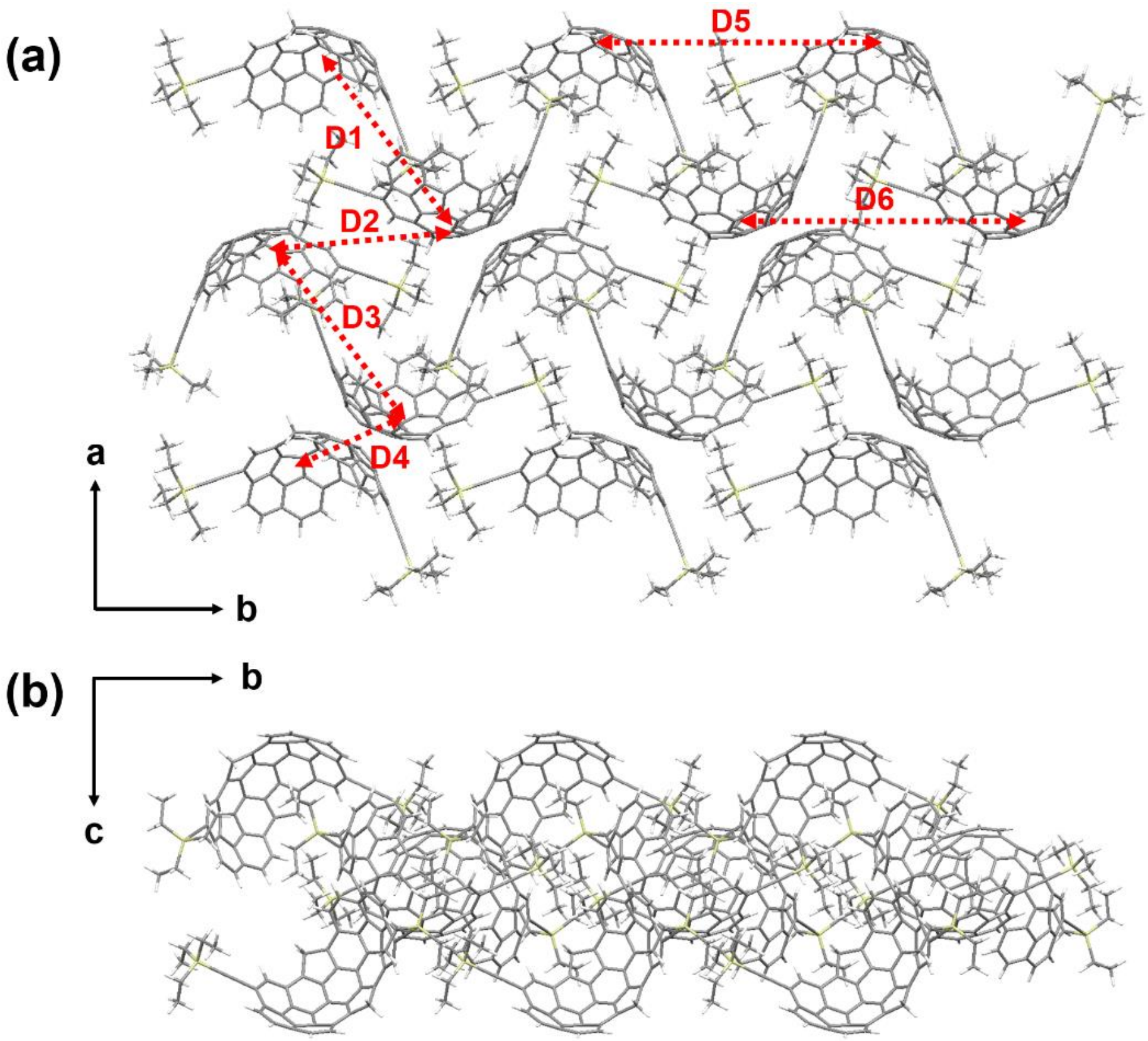

**Figure S22.** Crystal structure of DCPP9-TES showing different conducting channels: (a) side view and (b) top view.

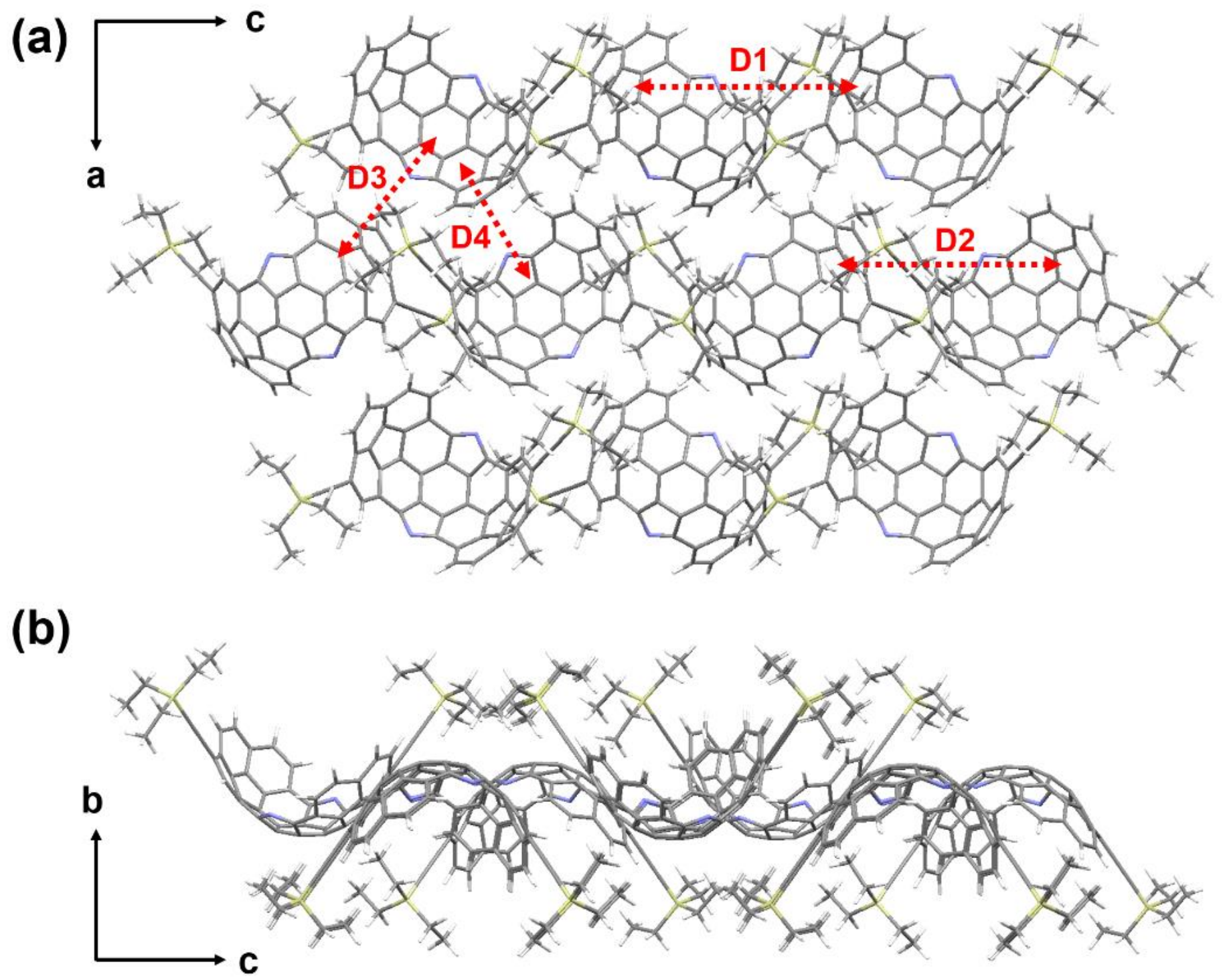

**Figure S23.** Crystal structure of DCPP10-TES showing different conducting channels: (a) side view and (b) top view.

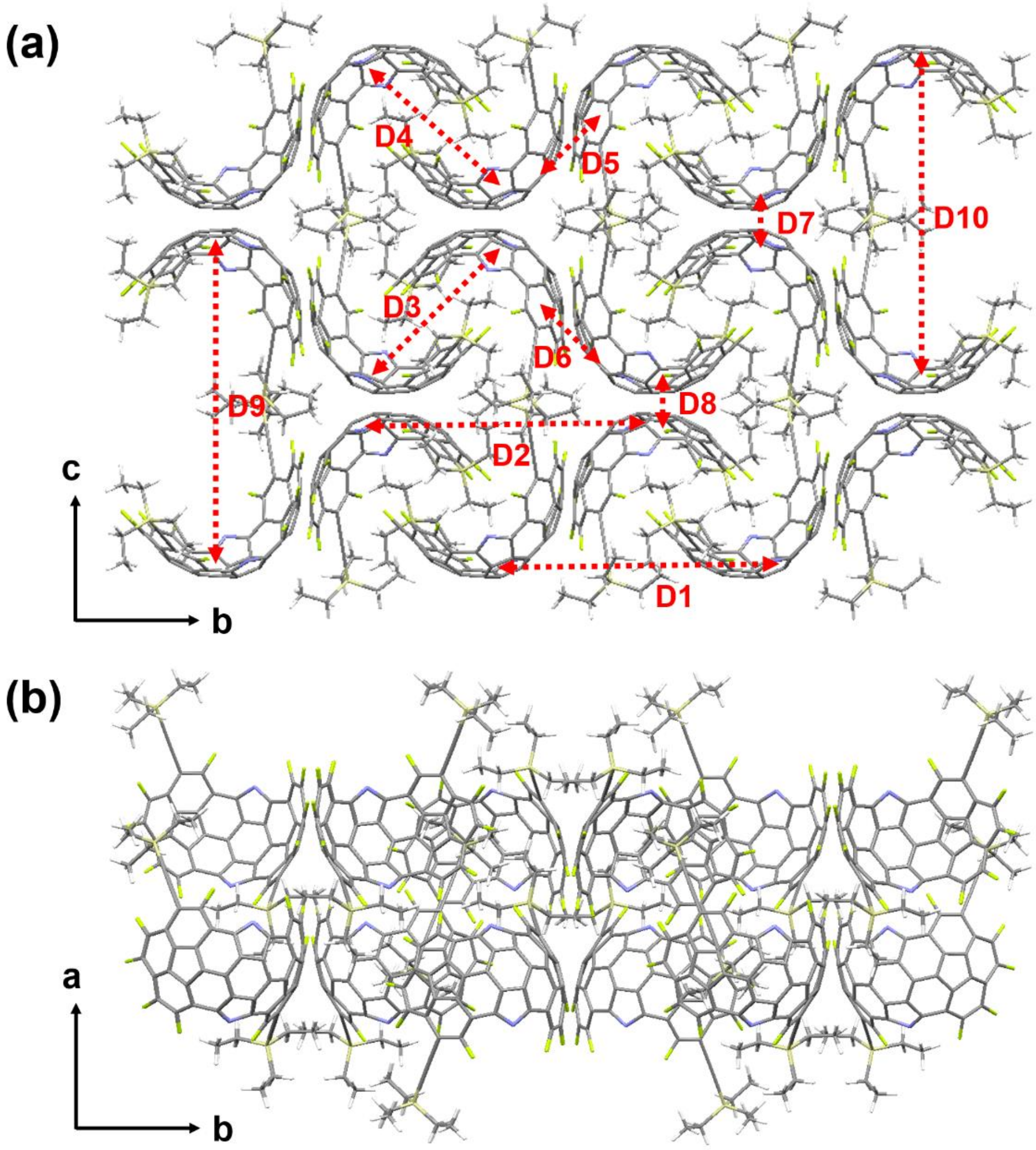

**Figure S24.** Crystal structure of DCPP11-TES showing different conducting channels: (a) side view and (b) top view.

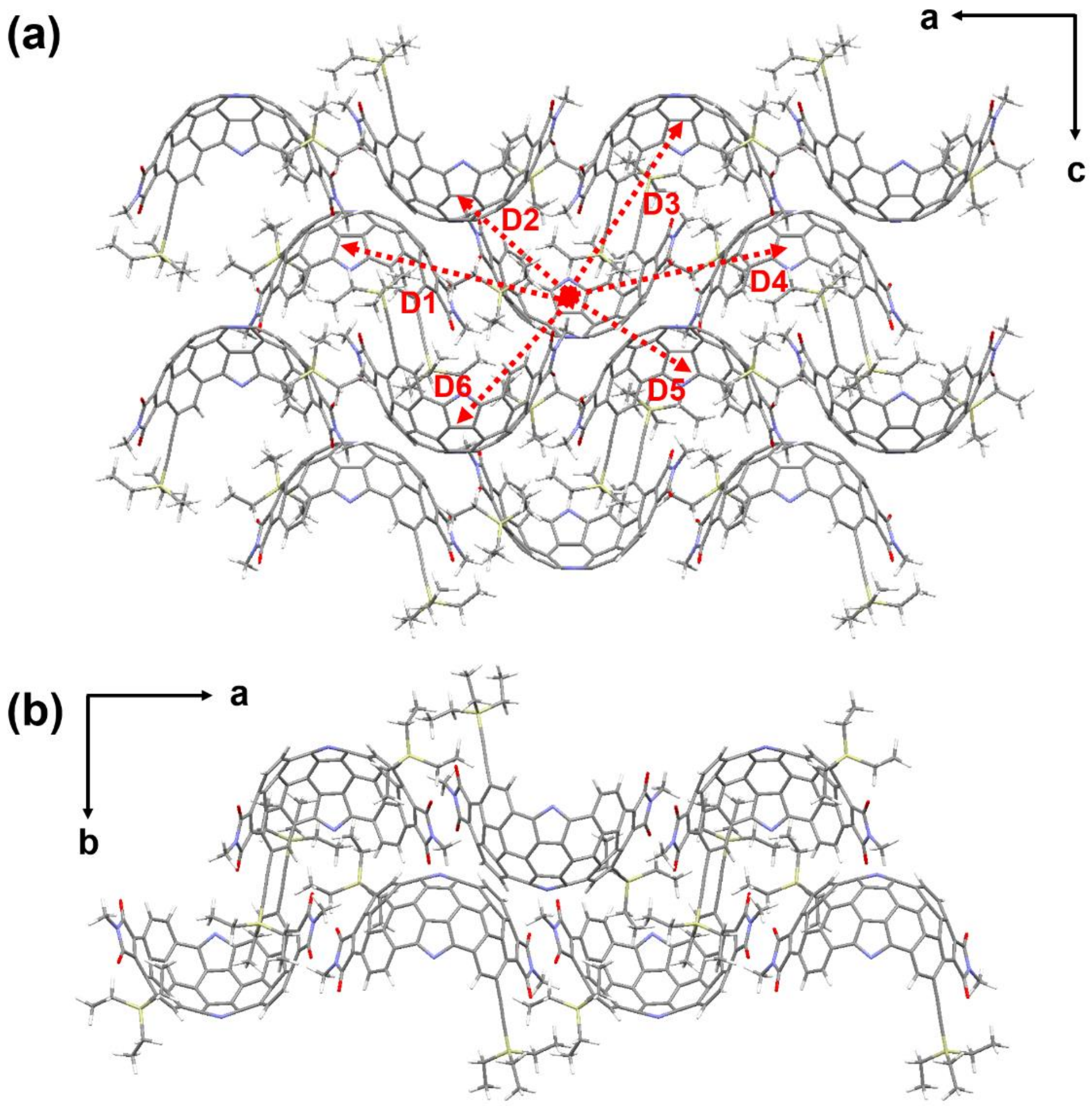

**Figure S25.** Crystal structure of DCPP12-TES showing different conducting channels: (a) side view and (b) top view.

**Table S8.** The space groups and unit cell parameters of the crystal structures of corresponding DCPP- and DCPP-TES-based compounds.

| Compound | Crystal System | Space group | Lattice parameters | | | | | |
|---|---|---|---|---|---|---|---|---|
| | | | a | b (Å) | c | α | β (Å) | γ |
| DCPP-1 | Orthorhombic | $Pna2_1$ | 7.34 | 18.99 | 9.94 | 90.0 | 90.0 | 90.0 |
| DCPP-2 | Orthorhombic | $Pna2_1$ | 7.59 | 14.51 | 15.46 | 90.0 | 90.0 | 90.0 |
| DCPP-3 | Orthorhombic | $P2_12_12_1$ | 14.52 | 17.18 | 8.03 | 90.0 | 90.0 | 90.0 |
| DCPP-4 | Orthorhombic | $Pna2_1$ | 8.06 | 17.99 | 15.39 | 90.0 | 90.0 | 90.0 |
| DCPP-5 | Orthorhombic | $Pna2_1$ | 8.18 | 13.65 | 24.52 | 90.0 | 90.0 | 90.0 |
| DCPP-6 | Monoclinic | $P2_1/c$ | 15.81 | 14.74 | 8.53 | 90.0 | 97.69 | 90.0 |
| DCPP-7 | Orthorhombic | *Pbcn* | 21.63 | 8.48 | 23.64 | 90.0 | 90.0 | 90.0 |
| DCPP-8 | Orthorhombic | $P2_12_12_1$ | 22.74 | 8.78 | 13.58 | 90.0 | 90.0 | 90.0 |
| DCPP-9 | Monoclinic | $P2_1/c$ | 13.37 | 22.92 | 8.87 | 90.0 | 110.68 | 90.0 |
| DCPP-10 | Orthorhombic | $Pna2_1$ | 12.33 | 11.09 | 17.57 | 90.0 | 90.0 | 90.0 |
| DCPP-11 | Orthorhombic | $Pna2_1$ | 13.05 | 11.71 | 18.15 | 90.0 | 90.0 | 90.0 |
| DCPP-12 | Orthorhombic | $Pna2_1$ | 26.46 | 11.23 | 10.91 | 90.0 | 90.0 | 90.0 |
| DCPP-TES-1 | Orthorhombic | *Pbca* | 41.19 | 21.03 | 8.72 | 90.0 | 90.0 | 90.0 |
| DCPP-TES-2 | Orthorhombic | *Pbca* | 8.03 | 54.84 | 18.54 | 90.0 | 90.0 | 90.0 |
| DCPP-TES-3 | Orthorhombic | *Pbca* | 32.97 | 34.16 | 8.34 | 90.0 | 90.0 | 90.0 |
| DCPP-TES-4 | Orthorhombic | *Pbca* | 7.60 | 18.07 | 73.32 | 90.0 | 90.0 | 90.0 |
| DCPP-TES-5 | Orthorhombic | $P2_12_12_1$ | 19.52 | 20.55 | 13.69 | 90.0 | 90.0 | 90.0 |
| DCPP-TES-6 | Orthorhombic | *Pbca* | 27.18 | 12.37 | 27.08 | 90.0 | 90.0 | 90.0 |
| DCPP-TES-7 | Orthorhombic | *Pbcn* | 39.57 | 15.87 | 17.38 | 90.0 | 90.0 | 90.0 |
| DCPP-TES-8 | Orthorhombic | $P2_12_12_1$ | 17.24 | 14.69 | 21.55 | 90.0 | 90.0 | 90.0 |
| DCPP-TES-9 | Orthorhombic | $P2_12_12_1$ | 25.80 | 17.61 | 12.40 | 90.0 | 90.0 | 90.0 |
| DCPP-TES-10 | Orthorhombic | *Pbca* | 18.13 | 23.50 | 24.12 | 90.0 | 90.0 | 90.0 |
| DCPP-TES-11 | Orthorhombic | *Pbca* | 15.90 | 30.68 | 21.51 | 90.0 | 90.0 | 90.0 |
| DCPP-TES-12 | Orthorhombic | $P2_12_12_1$ | 26.34 | 15.34 | 13.85 | 90.0 | 90.0 | 90.0 |

**Table S9.** The notable absorption wavelengths ($\lambda_{abs}$), oscillator strengths ($f$), excitation energies, and major spectral compositions of the DCPP and DCPP-TES compounds, simulated in the TD-DFT framework in dichloromethane solvent, using CAM-B3LYP/6-311G(d,p) level of theory.

| Compound | states | $\lambda_{abs}$ (nm) | $f$ | energy (eV) | Major compositions |
|---|---|---|---|---|---|
| DCPP-1 | $S_1$ | 398.0 | 0.440 | 3.115 | H→L (98%) |
| | $S_9$ | 235.6 | 1.323 | 5.264 | H-1→L (45%), H→L+2 (54%) |
| | $S_{12}$ | 206.4 | 0.528 | 6.007 | H-2→L+1 (59%), H-1→L+2 (25%) |
| DCPP-2 | $S_1$ | 450.1 | 0.662 | 2.754 | H→L (98%) |
| | $S_9$ | 254.7 | 0.551 | 4.867 | H-3→L (26%), H-1→L+1 (53%) |
| | $S_{18}$ | 209.4 | 0.786 | 5.922 | H-4→L+1 (54%), H-2→L+4 (17%) |
| DCPP-3 | $S_1$ | 490.2 | 0.860 | 2.530 | H→L (98%) |
| | $S_9$ | 276.7 | 0.584 | 4.481 | H-6→L (12%), H-2→L+4 (10%), H-1→L+2 (66%) |
| | $S_{15}$ | 242.8 | 0.801 | 5.107 | H-6→L (36%), H-1→L+3 (45%) |
| DCPP-4 | $S_1$ | 501.1 | 0.897 | 2.474 | H→L (98%) |
| | $S_6$ | 304.3 | 0.146 | 4.074 | H-4→L (69%) |
| | $S_{11}$ | 256.2 | 0.819 | 4.839 | H-6→L (22%), H-1→L+2 (51%) |
| DCPP-5 | $S_1$ | 521.6 | 1.116 | 2.377 | H→L (96%) |
| | $S_4$ | 353.3 | 0.114 | 3.509 | H→L+1 (80%) |
| | $S_6$ | 319.3 | 0.178 | 3.883 | H-3→L (66%), H→L+7 (10%) |
| | $S_{15}$ | 277.9 | 1.314 | 4.461 | H-1→L+2 (49%), H→L+3 (35%) |
| DCPP-6 | $S_3$ | 512.8 | 0.071 | 2.418 | H-4→L (22%), H-2→L (67%) |
| | $S_5$ | 412.1 | 0.361 | 3.009 | H-5→L (27%), H-4→L (51%) |
| | $S_9$ | 332.4 | 0.458 | 3.731 | H-8→L (17%), H→L+1 (59%) |
| | $S_{17}$ | 270.9 | 1.018 | 4.577 | H-1→L+3 (66%) |
| DCPP-7 | $S_3$ | 523.3 | 0.071 | 2.370 | H-3→L (24%), H-2→L (64%) |
| | $S_5$ | 417.5 | 0.370 | 2.970 | H-5→L (24%), H-3→L (51%) |
| | $S_{10}$ | 338.6 | 0.543 | 3.661 | H→L+1 (71%) |
| | $S_{17}$ | 265.9 | 1.052 | 4.663 | H-1→L+3 (71%) |
| DCPP-8 | $S_5$ | 414.4 | 0.177 | 2.992 | H-5→L (56%), H-4→L (30%) |
| | $S_7$ | 382.4 | 0.436 | 3.243 | H-11→L (33%), H-4→L (44%) |
| | $S_8$ | 357.8 | 0.592 | 3.466 | H-1→L+2 (10%), H→L+1 (62%) |
| | $S_{18}$ | 285.1 | 1.061 | 4.348 | H-13→L (16%), H-1→L+3 (50%) |
| DCPP-9 | $S_1$ | 424.4 | 0.576 | 2.922 | H→L (95%) |
| | $S_{16}$ | 280.3 | 0.822 | 4.423 | H-5→L+1 (50%), H-2→L+2 (33%) |
| | $S_{19}$ | 253.6 | 0.923 | 4.889 | H-1→L+4 (78%) |
| DCPP-10 | $S_3$ | 611.1 | 0.101 | 2.029 | H-2→L (83%) |
| | $S_6$ | 488.5 | 0.149 | 2.538 | H-7→L (16%), H-6→L (71%) |
| | $S_9$ | 377.5 | 0.178 | 3.284 | H-8→L (52%), H-7→L (29%), H-6→L (10%) |
| | $S_{13}$ | 323.6 | 0.614 | 3.831 | H-10→L (25%), H→L+1 (13%), H→L+2 (44%) |
| DCPP-11 | $S_5$ | 585.5 | 0.101 | 2.117 | H-4→L (87%) |
| | $S_6$ | 520.9 | 0.169 | 2.380 | H-6→L (78%), H-2→L (10%) |
| | $S_{14}$ | 331.0 | 0.281 | 3.746 | H-10→L (24%), H→L+1 (26%), H→L+2 (25%) |
| | $S_{16}$ | 317.4 | 0.572 | 3.906 | H-3→L+2 (14%), H-1→L+3 (22%), H-1→L+4 (10%), H→L+2 (27%) |
| DCPP-12 | $S_2$ | 876.8 | 0.035 | 1.414 | H-1→L (92%) |
| | $S_3$ | 614.4 | 0.107 | 2.018 | H-2→L (79%) |
| | $S_6$ | 496.1 | 0.174 | 2.499 | H-6→L (77%) |
| | $S_9$ | 377.0 | 0.519 | 3.289 | H-12→L (26%), H-7→L (25%), H→L+1 (27%) |
| | $S_{15}$ | 334.0 | 0.486 | 3.712 | H→L+3 (63%) |
| DCPP-TES-1 | $S_1$ | 403.7 | 0.406 | 3.072 | H→L (97%) |
| | $S_6$ | 267.3 | 2.841 | 4.638 | H-2→L (32%), H-1→L+1 (20%), H→L+2 (34%) |
| DCPP-TES-2 | $S_1$ | 457.6 | 0.604 | 2.710 | H→L (97%) |
| | $S_8$ | 277.1 | 1.460 | 4.475 | H-3→L (13%), H-2→L (10%), H-1→L+1 (16%), H→L+2 (14%), H→L+3 (25%) |
| | $S_{10}$ | 266.5 | 0.639 | 4.652 | H-4→L (13%), H-1→L+1 (25%), H→L+4 (33%) |
| DCPP-TES-3 | $S_1$ | 494.0 | 0.797 | 2.510 | H→L (97%) |
| | $S_{10}$ | 285.1 | 1.371 | 4.349 | H-2→L+1 (20%), H→L+3 (45%) |
| | $S_{19}$ | 245.8 | 1.024 | 5.045 | H-6→L (23%), H-1→L+4 (47%) |
| DCPP-TES-4 | $S_1$ | 506.8 | 0.846 | 2.446 | H→L (97%) |
| | $S_6$ | 332.8 | 0.432 | 3.725 | H-4→L (34%), H-3→L (39%), H-1→L+1 (14%) |
| | $S_9$ | 287.2 | 0.557 | 4.318 | H-8→L (23%), H-2→L+1 (21%), H→L+3 (24%) |
| DCPP-TES-5 | $S_1$ | 526.0 | 1.022 | 2.357 | H→L (96%) |
| | $S_6$ | 347.8 | 0.466 | 3.565 | H-3→L (55%), H-2→L (10%) |
| | $S_{16}$ | 281.9 | 0.857 | 4.398 | H-2→L+2 (10%), H→L+3 (27%), H→L+4 (27%) |
| DCPP-TES-6 | $S_3$ | 576.9 | 0.167 | 2.149 | H-4→L (17%), H-2→L (71%) |
| | $S_4$ | 482.2 | 0.531 | 2.571 | H-5→L (30%), H-4→L (42%), H-2→L (23%) |
| | $S_{10}$ | 353.9 | 0.716 | 3.504 | H→L+1 (63%) |
| | $S_{18}$ | 299.1 | 0.537 | 4.145 | H-16→L (34%), H-12→L (31%) |
| DCPP-TES-7 | $S_3$ | 596.0 | 0.212 | 2.080 | H-3→L (15%), H-2→L (72%) |
| | $S_4$ | 498.0 | 0.489 | 2.490 | H-5→L (30%), H-3→L (43%), H-2→L (21%) |
| | $S_{11}$ | 360.7 | 0.667 | 3.437 | H→L+1 (65%) |
| | $S_{19}$ | 295.6 | 0.728 | 4.194 | H-17→L (28%), H-13→L (17%), H-2→L+2 (16%) |
| DCPP-TES-8 | $S_3$ | 589.3 | 0.120 | 2.104 | H-3→L (12%), H-2→L (75%) |
| | $S_4$ | 490.8 | 0.506 | 2.526 | H-5→L (21%), H-3→L (45%), H-2→L (18%) |
| | $S_8$ | 377.1 | 0.755 | 3.288 | H→L+1 (68%) |
| | $S_{17}$ | 305.1 | 0.544 | 4.064 | H-9→L (32%), H-2→L+2 (13%), H-1→L+3 (11%) |
| DCPP-TES-9 | $S_1$ | 426.4 | 0.556 | 2.908 | H→L (93%) |
| | $S_7$ | 337.2 | 0.241 | 3.677 | H-6→L (12%), H-4→L (17%), H-3→L+1 (11%), H-2→L (16%), H-1→L+3 (11%) |
| | $S_{14}$ | 295.4 | 1.016 | 4.197 | H-5→L+1 (40%), H-4→L+2 (25%), H-2→L (12%) |
| DCPP-TES-10 | $S_3$ | 654.8 | 0.201 | 1.893 | H-2→L (76%), H-1→L (11%) |
| | $S_6$ | 516.1 | 0.244 | 2.403 | H-8→L (21%), H-6→L (60%) |
| | $S_{12}$ | 344.6 | 0.730 | 3.598 | H→L+1 (78%) |
| DCPP-TES-11 | $S_3$ | 694.6 | 0.182 | 1.785 | H-2→L (73%), H-1→L (14%) |
| | $S_5$ | 598.5 | 0.103 | 2.072 | H-4→L (87%) |
| | $S_6$ | 543.6 | 0.208 | 2.281 | H-10→L (12%), H-6→L (66%) |
| | $S_{12}$ | 352.6 | 0.499 | 3.516 | H-16→L (23%), H-1→L+3 (12%), H→L+1 (15%), H→L+2 (15%) |
| DCPP-TES-12 | $S_3$ | 688.8 | 0.238 | 1.800 | H-2→L (66%), H-1→L (20%) |
| | $S_6$ | 527.3 | 0.180 | 2.351 | H-5→L (30%), H-4→L (56%) |
| | $S_{11}$ | 382.6 | 0.538 | 3.241 | H-1→L+2 (16%), H→L+1 (64%) |

H: Highest occupied molecular orbital (HOMO); L: lowest unoccupied molecular orbital (LUMO).